\documentclass[12pt]{article}

\usepackage[margin=1in]{geometry}
\usepackage{setspace}
\usepackage{amsmath, amssymb, amsfonts, bm, bbm}
\usepackage{amsthm}
\usepackage{natbib}
\usepackage{float}

\usepackage{graphicx}
\graphicspath{{Fig/}}
\usepackage{booktabs}
\usepackage{multirow}
\usepackage{dcolumn}
\newcolumntype{d}[1]{D{.}{.}{#1}}
\usepackage{hhline}
\usepackage{colortbl}
\usepackage{pdflscape}
\usepackage{multirow}
\usepackage{makecell}
\usepackage{xcolor}
\usepackage{verbatim}
\usepackage{enumerate}
\usepackage{url}
\usepackage[hidelinks]{hyperref}
\usepackage{xurl}           % lets URLs break almost anywhere
\hypersetup{breaklinks=true}
\usepackage{pifont}
\usepackage[normalem]{ulem}
\usepackage{times}

\usepackage[english]{babel}
\usepackage[hidelinks]{hyperref}
\usepackage{cleveref}

\theoremstyle{plain}
\newtheorem{theorem}{Theorem}[section]
\newtheorem{proposition}[theorem]{Proposition}
\newtheorem{lemma}{Lemma}
\newtheorem{corollary}[theorem]{Corollary}
\theoremstyle{definition}
\newtheorem{definition}[theorem]{Definition}

\theoremstyle{remark}
\newtheorem{condition}{Condition}
\newcommand\norm[1]{\left\lVert#1\right\rVert}
\newtheorem{remark}[theorem]{Remark}

\DeclareMathOperator{\expit}{expit}

\newcommand{\simiid}{\,{\buildrel \text{iid} \over \sim}\,}

\title{\Large \bf Towards Robust Matched Observational Studies with General Treatment Types: Consistency, Efficiency, and Adaptivity}

\author{%
  \large
  Siyu Heng$^{1}$\thanks{Siyu Heng and Elaine K. Chiu contributed equally to this work.}%
  \thanks{Address for Correspondence: Siyu Heng, Department of Biostatistics, New York University, New York, NY 10003, U.S.A. (Email: siyuheng@nyu.edu); Hyunseung Kang, Department of Statistics, University of Wisconsin-Madison, Madison, WI 53706, U.S.A. (Email: \href{mailto:hyunseung@stat.wisc.edu}{hyunseung@stat.wisc.edu}).},\;
  Elaine K. Chiu$^{2}$\footnotemark[1],\;
  Hyunseung Kang$^{2}$\footnotemark[2] \\[0.5em]
  \textit{$^{1}$Department of Biostatistics, New York University, U.S.A.} \\
  \textit{$^{2}$Department of Statistics, University of Wisconsin-Madison, U.S.A.}
}

\date{} % empty date

\begin{document}
\maketitle
\vspace{-2em}

\begin{abstract}
To ensure reliable causal conclusions from observational studies, researchers routinely conduct sensitivity analysis to assess robustness to unmeasured confounding. In matched observational studies (one of the most popular observational study designs), two foundational concepts, design sensitivity and Bahadur-Rosenbaum efficiency, are used to quantify the robustness of test statistics and study designs in sensitivity analyses. Unfortunately, these measures of robustness are unavailable for non-binary treatments (e.g., continuous treatments) and consequently, prevailing recommendations about robust tests may be misleading. In this work, we provide a unified framework to quantify the robustness of test statistics and study designs for any treatment type. We first present a negative result about a popular, ad-hoc approach based on dichotomizing the treatment variable. Next, we consider two complementary parameterizations of the sensitivity parameter that apply to arbitrary treatment types and establish a one-to-one correspondence between them. Using these equivalent parameterizations, we generalize design sensitivity and Bahadur-Rosenbaum efficiency and derive unified formulas for both quantities under general treatment types. We also propose a general data-adaptive approach that combines candidate test statistics to enhance robustness against unmeasured confounding. Our results yield new insights into the robustness of tests and study designs in matched observational studies with non-binary treatments.
\end{abstract}


\section{Introduction}\label{sec: introduction}

\subsection{Background: Matched Observational Studies and Measuring Robustness Under Unmeasured Confounding With Test Consistency and Efficiency}\label{sec: background}

In observational (non-randomized) studies, unmeasured confounders are often present, making it generally unrealistic to assume that treatment assignment is random, even after adjusting for measured confounders through methods such as matching or stratification. More importantly, if the bias from unmeasured confounding is inevitable, it is critical to choose a statistical procedure (e.g., a test statistic in a hypothesis testing procedure) that is robust to such bias. In matched observational studies (one of the most widely used observational study designs), there is an extensive literature on measuring robustness against unmeasured confounding through the lens of sensitivity analysis. Briefly, sensitivity analysis seeks to study the impact of unmeasured confounding bias on estimates or tests of the treatment effect. We focus on two measures of this impact: \textit{design sensitivity} \citep{rosenbaum2004design} and \textit{Bahadur-Rosenbaum efficiency} \citep{rosenbaum2015bahadur}. Specifically, the design sensitivity of a test statistic is the largest level of unmeasured confounding under which the test remains consistent (i.e., its power tends to one under the alternative). The Bahadur-Rosenbaum efficiency (exact slope) measures the exponential rate at which the worst-case sensitivity analysis $p$-value tends to zero under a fixed alternative as the number of matched pairs increases; tests can be compared via the Bahadur-Rosenbaum relative efficiency, defined as the ratio of their exact slopes. See Section~\ref{subsec: design sensitivity and efficiency in binary treatment} for details and Figure~\ref{fig:contribution_figure} for a visual illustration. Loosely speaking, a test statistic with a high design sensitivity is preferred as the test can detect the presence of a nonzero treatment effect even with a strong unmeasured confounder. Similarly, a test statistic with higher Bahadur-Rosenbaum efficiency is preferred as it can detect the nonzero treatment effect with a small sample size.

These two measures of robustness have been used in several works to choose tests that are less sensitive to unmeasured confounding (e.g., \citealp{heller2009split, stuart2013commentary, zubizarreta2013designing, hansen2014clustered, ertefaie2018quantitative, zhao2019sensitivityvalue, rosenbaum2020design, fogarty2021biased, howard2021uniform, karmakar2019integrating, zhang2021selecting, zhang2024sensitivity, ye2025combining}). For instance, under binary treatments, existing work \citep{Rosenbaum2011Ustatistics, rosenbaum2020design} has shown that test statistics with superior power under no unmeasured confounding, such as the paired $t$-test, may have poor design sensitivity and thus poor robustness against unmeasured confounding. These insights have shaped recommendations for test selection and are embedded in commonly used \textsf{R} packages \citep{rosenbaum2015two}.

\subsection{Motivation: Sensitivity Analysis with Non-Binary Treatments and Failure of Ad-Hoc Procedures Based on Dichotomization} 

Despite hundreds of works on design sensitivity and Bahadur-Rosenbaum efficiency, to the best of our knowledge, all existing works have been confined to binary treatments; the one exception is the work by \citet{zhang2024sensitivity} (see Table~\ref{tab:contribution}). We believe one major reason for the paucity of works on non-binary treatment is due to an important technical hurdle in sensitivity analysis that is only present in non-binary treatments. Specifically, under non-binary (e.g., continuous or ordinal) treatments, an individual's propensity to take one dose over another may depend on her/his unmeasured confounder, the magnitude of the difference between the two doses, or both. In contrast, under binary treatments, the magnitude of the difference between the two doses is identical (i.e., $1$), and thus the propensity only depends on the unmeasured confounder after adjusting for measured confounders. Thus, under binary treatment, the ``worst-case'' data-generating process depends only on the unmeasured confounder, while under non-binary treatment it also depends on dose differences between units; see Section~\ref{sec:review rosenbaum} for details.

In practice, a common workaround for non-binary treatments (e.g., ordinal or continuous treatments) is to dichotomize the treatment and then apply existing methods for binary treatments, such as the classical Rosenbaum sensitivity bounds in (\ref{eqn: binary-treatment Rosenbaum bounds}). However, dichotomization discards treatment dose information and generally cannot recover the sharp pair-specific sensitivity bounds (see Theorem~\ref{prop: impossible} for details). Consequently, a sensitivity analysis based on the dichotomized treatment may either be invalid (i.e., not guaranteed to control the Type I error rate) or, when the common sensitivity parameter is chosen sufficiently large to ensure validity, be overly conservative and have unnecessarily low statistical power. Either case may lead to misleading conclusions.

For example, sensitivity analyses for matched encouragement designs with a continuous encouragement variable have often dichotomized the encouragement and, based on the resulting analyses, concluded that increasing the within-pair encouragement dose difference would almost always improve robustness. By discarding the treatment dose information and its association with unmeasured confounding, these analyses fail to account for the possibility that the magnitude of within-pair unmeasured confounding may vary with the encouragement dose difference. In practice, a larger encouragement dose difference may also be associated with a greater within-pair imbalance in unmeasured confounders. Increasing the encouragement dose difference therefore improves robustness only when the benefit of strengthening the within-pair encouragement dose contrast outweighs the accompanying increase in unmeasured confounding bias \citep{heng2023instrumental}; see Appendix D.6 for a detailed discussion.

The lack of formal methods to assess unmeasured confounding bias with non-binary treatment has critical, practical implications, and we highlight one concrete example. Historically, many existing works (e.g., \citealp{rosenbaum1997signed, rosenbaum2004design, zhang2022bridging, zhang2023statistical}) have advocated for incorporating treatment dose into test statistics (e.g., dose-weighted statistics) to improve robustness under non-binary treatments. However, to the best of our knowledge, this recommendation lacked theoretical justification. In particular, there was no theory quantifying the gains in robustness of a test that uses treatment dose information relative to one that does not. With our newly developed tools below, we can now make formal comparisons between such tests and show that the prevailing advice may fail to improve robustness. In particular, for a dose-response curve where its derivative is decreasing as the dose increases, it is actually harmful to incorporate treatment dose in a test where such a test is more sensitive to unmeasured confounding biases than a test that does not include treatment dose; see Section~\ref{sec:design_sensitivity_simulation_studies}.

\subsection{Our Contributions: A Unified Sensitivity Analysis for Binary and Non-Binary Treatments}

The main goal of the paper is to address this critical gap in evaluating the robustness of tests in matched observational studies with non-binary treatments. Specifically, we generalize design sensitivity and Bahadur-Rosenbaum efficiency to accommodate arbitrary treatment types, including binary, ordinal, and continuous treatments. In other words, we propose a unified, all-in-one solution to evaluate robustness against unmeasured confounding that is agnostic to treatment type. Along the way, we state auxiliary results that complement our study on robustness. Our main contributions are as follows:
\begin{itemize}
\item First, we formally show that the common practice of dichotomizing a non-binary treatment into a binary variable leads to sensitivity bounds that are incompatible for a family of allowable distributions under a sensitivity analysis with continuous treatment; see Section \ref{subsec: Motivation: Inapplicability of the original Rosenbaum bounds} and Theorem \ref{prop: impossible}. Practically, this means that the sensitivity bounds based on the derived binary variables may not encapsulate all the effects that an unmeasured confounder may have on the treatment.
\item Second, motivated by the negative result above, we consider two complementary sensitivity parameterizations, $\gamma$ and $\overline{\Gamma}$, that apply to both binary and non-binary treatments; see Section~\ref{sec: design sensitivity} and Proposition~\ref{prop:sufficiency}. Their one-to-one correspondence allows the robustness of tests to be quantified on both the population-level $\gamma$-scale and the post-matching (data-specific) $\overline{\Gamma}$-scale, regardless of treatment type.
\item Third, building on these complementary sensitivity parameterizations and their one-to-one correspondence, we generalize design sensitivity and Bahadur-Rosenbaum efficiency (illustrated in Figure~\ref{fig:contribution_figure}) to accommodate arbitrary treatment types; see Sections~\ref{sec: design sensitivity} and~\ref{sec:bahadur}. Equally important, we develop novel techniques that, for the first time, yield explicit formulas for design sensitivity and Bahadur-Rosenbaum efficiency across a wide range of treatment types (binary, ordinal, and continuous) and outcome types; see Theorems~\ref{thm: design sensitivity of the general signed rank test} and~\ref{thm: Bahadur slope}. Our results also substantially extend those of \citet{zhang2024sensitivity}, who derived a design sensitivity formula only for continuous treatments with binary outcomes; see Table~\ref{tab:contribution} for a summary of the comparison. The main text focuses on continuous treatments, arguably the most challenging and underexplored setting, while the corresponding results for discrete treatments (e.g., ordinal treatments) are provided in Appendix~B.2.

\begin{figure}[ht]
\centering
\caption{An illustration of the generalized design sensitivity and Bahadur-Rosenbaum relative efficiency in sensitivity analysis. Figure (a) illustrates the generalized design sensitivities of two tests where test statistic $T_2$ has a higher generalized design sensitivity than test statistic $T_1$ (i.e., $\overline{\Gamma}_{*,1} < \overline{\Gamma}_{*, 2}$). Figure (b) illustrates the generalized Bahadur-Rosenbaum relative efficiency $\Upsilon_2/\Upsilon_1$ for a fixed sensitivity parameter $\overline{\Gamma}$, where $T_2$ is more efficient than $T_1$ and the generalized Bahadur-Rosenbaum relative efficiency captures the ratio of the minimal required numbers of matched pairs for $T_1$ or $T_2$ to achieve a specific power $\beta$ under some significance level $\alpha$.}
\includegraphics[width=1\textwidth]{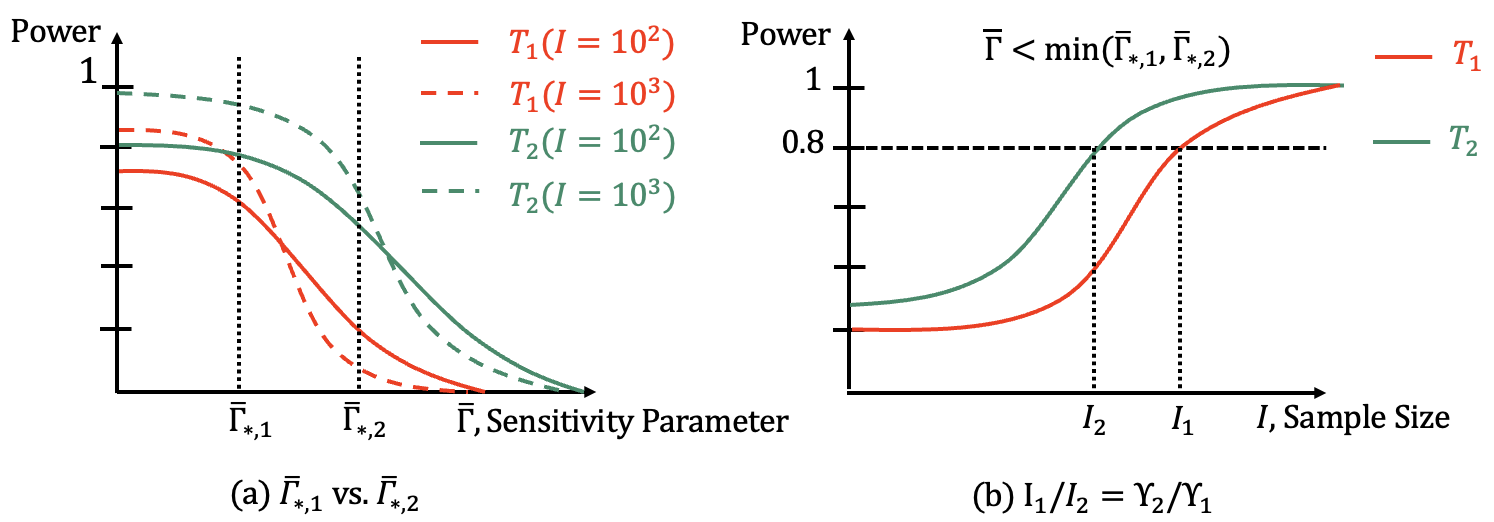}
\label{fig:contribution_figure}
\end{figure}

\begin{table}[ht]
\centering
\caption{The scope and contribution of existing works and our work in matched observational studies. In sensitivity analysis, consistency is measured by design sensitivity, and efficiency is measured by Bahadur-Rosenbaum efficiency.}
\footnotesize
\begin{tabular}{c| c c c}
\hline
\multirowcell{2}{Property of Tests} & \multirowcell{2}{Binary Treatments} & \multicolumn{2}{c}{Non-Binary Treatments}  \\ 
\cline{3-4} 
&   &  Binary Outcomes & Non-Binary Outcomes \\
\hline
Consistency  &  Rosenbaum (2004; \textit{Biometrika})  & Zhang et al. (2024; \textit{Biometrika})  & This work\\
Efficiency & Rosenbaum (2015; \textit{JASA})  & This work  &  This work \\
Adaptivity  & Rosenbaum (2012; \textit{Biometrika})  & Zhang et al. (2024; \textit{Biometrika})  &  This work \\
\hline
\end{tabular}
\label{tab:contribution}
\end{table}

\item Fourth, we use our formulas to generate new insights into the robustness of competing test statistics under a wide variety of data-generating processes. For instance, our results reveal that incorporating treatment dose into test statistics does not necessarily improve robustness in sensitivity analyses, which is contrary to common intuition. For more examples, see Sections~\ref{sec:design_sensitivity_simulation_studies}, \ref{sec:adaptive_simulation_studies}, and \ref{sec:data_analysis}, and Appendix C.

\item Fifth, to mitigate the loss of robustness from selecting the less robust of two candidate tests, we extend the binary-treatment procedure of \citet{rosenbaum2012testing} to general treatment types, developing an adaptive approach that combines two candidate tests (e.g., the original and dose-weighted versions); see Sections~\ref{sec: adaptive test method} and~\ref{sec:adaptive_simulation_studies}. We also develop a new coupling argument that rigorously establishes the asymptotic validity of the adaptive procedure and, as a special case, fills a technical gap in the original binary-treatment validity argument; see Appendices~A.5 and~D.7. We show that the adaptive procedure attains the larger of the design sensitivities and the larger of the Bahadur-Rosenbaum efficiencies of its component tests, regardless of the underlying data-generating process; see Theorems~\ref{prop: combine test valid} and~\ref{prop: adaptivity}. Simulations show that its power closely tracks that of the better-performing component test across a range of settings.
\end{itemize}

\section{Review}\label{sec: review}

\subsection{Matched Observational Studies}

We start with a review of matched observational studies for binary and non-binary treatments (e.g., \citealp{Rosenbaum1987, rosenbaum1989sensitivity, rosenbaum2002observational, gastwirth1998dual}). Let there be $I$ independent matched pairs. For individual $j=1,2$ in matched pair $i=1,\ldots,I$ (henceforth referred to as individual $ij$), let $Z_{ij}$ be the treatment variable (binary, ordinal, or continuous), $\mathbf{x}_{ij}$ be the measured confounders, $u_{ij}$ be the unmeasured confounder, and $Y_{ij}$ be the observed outcome. In a matched-pair design, two individuals $i1$ and $i2$ have the same or similar measured confounders (i.e., $\mathbf{x}_{i1}=\mathbf{x}_{i2}$ or $\mathbf{x}_{i1}\approx \mathbf{x}_{i2}$) but different treatment doses (i.e., $Z_{i1}\neq Z_{i2}$). For each matched pair $i$, let $z_{i}^{**}=\max\{Z_{i1}, Z_{i2}\}$ be the higher treatment dose and $z_{i}^{*}=\min\{Z_{i1}, Z_{i2}\}$ be the lower treatment dose. Note that under binary treatments, we have $z_{i}^{**} = 1$ and $z_{i}^{*}=0$.

Next, we review the randomization-based (design-based) inference for matched observational studies under no unmeasured confounding. Let $Y_{ij}(z)$ denote the potential outcome of individual $ij$ under the treatment dose $z$. Consider a general class of hypotheses: $H_{\boldsymbol{\beta}}: Y_{ij}(z_{i}^{**})=Y_{ij}(z^{*}_{i})+h(\boldsymbol{\beta}, z^{*}_{i}, z_{i}^{**}, \mathbf{x}_{ij})$ for all $i$ and $j$. The vector $\boldsymbol{\beta}\in \mathbb{R}^{p}$ represents some prespecified causal parameter(s) of interest, and $h$ can be any prespecified treatment effect function. When $h = 0$, $H_{\boldsymbol{\beta}}$ reduces to Fisher's sharp null of no effect, i.e., $H_{0}: Y_{ij}(z_{i}^{**})=Y_{ij}(z^{*}_{i})$ for all $i$ and $j$. When $h=\beta(z_{i}^{**}-z_{i}^{*})$, $H_{\beta}$ is testing for linear treatment effects. When $h$ contains interaction terms between $\boldsymbol{\beta}$ and $\mathbf{x}_{ij}$, $H_{\boldsymbol{\beta}}$ is testing for heterogeneous/individual treatment effects. Let $\mathbf{Z}=(Z_{11}, \dots, Z_{I2})$ and $\mathbf{Y}=(Y_{11}, \dots, Y_{I2})$ denote the vector of treatment doses and observed outcomes, respectively. Let $\mathcal{Z} = \{\mathbf{Z}: (Z_{i1}, Z_{i2})=(z_{i}^{*}, z_{i}^{**}) \text{ or } (Z_{i1}, Z_{i2})=(z_{i}^{**}, z_{i}^{*})\}$ denote all possible post-matching treatment dose assignments and let $\mathcal{F}_{0}=\{(Y_{ij}(z_{i}^{*}), Y_{ij}(z^{**}_{i}), \mathbf{x}_{ij}): i=1,\dots, I, j=1,2\}$ denote the set of potential outcomes and measured confounders. 

Under no unmeasured confounding and conditional on matching on measured confounders, the two matched individuals in pair $i$ have an equal chance of receiving the higher (or lower) dose: $P(Z_{i1}=z_{i}^{**}, Z_{i2}=z^{*}_{i}\mid \mathcal{F}_{0}, \mathcal{Z})= P(Z_{i1}=z^{*}_{i}, Z_{i2}=z_{i}^{**}\mid \mathcal{F}_{0}, \mathcal{Z})=1/2$. Furthermore, under $H_{0}$, this ``design-based'' distribution of $Z$ leads to a randomization-based $p$-value of a test statistic $T(\mathbf{Z}, \mathbf{Y})$ via $P(T\geq t \mid \mathcal{F}_{0}, \mathcal{Z}, H_{0})=|\{\mathbf{z}\in \mathcal{Z}: T(\mathbf{Z}=\mathbf{z}, \mathbf{Y})\geq t\}|/|\mathcal{Z}|$, where the cardinality $|\mathcal{Z}|=2^{I}$ and $t$ is the observed value of the test statistic $T$. For testing the general null $H_{\boldsymbol{\beta}}$, we can replace $\mathbf{Y}$ with the adjusted outcomes $\mathbf{Y}^{\text{adj}}=(Y_{11}^{\text{adj}},\dots, Y_{I2}^{\text{adj}})$, where $Y_{ij}^{\text{adj}}=Y_{ij}-\mathbbm{1}\{Z_{ij}=z_{i}^{**}\} h(\boldsymbol{\beta}, z^{*}_{i}, z_{i}^{**}, \mathbf{x}_{ij})$. In other words, testing $H_{\boldsymbol{\beta}}$ with $\mathbf{Y}$ is procedurally equivalent to testing $H_{0}$ with $\mathbf{Y}^{\text{adj}}$ and thus, without loss of generality, our arguments focus on $H_{0}$.

\subsection{Sensitivity Analysis Framework in Matched Observational Studies}
\label{sec:review rosenbaum}

Finally, we review sensitivity analysis for unmeasured confounding in matched observational studies. Formally, consider the widely used Rosenbaum sensitivity model for general treatment $Z$ \citep{Rosenbaum1987, rosenbaum1989sensitivity,  rosenbaum2002observational, gastwirth1998dual}:
\begin{equation}\label{eqn: rosenbaum sensitivity analysis model}
    P(Z=z \mid \mathbf{x}, u)=\xi(\mathbf{x}, u) \eta (z,\mathbf{x}) \exp\{\gamma \phi(z) u \}, \quad{} u \in [0,1], \quad{} \gamma \geq 0,
\end{equation}
where $\xi(\mathbf{x},u)$ and $\eta(z,\mathbf{x})$ are unspecified nuisance functions that need not be modeled or estimated, $\phi$ is a prespecified dose link function, and the unmeasured confounder $u$ is normalized to the unit interval to make the sensitivity parameter $\gamma$ more interpretable. In practice, $\phi(z)=z$ is a natural default when the influence of unmeasured confounding on dose assignment is believed to be approximately linear on the original dose scale, as in \citet{rosenbaum1989sensitivity} and \citet{gastwirth1998dual}. However, a nonlinear transformation may be more appropriate based on subject-matter knowledge. More broadly, we recommend choosing $\phi$ based on subject-matter knowledge or using a measured confounder that serves as a proxy for the unmeasured confounder to inform this choice. When several choices of $\phi$ are plausible, sensitivity analyses should be reported under each. Neither our methods nor our theoretical results depend on any particular choice of $\phi$; see Appendix~D.1 for further discussion. For notational simplicity, we assume throughout the paper that $\phi$ is bijective and monotonic. Most of the arguments and results in the paper require neither bijectivity nor monotonicity of $\phi$, although additional mild conditions on $\phi$ are needed in certain settings to avoid degenerate cases; see Appendix~D.3 for details.

In the binary treatment case, model (\ref{eqn: rosenbaum sensitivity analysis model}) reduces to the familiar Rosenbaum biased propensity score model \citep{Rosenbaum1987, rosenbaum2002observational}:
\begin{equation}\label{eqn: rosenbaum biased treatment model}
   \text{logit} P(Z=1\mid \mathbf{x}, u)=g(\mathbf{x})+\gamma \cdot u, \quad{} u \in [0,1], \quad{} \gamma \geq 0.
\end{equation}
In the continuous treatment case, by setting $\phi(z)=z$, model (\ref{eqn: rosenbaum sensitivity analysis model}) reduces to the generalized partially linear sensitivity model considered in \citet{rosenbaum1989sensitivity}, \citet{gastwirth1998dual}, and \citet{zhang2024sensitivity}, where we have $-\gamma(z_{i}^{**}-z_{i}^{*}) \leq \text{logit} \{ P(Z_{i 1}=z_{i}^{* *}, Z_{i 2}=z_{i}^{*} \mid \mathcal{F}, \mathcal{Z})\} \leq \gamma(z_{i}^{**}-z_{i}^{*})$ and $\mathcal{F}=\mathcal{F}_{0} \cup\left\{u_{11}, \ldots, u_{I2}\right\}$. In other words, the sensitivity parameter $\gamma\geq 0$ quantifies how unmeasured confounding would bias the logit (i.e., log odds) of receiving the higher (or lower) dose after matching, where the bias is proportional to the difference in the paired treatment doses $z_{i}^{**}-z_{i}^{*}$. 

For each matched pair $i$, let $\Gamma_{i} = \exp\{\gamma|\phi(z_{i}^{**})-\phi(z_{i}^{*})|\}$. The Rosenbaum sensitivity model in equation (\ref{eqn: rosenbaum sensitivity analysis model}) implies the following bounds on post-matching treatment assignment probabilities (henceforth referred to as the \textit{generalized Rosenbaum sensitivity bounds}): 
\begin{equation}\label{eqn: Rosenbaum bounds}
\frac{1}{1+\Gamma_{i}} \leq P(Z_{i 1}=z_{i}^{* *}, Z_{i 2}=z_{i}^{*} \mid \mathcal{F}, \mathcal{Z}) \leq \frac{\Gamma_{i}}{1+\Gamma_{i}}, \quad{} \Gamma_i \geq 1.
\end{equation}
Each $\Gamma_i$ represents the bias in treatment assignment within matched pair $i$. If $\Gamma_i = 1$, $\gamma=0$ and the upper and lower bounds in \eqref{eqn: Rosenbaum bounds} collapse to $1/2$, i.e., the ``bias-free'' randomization distribution of $(Z_{i1},Z_{i2})$ under no unmeasured confounding. As $\Gamma_i$ deviates from $1$, $\gamma$ deviates from $0$, and unmeasured confounding induces more bias in the distribution of the treatment variables. When the treatment is binary, under the Rosenbaum sensitivity model (\ref{eqn: rosenbaum sensitivity analysis model}), the generalized Rosenbaum sensitivity bounds (\ref{eqn: Rosenbaum bounds}) reduce to the \textit{classical Rosenbaum sensitivity bounds:}
\begin{equation}\label{eqn: binary-treatment Rosenbaum bounds}
\frac{1}{1+\Gamma } \leq P(Z_{i 1}=1, Z_{i 2}=0 \mid \mathcal{F}, \mathcal{Z}) \leq \frac{\Gamma}{1+\Gamma}
\end{equation}
by setting $\phi$ as the identity function and $\Gamma=\exp(\gamma) \geq 1$, where $\Gamma$ is the well-known Rosenbaum sensitivity parameter (\citealp{Rosenbaum1987, rosenbaum2002observational, rosenbaum2020design}).

A key distinction between the classical Rosenbaum sensitivity bounds \eqref{eqn: binary-treatment Rosenbaum bounds} and the generalized Rosenbaum sensitivity bounds \eqref{eqn: Rosenbaum bounds} is that the former impose the same upper bound, indexed by a single parameter $\Gamma$, on the magnitude of hidden bias across all matched pairs, while the latter allow this upper bound to vary with the treatment doses. In particular, each pair-specific parameter $\Gamma_i$ in \eqref{eqn: Rosenbaum bounds} can incorporate the treatment dose contrast within matched pair $i$. A large difference in treatment dose can amplify the potential impact of unmeasured confounding, while a small difference in dose can attenuate it. Consequently, a central technical goal of our work is to develop a framework for comparing test statistics when the strength of hidden bias varies with treatment doses.

In a sensitivity analysis, under each prespecified sensitivity parameter $\gamma\geq 0$, researchers typically report the worst-case (upper-bound) $p$-value, which is defined as the largest $p$-value under all possible allocations of unmeasured confounding subject to the constraints in equation (\ref{eqn: Rosenbaum bounds}). Specifically, consider a family of sign-score test statistics $T(\mathbf{Z}, \mathbf{Y})=\sum_{i=1}^{I} q_{I,i}  \mathbbm{1}\{D_{i}>0\}$, where $D_{i}=(Z_{i 1}-Z_{i 2})(Y_{i 1}-Y_{i 2})$, and $q_{I,i}\geq 0$ can be any prespecified function (score) of the absolute paired differences in treatment doses $\mathbf{Z}^{\delta}=\left(\left|Z_{11}-Z_{12}\right|, \ldots,\left|Z_{I 1}-Z_{I 2}\right|\right)$ and the absolute paired differences in outcomes $\mathbf{Y}^{\delta}=\left(\left|Y_{11}-Y_{12}\right|, \ldots,\left|Y_{I 1}-Y_{I 2}\right|\right)$. The family of sign-score tests includes many widely used tests as specific examples, such as the Wilcoxon signed rank test, the permutation $t$-test, the Mantel-Haenszel test, McNemar's test, Huber's $m$-test, and the U-statistics; see \cite{rosenbaum2002observational} and Section \ref{sec: design sensitivity} for more examples. As shown in a series of seminal works \citep{Rosenbaum1987, rosenbaum1989sensitivity, rosenbaum2002observational}, the worst-case (upper-bound) $p$-value equals $P(T^{+}\geq t\mid\mathcal{F},\mathcal{Z})$, where $t$ is the observed value of $T$. The bounding test statistic $T^{+}$ is the sum of $I$ independent random variables $T_{i}^{+}$. For pair $i$, $T_{i}^{+}$ equals $q_{I,i}$ with probability $p_{I,i}^{+}=\Gamma_i/(1+\Gamma_i)$ and zero with probability $p_{I,i}^{-}=1/(1+\Gamma_i)$ when $Y_{i1}\neq Y_{i2}$; when $Y_{i1}=Y_{i2}$, $T_{i}^{+}$ equals zero with probability one. Then, under $H_{0}$ and assuming for notational simplicity that there are no outcome ties, the worst-case $p$-value for any sign-score test statistic $T$ can be approximated as:
\begin{equation}\label{eqn: worst-case $p$-value}
    \max_{\mathbf{u} }P(T\geq t\mid \mathcal{F}, \mathcal{Z}, H_{0})\simeq 1-\Phi\left(\frac{t-\sum_{i=1}^{I}q_{I,i}p_{I,i}^{+} }{\sqrt{\sum_{i=1}^{I}q_{I,i}^{2}p_{I,i}^{+}(1-p_{I,i}^{+})}  }\right),
\end{equation}
where ``$\simeq$'' denotes asymptotically equal as $I \rightarrow \infty$ and $\Phi$ is the distribution function of the standard normal distribution. Given an observed dataset, researchers typically conduct the sensitivity analysis over an increasing sequence of sensitivity parameter values, such as $\gamma$ under the general Rosenbaum sensitivity model~(\ref{eqn: rosenbaum sensitivity analysis model}) or $\Gamma$ under the classical Rosenbaum bounds~(\ref{eqn: binary-treatment Rosenbaum bounds}). For each value, they report the corresponding worst-case $p$-value and determine the largest sensitivity parameter value for which the worst-case $p$-value remains below the prespecified significance level. This changepoint is referred to as the \textit{sensitivity value} \citep{zhao2019sensitivityvalue}. It quantifies the magnitude of unmeasured confounding bias that the evidence against the null hypothesis can withstand before the conclusion is no longer statistically significant. 

However, the worst-case $p$-value or, more generally, the null distribution of $T$, does not provide a way to assess which test statistic $T$ is more robust to unmeasured confounding. To answer this, we need to assess the statistical power of a test $T$ in a sensitivity analysis, and the next section summarizes the key concepts.

\subsection{Power of a Sensitivity Analysis, Design Sensitivity, and Bahadur-Rosenbaum Efficiency Under Binary Treatments}\label{subsec: design sensitivity and efficiency in binary treatment}

In randomized experiments where there is no unmeasured confounding, a power analysis provides valuable information about the design and analysis of the experiments. Similarly, in observational studies where unmeasured confounding bias is unavoidable, a power analysis in a sensitivity analysis provides valuable insights about the design and analysis of observational studies. Specifically, in a sensitivity analysis, to compare tests or designs, we evaluate power, the probability of correctly rejecting the no-treatment-effect null, under a prespecified data-generating model with a nonzero treatment effect, while accounting for a prespecified degree of unmeasured confounding \citep{rosenbaum2010design, rosenbaum2020design}. Akin to a power analysis in randomized experiments, a test statistic with a high power in a sensitivity analysis is preferred in observational studies; see Chapter 15 of \citet{rosenbaum2020design} for more discussion. 

In this paper, we focus on two aspects of power in a sensitivity analysis, \emph{design sensitivity} and \emph{Bahadur-Rosenbaum efficiency}. These two concepts are among the most widely used robustness measures in the design of observational studies \citep{rosenbaum2020design}, and are also closely related to other important robustness measures such as \emph{sensitivity value} \citep{zhao2019sensitivityvalue}. \citet{rosenbaum2004design} proposed \textit{design sensitivity} in matched observational studies under binary treatments. Formally, given an anticipated model with a nonzero treatment effect, the design sensitivity $\Gamma_*$ of a test is a positive, scalar number where for all $\Gamma < \Gamma_*$ (recall that $\Gamma$ is the classic Rosenbaum sensitivity parameter in the binary treatment case), the power of sensitivity analysis converges to one as sample size increases (i.e., the test is consistent) and for all $\Gamma > \Gamma_*$, the power of the sensitivity analysis converges to zero (i.e., the test is inconsistent). In other words, $\Gamma_{*}$ measures the level of consistency of a test, as it represents the threshold level of unmeasured confounding at which the test becomes inconsistent. A test with a large $\Gamma_{*}$ is preferred since the test remains consistent to detect nonzero treatment effects for large magnitudes of unmeasured confounding bias $\Gamma$. For various applications of design sensitivity, see \citet{rosenbaum2004design, rosenbaum2020design}, \citet{heller2009split}, \citet{zubizarreta2013designing}, \citet{ertefaie2018quantitative}, \citet{karmakar2019integrating}, \citet{zhao2019sensitivityvalue}, \citet{fogarty2021biased}, \citet{howard2021uniform}, \citet{zhang2021selecting, zhang2024sensitivity}, and \citet{ye2025combining}, among many others.

While design sensitivity provides a convenient, scalar measure of a test's statistical power under unmeasured confounding bias, as noted by \citet{rosenbaum2015bahadur}, it is a coarse measure since it only measures a test's consistency in sensitivity analysis and does not measure efficiency. Concretely, consider two tests $T_{1}$ and $T_{2}$ with design sensitivities $\Gamma_{*, 1}$ and $\Gamma_{*, 2}$, respectively, where $\Gamma_{*, 1} < \Gamma_{*,2}$. According to the definition of design sensitivity, when conducting sensitivity analysis for $\Gamma$ that is between the two design sensitivities (i.e., $\Gamma_{*, 1} < \Gamma < \Gamma_{*,2}$), $T_2$ is preferred over $T_1$ since $T_2$ is consistent while $T_1$ is not. However, when conducting sensitivity analysis for small to moderate $\Gamma$, specifically when $\Gamma < \min(\Gamma_{*, 1}, \Gamma_{*,2})$, both tests are consistent, and it is unclear which test should be preferred. In matched observational studies with binary treatments, \citet{rosenbaum2015bahadur} proposed what we refer to as Bahadur-Rosenbaum efficiency to provide a more refined comparison of tests' performance in a sensitivity analysis. Broadly speaking, Bahadur-Rosenbaum efficiency extends Bahadur efficiency (\citealp{bahadur1960stochastic}) from randomized experiments to observational studies with unmeasured confounding by deriving the Bahadur exact slope under unmeasured confounding bias. For various applications of Bahadur-Rosenbaum efficiency, see \citet{rosenbaum2015bahadur, rosenbaum2020design, rosenbaum2024bahadur}, \citet{ertefaie2018quantitative}, \citet{karmakar2019integrating}, \citet{heng2020finding}, and \citet{yu2024using}, among others.

\section{A Universal Framework for Demystifying Robustness of Matched Observational Studies under General Treatments}\label{sec: framework}

\subsection{A Negative Result: Inconsistency of Sensitivity Bounds Based on Dichotomization} \label{subsec: Motivation: Inapplicability of the original Rosenbaum bounds}

While the robustness of a test statistic or study design in matched observational studies has been well-understood under binary treatment, there is a paucity of work under non-binary treatment (e.g., ordinal or continuous). In particular, the original design sensitivity and Bahadur-Rosenbaum efficiency are not well-defined for non-binary treatments because, as mentioned in Section \ref{sec:review rosenbaum}, $\Gamma_i$ varies between matched sets. A simple workaround solution is to dichotomize the non-binary treatment variable, say by using the median of a continuous treatment as a cut-off point and then applying the classic design sensitivity and Bahadur-Rosenbaum efficiency based on the dichotomized treatment variable with the classic, binary-treatment sensitivity parameter $\Gamma$ (i.e., all $\Gamma_{i}$ are assumed to collapse to a single parameter $\Gamma$ after dichotomization). Although this workaround may sound appealing, we show in this section that it can produce misleading sensitivity analyses.

To formalize our result, let $P_{Z\mid\mathbf{x},u}$ denote the conditional distribution of the treatment dose $Z$ given the measured confounders $\mathbf{x}$ and the unmeasured confounder $u$, and let $\mathcal{S}$ denote the common conditional support of $Z$ given $(\mathbf{x},u)$. We define the following family of conditional treatment dose distributions:
$$
\mathcal{L}=\left\{P_{Z\mid\mathbf{x},u}: P(Z=z\mid\mathbf{x},u) \propto \eta(z,\mathbf{x})\zeta(z,u),\ z\in\mathcal{S}
\right\},$$
where $\eta(z,\mathbf{x})>0$ for every $z\in\mathcal{S}$ and every relevant value of $\mathbf{x}$, function $\zeta:\mathcal{S}\times[0,1]\to(0,\infty)$ is an unknown $C^2$ function (i.e., has continuous second-order derivatives), and $\propto$ denotes proportional to. $\mathcal{L}$ includes the widely used Rosenbaum sensitivity model (\ref{eqn: rosenbaum sensitivity analysis model}) as a specific example, and is the most general family of treatment dose models that retains the most important characteristic of the Rosenbaum sensitivity model: there are no interactions between $\mathbf{x}$ and $u$ (i.e., no $X$-$U$ interactions). The exclusion of $X$-$U$ interactions is primarily intended to isolate heterogeneity arising from treatment dose and to keep the statement and proof of Theorem~\ref{prop: impossible} transparent; it should not be interpreted as suggesting that the theorem's qualitative conclusion fails outside $\mathcal{L}$. When $X$-$U$ interactions are present, however, a sensitivity bound that is both valid and sharp generally must depend not only on the paired treatment doses but also on the measured confounders involved in those $X$-$U$ interactions \citep{heng2021sharpening}. Consequently, the classical Rosenbaum bounds in (\ref{eqn: binary-treatment Rosenbaum bounds}), which impose a common $\Gamma$ across matched pairs and account for neither heterogeneity in dose differences nor heterogeneity arising from $X$-$U$ interactions, generally cannot be both valid and sharp; see Remark~S.3 in Appendix~A.1 for further discussion.

Suppose that an investigator dichotomizes $Z$ and applies the classical binary-treatment Rosenbaum sensitivity bounds to the resulting binary treatment variable by setting all $\Gamma_i$ in \eqref{eqn: Rosenbaum bounds} equal to a common value $\Gamma$. The following result shows that sensitivity bounds based on such dichotomization cannot generally be simultaneously valid and sharp across all possible matched pairs. Specifically, at a prespecified value of the sensitivity parameter, we call sensitivity bounds \emph{valid} if they hold for every matched pair under every possible allocation of the unmeasured confounders permitted by the sensitivity model. We call the sensitivity bounds \emph{sharp} if, for every matched pair, they coincide with the tightest bounds on the post-matching treatment assignment probabilities implied by the sensitivity model.

\begin{theorem}\label{prop: impossible}
{(Impossibility of Simultaneously Valid and Sharp Classical Rosenbaum Bounds Uniformly Across All Matched Pairs After Dichotomizing a Continuous Treatment)} Let $\mathcal{S}$ denote the support of the continuous treatment $Z$, and suppose that $\mathcal{S}$ is a connected interval with nonempty interior. Fix any $P_{Z\mid\mathbf{x},u}\in\mathcal{L}$ under which $Z$ is subject to confounding by an unmeasured confounder $u$, and any prespecified $\Gamma\geq 1$. Suppose that an investigator dichotomizes $Z$ and applies the classical Rosenbaum sensitivity bounds by setting all $\Gamma_i$ in \eqref{eqn: Rosenbaum bounds} equal to the common value $\Gamma$, thereby discarding the original treatment dose information. Then the following two statements cannot both hold: (i) the sensitivity bounds \eqref{eqn: Rosenbaum bounds} are valid for all matched pairs with treatment doses $z^{*}<z^{**}$ in $\mathcal{S}$; and (ii) the sensitivity bounds \eqref{eqn: Rosenbaum bounds} are sharp for all matched pairs with treatment doses $z^{*}<z^{**}$ in $\mathcal{S}$.
\end{theorem}

All the technical proofs in this paper can be found in Appendix A, and we here give some intuition of Theorem~\ref{prop: impossible}. If an unmeasured confounder $u$ exists, it is correlated with the treatment $Z$. Furthermore, the absolute paired difference in treatment doses $|Z_{i1}-Z_{i2}|$ should be correlated with the absolute paired difference in the unmeasured confounder $|u_{i1} -u_{i2}|$. To ensure that $u$ is correlated with $Z$ for all values of $Z$, a sensitivity model must incorporate dose information. However, dichotomization discards this dose information and replaces the heterogeneous pair-specific sensitivity bounds with the classical Rosenbaum sensitivity bounds based on a common sensitivity parameter. Consequently, the resulting sensitivity bounds cannot generally be simultaneously valid and sharp uniformly over all possible matched pairs, and the resulting sensitivity analysis may either lack guaranteed Type I error rate control or be overly conservative.

\subsection{Generalized Design Sensitivity: Concept and Formulas}\label{sec: design sensitivity}

To extend design sensitivity and Bahadur-Rosenbaum efficiency to treatments of arbitrary types, we first distinguish between two parameterizations of hidden bias that will be used throughout our framework. The first is the parameter $\gamma\geq 0$ in the population-level (pre-matching) Rosenbaum sensitivity model (\ref{eqn: rosenbaum sensitivity analysis model}). The second is defined as $\overline{\Gamma}=I^{-1}\sum_{i=1}^{I}\Gamma_{i}\geq 1$, which represents the average of the pair-specific bias parameters across the $I$ matched pairs and provides a natural extension of the classical Rosenbaum sensitivity parameter $\Gamma$. For example, when $\overline{\Gamma}=1$, we have $\Gamma_i=1$ for all $i$, yielding the bias-free randomization distribution described above. As $\overline{\Gamma}$ increases above $1$, the individual $\Gamma_i$ values depart from $1$ according to their respective dose differences. Given the observed dataset, any specified value of $\overline{\Gamma}$ uniquely determines the full bias vector $(\Gamma_{1}, \dots, \Gamma_{I})$; see Appendix~A.2 for the explicit construction. This bias vector can then be used to calculate the worst-case (upper-bound) $p$-value, denoted by $p_{\overline{\Gamma}, I}$. Under binary treatment, we have $\overline{\Gamma} = \Gamma$. Therefore, in the binary-treatment setting, the proposed scalar summary $\overline{\Gamma}$ coincides with the classical Rosenbaum sensitivity parameter $\Gamma$ and retains its usual interpretation.

The following result establishes that, for a given observed dataset, the sensitivity parameters $\gamma$ and $\overline{\Gamma}$ are in one-to-one correspondence:
\begin{proposition}[One-to-One Correspondence Between the Sensitivity Parameters $\gamma$ and $\overline{\Gamma}$] \label{prop:sufficiency}
Under the Rosenbaum sensitivity model (\ref{eqn: rosenbaum sensitivity analysis model}), there is a bijection between the original sensitivity parameter $\gamma$ in (\ref{eqn: rosenbaum sensitivity analysis model}) and the induced sensitivity parameter $\overline{\Gamma}$. An explicit construction of this bijection is provided in Appendix~A.2.
\end{proposition}

Proposition~\ref{prop:sufficiency} implies that, given an observed dataset, researchers may conduct a sensitivity analysis for general treatments using either $\gamma$ or $\overline{\Gamma}$ as the sensitivity parameter. For the theoretical results developed in the remainder of this paper, we therefore provide two parallel formulations: one in terms of the pre-matching (population-level) sensitivity parameter $\gamma$ and the other in terms of the post-matching (data-specific) sensitivity parameter $\overline{\Gamma}$. These two formulations are equivalent by Proposition~\ref{prop:sufficiency}.

In practice, to provide a more comprehensive interpretation of the sensitivity analysis, we recommend reporting results in terms of both $\gamma$ and $\overline{\Gamma}$. Researchers may either first specify $\gamma$ and then calculate the corresponding $\overline{\Gamma}$ and worst-case $p$-value, or first specify $\overline{\Gamma}$ and then calculate the corresponding $\gamma$ and worst-case $p$-value. Reporting $\gamma$ alone characterizes the strength of hidden bias in the population-level sensitivity model. However, because $\gamma$ does not incorporate the realized treatment dose information, it does not by itself convey the magnitude of the pair-specific hidden bias allowed within the matched sample. The parameter $\overline{\Gamma}$ provides a complementary, data-specific summary by representing the average post-matching hidden bias parameter across the $I$ matched pairs. Conversely, because $\overline{\Gamma}$ depends on the realized dose differences, its numerical value should not be compared across studies without considering this data-specific treatment dose information. Researchers who directly specify $\overline{\Gamma}$ should therefore also report the corresponding value of $\gamma$, which places the data-specific value of $\overline{\Gamma}$ within the population-level Rosenbaum sensitivity model.

Using the two sensitivity parameters $\gamma$ and $\overline{\Gamma}$, we define generalized design sensitivity on both scales as follows:

\begin{definition}[Generalized Design Sensitivity]\label{def: generalized design sensitivity}
Consider a matched observational study with an arbitrary treatment type (e.g., binary, ordinal, or continuous) and a sensitivity analysis based on a specified test statistic under a fixed data-generating process. Under mild regularity conditions (see, for example, Theorem~\ref{thm: design sensitivity of the general signed rank test}), there exist two corresponding thresholds, $\gamma_{*}$ and $\overline{\Gamma}_{*}$, called the \textit{generalized design sensitivities} on the $\gamma$-scale and the $\overline{\Gamma}$-scale, respectively. As the sample size $I$ tends to infinity, the power of the sensitivity analysis converges to one for every $\gamma<\gamma_{*}$ and converges to zero for every $\gamma>\gamma_{*}$. Also, the power converges to one for every $\overline{\Gamma}<\overline{\Gamma}_{*}$ and converges to zero for every $\overline{\Gamma}>\overline{\Gamma}_{*}$. The thresholds $\gamma_{*}$ and $\overline{\Gamma}_{*}$ are linked through the asymptotic one-to-one correspondence between $\gamma$ and $\overline{\Gamma}$ induced by the Rosenbaum sensitivity model.
\end{definition}

Analogous to the original design sensitivity, the generalized design sensitivities $\gamma_{*}$ and $\overline{\Gamma}_{*}$ characterize the largest magnitude of unmeasured confounding under which a test remains consistent in sensitivity analysis, expressed on the population-level $\gamma$-scale and the post-matching $\overline{\Gamma}$-scale, respectively. Because the two thresholds are equivalent parameterizations of the same robustness boundary, a test statistic with a larger $\gamma_{*}$, or equivalently a larger $\overline{\Gamma}_{*}$, is preferred. When the treatment is binary, $\overline{\Gamma}_{*}$ reduces to the original design sensitivity of \citet{rosenbaum2004design}.

For example, consider the following widely used class of rank-based tests for matched-pair designs with continuous treatments \citep{rosenbaum1989sensitivity, gastwirth1998dual, zhang2022bridging}:
$$
T_{\psi}(\mathbf{Z}, \mathbf{Y})=\sum_{i=1}^{I}\psi(r_{I, i}^{z}, r_{I, i}^{y})  \mathbbm{1}\{D_{i}>0 \}, 
\begin{cases}
r_{I, i}^{z}=I^{-1}\sum_{i^{\prime}=1}^{I}\mathbbm{1}\{|Z_{i1}-Z_{i2}|\geq |Z_{i^{\prime}1}-Z_{i^{\prime}2}|\} \\
r_{I, i}^{y}=I^{-1}\sum_{i^{\prime}=1}^{I}\mathbbm{1}\{|Y_{i1}-Y_{i2}|\geq |Y_{i^{\prime}1}-Y_{i^{\prime}2}|\}.  
\end{cases}
$$
The function $\psi: [0,1]^2 \to \mathbb{R}$ is some non-negative bounded score function. The test $T_{\psi}$ can represent any rank-based test statistic that involves the rank of treatment differences $r_{I, i}^{z}$, the rank of outcome differences $r_{I, i}^{y}$, or both. Specifically, if $\psi = 1$, $T_{\psi}$ is McNemar's test (under binary outcomes) or the sign test (under continuous outcomes). If $\psi= r_{I,i}^{z}$, $T_{\psi}$ is the dose-weighted McNemar's test (under binary outcomes) or the dose-weighted sign test (under continuous outcomes) (\citealp{zhang2024sensitivity}). If $\psi=r_{I,i}^{y}$, $T_{\psi}$ is the Wilcoxon signed rank test. If $\psi=r^{z}_{I,i}r^{y}_{I,i}$, $T_{\psi}$ is the dose-weighted Wilcoxon signed rank test 
(\citealp{rosenbaum1989sensitivity, gastwirth1998dual}). If $\psi$ is a polynomial of $r_{I,i}^{z}$ and $r_{I,i}^{y}$ (including their interactions), $T_{\psi}$ contains a family of polynomial rank tests (\citealp{rosenbaum2007confidence}). If $\psi=\sum_{l=\underline{m}}^{\overline{m}}l \binom{m}{l}(r_{I,i}^y)^{l-1}(1-r_{I,i}^y)^{m-l}$ for some integers $(m,\underline{m},\overline{m})$,  $T_{\psi}$ contains a family of U-statistics indexed by $(m,\underline{m},\overline{m})$ (\citealp{Rosenbaum2011Ustatistics}). Accordingly, we refer to the test taking $\psi=\sum_{l=\underline{m}}^{\overline{m}}l \binom{m}{l}(r_{I,i}^zr_{I,i}^y)^{l-1}(1-r_{I,i}^zr_{I,i}^y)^{m-l}$ as the dose-weighted U-statistics.  

Theorem~\ref{thm: design sensitivity of the general signed rank test} presents the formula for computing generalized design sensitivity with continuous treatments and any outcome types. As mentioned earlier,  \citet{zhang2024sensitivity} derived a formula for design sensitivity under continuous treatments, but restricted to binary outcomes. In contrast, our formula in equation \eqref{eqn: formula of the generalized design sensitivity} works for any binary, discrete, and continuous outcomes. Also, Appendix B shows two complementary results: (i) the design sensitivity formula for other types of treatment variables (e.g., binary or ordinal); and (ii) a formula to compute the generalized design sensitivity for a family of test statistics that do not involve ranks (e.g., the permutation $t$-test and its extensions).
\begin{theorem}[Generalized Design Sensitivity Formula Under Continuous Treatments]\label{thm: design sensitivity of the general signed rank test}
Suppose that $(Z_{i1}, Z_{i2}, Y_{i1}, Y_{i2})$ are independent and identically distributed (iid) realizations from a multivariate distribution $F$ in which $(Z_{i1}, Z_{i2})$ are continuous. Let $F_{Z}$ and $F_{Y}$ denote the marginal distribution functions of $|Z_{i1}-Z_{i2}|$ and $|Y_{i1}-Y_{i2}|$, respectively, and let $\psi^{*}_{i}=\psi( F_{Z}(|Z_{i1}-Z_{i2}|), F_{Y}(|Y_{i1}-Y_{i2}|))$ and $\mu=E\big [\psi^{*}_{i} \mathbbm{1}\{D_{i}>0\}\big]$. Under some mild regularity conditions (see Appendix A.3 for details), the following equation for $\gamma$ (in which the function $\expit(x)=\exp(x)/\{1+\exp(x)\}$),
\begin{equation}\label{eqn: formula of the generalized design sensitivity}
    E\big[\psi^{*}_{i} \expit\{\gamma |\phi(Z_{i1})-\phi(Z_{i2})|\} \big]= \mu, 
\end{equation}
has (i) a unique solution $\gamma_{*}\in (0, +\infty)$ and (ii) the left-hand-side of \eqref{eqn: formula of the generalized design sensitivity} is strictly monotonically increasing in $\gamma$. Then, $\gamma_{*}$ is the generalized design sensitivity of $T_{\psi}$ on the $\gamma$-scale, and
$\overline{\Gamma}_{*}=E\big[\exp\{\gamma_{*}|\phi(Z_{i1})-\phi(Z_{i2})|\}\big]$
is the corresponding generalized design sensitivity on the $\overline{\Gamma}$-scale. 
\end{theorem}

Similar to the formula for the original design sensitivity in \citet{rosenbaum2004design}, equation~\eqref{eqn: formula of the generalized design sensitivity}, which determines $\gamma_{*}$ and hence the corresponding $\overline{\Gamma}_{*}$, depends only on (i) the data distribution $F$ under the alternative and (ii) the test statistic $T_\psi$. Because the left-hand side of \eqref{eqn: formula of the generalized design sensitivity} is strictly increasing in $\gamma$, its unique solution $\gamma_{*}$ can be obtained using the bisection method. The expectation in \eqref{eqn: formula of the generalized design sensitivity} can be readily approximated using Monte Carlo draws of $(Z_{i1}, Z_{i2}, Y_{i1}, Y_{i2})$ from $F$ under the alternative. Similarly, the corresponding generalized design sensitivity on the $\overline{\Gamma}$-scale can be calculated by applying the Monte Carlo method to $E\big[\exp\{\gamma_{*}|\phi(Z_{i1})-\phi(Z_{i2})|\}\big]$. For other approaches to evaluating expectations, see \citet[Chapter 7]{mcbook}. More broadly, Theorem~\ref{thm: design sensitivity of the general signed rank test}, together with the complementary results in Appendix~B, shows that generalized design sensitivity can accommodate general treatment and outcome types, as well as a broad range of test statistics, whether rank-based or not, in matched observational studies.

We conclude by highlighting a concrete practical implication of Theorem~\ref{thm: design sensitivity of the general signed rank test}. For notational simplicity, we formulate the argument in terms of $\overline{\Gamma}$; the parallel argument on the $\gamma$-scale follows immediately. Suppose that a researcher is considering two candidate test statistics, $T_{1}$ and $T_{2}$, whose generalized design sensitivities are $\overline{\Gamma}_{*, 1}$ and $\overline{\Gamma}_{*, 2}$, respectively. Without loss of generality, suppose that $\overline{\Gamma}_{*, 1}<\overline{\Gamma}_{*, 2}$. Let $p_{\overline{\Gamma}, I, 1}$ denote the worst-case $p$-value produced by $T_{1}$ under sensitivity parameter $\overline{\Gamma}$ and sample size $I$, and define $p_{\overline{\Gamma}, I, 2}$ analogously for $T_{2}$. Then, Theorem~\ref{thm: design sensitivity of the general signed rank test} implies that, for any $\overline{\Gamma}\in (\overline{\Gamma}_{*, 1}, \overline{\Gamma}_{*, 2})$ and any prespecified significance level $\alpha$, $P(p_{\overline{\Gamma}, I, 1} <\alpha \text{ or } p_{\overline{\Gamma}, I, 2} \geq \alpha)\rightarrow 0$ as $I \rightarrow \infty$. Equivalently, with probability tending to one, the test statistic $T_{1}$ with the smaller generalized design sensitivity fails to reject the null hypothesis, while the test statistic $T_{2}$ with the larger generalized design sensitivity rejects it. In particular, it is asymptotically unlikely that $T_{1}$ rejects the null hypothesis while $T_{2}$ fails to reject it. Thus, when $\overline{\Gamma}\in (\overline{\Gamma}_{*, 1}, \overline{\Gamma}_{*, 2})$, the ordering of the generalized design sensitivities determines the asymptotic performance of the competing test statistics.

\begin{remark}
The proofs of Theorems~\ref{thm: design sensitivity of the general signed rank test} and~\ref{thm: Bahadur slope} require techniques beyond those used in the binary treatment setting. Under binary treatments, the classical Rosenbaum sensitivity bounds impose a common worst-case treatment assignment probability, indexed by the common sensitivity parameter $\Gamma$, across matched pairs. Under non-binary treatments, by contrast, these probabilities vary across pairs according to their treatment-dose contrasts. Rank-based statistics introduce an additional complication because each pair's score depends on empirical ranks of the treatment dose differences and outcome differences computed from the full sample, thereby inducing dependence across pair-specific scores. We address this challenge using uniform convergence arguments based on the Glivenko-Cantelli theorem. These arguments yield the generalized design sensitivity formula, while combining them with cumulant-generating-function and large-deviation arguments yields the generalized Bahadur-Rosenbaum efficiency. To our knowledge, this combination of techniques is new to sensitivity analysis for matched observational studies with non-binary treatments; see Appendix~A for details.
\end{remark}

\subsection{Generalized Bahadur-Rosenbaum Efficiency: Concept and Formulas} \label{sec:bahadur}

Next, we extend Bahadur-Rosenbaum efficiency to accommodate arbitrary treatment types. Throughout this section and the remainder of the paper, we formulate our arguments and results primarily in terms of the post-matching sensitivity parameter $\overline{\Gamma}$ for notational simplicity. By Proposition~\ref{prop:sufficiency}, the corresponding arguments and results can be expressed equivalently in terms of the population-level sensitivity parameter $\gamma$ through the one-to-one correspondence between $\gamma$ and $\overline{\Gamma}$.

\begin{theorem}[Generalized Bahadur-Rosenbaum Efficiency]\label{thm: Bahadur slope} Consider the same setting as in Theorem~\ref{thm: design sensitivity of the general signed rank test}. Let $\omega_{0}(t)=E\Big[\log \big \{ \widetilde{p}_{i}^{+}\exp(\psi^{*}_{i}t)+(1-\widetilde{p}_{i}^{+})\big\}\Big]$ where $\widetilde{p}_{i}^{+}=\expit\{\widetilde{\gamma}|\phi(Z_{i1})-\phi(Z_{i2})|\}$ and $\widetilde{\gamma}$ is the unique value such that $E[\exp\{\widetilde{\gamma}|\phi(Z_{i1})-\phi(Z_{i2})|\}]=\overline{\Gamma}$. Also, let $t=\widetilde{t}$ denote the unique solution to $\mu=\omega_{1}(t)$ where
\begin{equation*}
    \omega_{1}(t)=E\Big[\frac{\widetilde{p}_{i}^{+}\psi^{*}_{i}\exp(\psi^{*}_{i}t)}{\widetilde{p}_{i}^{+}\exp(\psi^{*}_{i}t)+(1-\widetilde{p}_{i}^{+})} \Big].
\end{equation*}
Then, for any fixed $\overline{\Gamma}$ where $1 \leq \overline{\Gamma} < \overline{\Gamma}_{*}$, the worst-case $p$-value $p_{\overline{\Gamma}, I}$ satisfies 
\begin{equation*}
    -I^{-1}\log (p_{\overline{\Gamma}, I}) \xrightarrow{p}   \mu  \widetilde{t}-\omega_{0}(\widetilde{t}) \ \text{ as $I\rightarrow \infty$.}
\end{equation*}
Also, $\Upsilon= 2\{  \mu \widetilde{t} - \omega_{0}(\widetilde{t})\}$ is the generalized Bahadur-Rosenbaum exact slope (efficiency).
\end{theorem}

Similar to Theorem \ref{thm: design sensitivity of the general signed rank test}, the expectation terms in the formulas for the generalized Bahadur-Rosenbaum exact slope can be calculated via the Monte-Carlo method. Specifically, under each data-generating process of $(Z_{i1}, Z_{i2}, Y_{i1}, Y_{i2})$ and each test statistic, we can calculate $\mu=E\big [\psi^{*}_{i} \mathbbm{1}\{D_{i}>0\}\big]$ and $\omega_{1}(t)$ for each fixed $t$. Then, according to Lemma~S.9 in Appendix A.4, $\omega_{1}(t)$ is strictly monotonically increasing for $t$. Therefore, we can use the bisection method to find the solution of the equation $\mu=\omega_{1}(t)$ in Theorem \ref{thm: Bahadur slope}, denoted as $\widetilde{t}$. Finally, given $\widetilde{t}$, we can use the Monte-Carlo method to calculate the generalized Bahadur-Rosenbaum exact slope $\Upsilon= 2\{  \mu \widetilde{t} - \omega_{0}(\widetilde{t})\}$. 

An important design-based implication from Theorem~\ref{thm: Bahadur slope} (or the complementary results in Appendix B) is that we can compute the relative efficiency between two tests in a sensitivity analysis. Specifically, Corollary \ref{cor: efficiency} shows that the ratio of the minimum sample sizes necessary to achieve a certain level of power from the two tests in a sensitivity analysis can be asymptotically approximated by the inverse ratio of the generalized Bahadur-Rosenbaum exact slopes of the two tests. More broadly, for small to moderate levels of unmeasured confounding bias $\overline{\Gamma} < \min \{\overline{\Gamma}_{*,1},\overline{\Gamma}_{*,2}\}$, the relative change in the generalized Bahadur-Rosenbaum exact slopes is asymptotically equivalent to a relative comparison of power curves between two tests.

\begin{corollary}[Generalized Bahadur-Rosenbaum Relative Efficiency]\label{cor: efficiency}
Consider two competing test statistics $T_{1}$ and $T_{2}$. Let $I_{1}(\alpha, \beta)$ (or $I_{2}(\alpha, \beta)$) denote the minimal required number of matched pairs for $T_{1}$ (or $T_{2}$) to achieve power of at least $\beta\in (\alpha,1)$ in a sensitivity analysis at the significance level $\alpha\in (0,1/2)$ and sensitivity parameter $\overline{\Gamma}$. Let $\Upsilon_{1}$ (or $\Upsilon_{2}$) denote the generalized Bahadur-Rosenbaum exact slope of $T_{1}$ (or $T_{2}$), and let $\overline{\Gamma}_{*, 1}$ (or $\overline{\Gamma}_{*, 2}$) denote the generalized design sensitivity of $T_{1}$ (or $T_{2}$). Let $\alpha_{v}$ ($v=1,2,\dots)$ be a sequence of significance levels that converges to zero. Under the setting in Theorem~\ref{thm: Bahadur slope}, for $\overline{\Gamma}<\min\{\overline{\Gamma}_{*,1}, \overline{\Gamma}_{*,2}\}$, we have $\lim_{v\rightarrow \infty}  I_{1}(\alpha_{v}, \beta)/I_{2}(\alpha_{v}, \beta)=\Upsilon_{2}/\Upsilon_{1}$.
\end{corollary}

\subsection{Simulation Studies: Robust Designs for Observational Studies With Continuous Treatments}\label{sec:design_sensitivity_simulation_studies}

Many existing works advocate incorporating treatment dose information into test statistics in matched observational studies with non-binary treatments (e.g., \citealp{rosenbaum1997signed,rosenbaum2004design,rosenbaum2020design, zhang2022bridging, zhang2023statistical}). However, this recommendation lacks a rigorous statistical justification because of the absence of theoretical tools to compare test statistics under continuous treatments, and it is unclear whether incorporating treatment doses would \emph{always} strengthen a test in the presence of unmeasured confounding. 

In this section, we apply our new results to study this problem by examining the generalized design sensitivity and generalized Bahadur-Rosenbaum efficiency of four key test statistics (as reviewed in Section~\ref{sec: design sensitivity}): the Wilcoxon signed rank statistic (Wilcox), the dose-weighted Wilcoxon signed rank statistic (D-Wilcox), the U-statistic (U) with $(m,\underline{m},\overline{m}) = (8,7,8)$, and the dose-weighted U-statistic (D-U) with $(m,\underline{m},\overline{m}) = (8,7,8)$. Each test is evaluated under six different dose-response curves. Specifically, we generate the within-pair low dose $z_i^{*} \simiid$ Unif(0.1,1) and the high-dose-minus-low-dose difference $z_{i}^{**} - z_i^{*} \simiid$ Unif(0.1,1). Following \citet{rosenbaum1989sensitivity} and \citet{gastwirth1998dual}, we set $\phi(z)=z$ in the Rosenbaum sensitivity model \eqref{eqn: rosenbaum sensitivity analysis model}. Outcomes are then generated according to $Y_{ij}(z_{ij}) = f(z_{ij})+ \epsilon_{ij}$, where $\epsilon_{ij} \simiid \mathcal{N}(0,1)$ and $f(\cdot)$ is one of the six dose-response curves: (i) Square Dose-Response Curve: $f(z) = 0.5z^{2}$; (ii) Kink Dose-Response Curve: $f(z) = 1.5 \times \mathbbm{1}\{z \geq 0.8\}(z-0.8)$; (iii) Linear Dose-Response Curve: $f(z) = z-1.2$; (iv) Square Root Dose-Response Curve: $f(z) = 1.6 \sqrt{z} - 1.8$; (v) Flat Dose-Response Curve: $f(z) = \mathbbm{1}\{z \leq 1.2\}\times 1.2 \times z + \mathbbm{1}\{z > 1.2\} \times 1.2 \times 1.2 -1.5$; (vi) Log Dose-Response Curve: $f(z) = 0.75 \times \log (z) - 0.8$. 

The top row of Table~\ref{tb:design_sensitivity_and_simulated_power_1} reports the design sensitivities calculated from Theorem~\ref{thm: design sensitivity of the general signed rank test}, with the remaining rows showing the simulated power of sensitivity analysis. This table first serves to verify the correctness of Theorem~\ref{thm: design sensitivity of the general signed rank test}. As predicted by the proposed theory, simulated powers from the Wilcox, D-Wilcox, U, and D-U tests decrease sharply as $\overline{\Gamma}$ approaches the generalized design sensitivity $\overline{\Gamma}_*$ and become zero for $\overline{\Gamma} > \overline{\Gamma}_*$. The simulated powers in Table~\ref{tb:design_sensitivity_and_simulated_power_1} are computed with $I=5000$ matched pairs to better reflect the asymptotic behavior. Appendix~C.2 extends this comparison to $I=100, 500, 1000,$ and $5000$. The results show that, when the competing design sensitivities are well separated, the ordering induced by $\overline{\Gamma}_{*}$ generally agrees with the ordering of simulated power once $I\geq 500$. When the design sensitivities are close, substantially larger samples may be required for the anticipated ordering to become apparent. In addition, for small or moderate values of the sensitivity parameter $\overline{\Gamma}$, several tests may already have power close to one at $I=500$ or $1000$ under a fixed alternative and at a fixed significance level, making their relative performance difficult to distinguish. As the subsequent simulations illustrate, Bahadur-Rosenbaum efficiency can complement design sensitivity in such settings by providing a finer asymptotic comparison of the competing test statistics.

\begin{figure}[ht]
\centering
\caption{The dose-response curves and the corresponding generalized design sensitivities across the four competing test statistics.}
\includegraphics[width=1\textwidth]{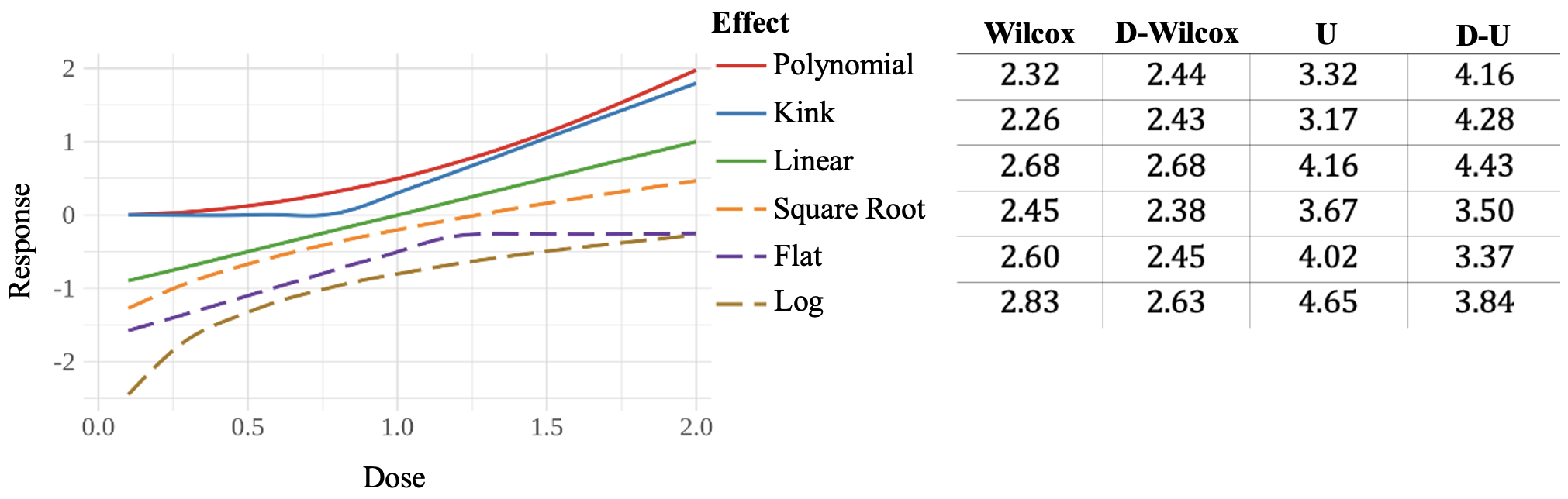}
\label{DoseResponseCurve}
\end{figure}

Next, Figure~\ref{DoseResponseCurve} compares the generalized design sensitivities across different dose-response alternatives. It reveals that the benefits of incorporating treatment dose information into tests depend critically on the shape of the dose-response curve. Tests that incorporate doses (D-Wilcox and D-U) are more robust than their non-dose counterparts (Wilcox and U) when the dose-response curve steepens with dose (e.g., the square, kink, and linear dose-response curves). Conversely, when the dose-response curve becomes flatter as the dose increases, incorporating doses leads to less robust tests, exemplified by the square root, flat, and log dose-response curves. Our data analysis in Section~\ref{sec:data_analysis} reaffirms this insight with a real data example. The intuition follows from the worst-case $p$-value formula in (\ref{eqn: worst-case $p$-value}): $\max_{\mathbf{u} }P(T\geq t\mid H_{0},\mathcal{F}, \mathcal{Z})\simeq 1-\Phi\left(\frac{t-\sum_{i=1}^{I}q_{I,i}p_{I,i}^{+} }{\sqrt{\sum_{i=1}^{I}q_{I,i}^{2}p_{I,i}^{+}(1-p_{I,i}^{+})}  }\right)$. Specifically, the worst-case $p$-value decreases as the following normalized test score increases: $\frac{t-\sum_{i=1}^{I}q_{I,i}p_{I,i}^{+} }{\sqrt{\sum_{i=1}^{I}q_{I,i}^{2}p_{I,i}^{+}(1-p_{I,i}^{+})}}$. Substituting the observed value $t=\sum_{i=1}^{I}q_{I,i}\mathbbm{1}\{D_{i}>0\}$, the normalized test score becomes $\frac{\sum_{i=1}^{I}q_{I,i}(\mathbbm{1}\{D_{i}>0\}-p_{I,i}^{+})}{\sqrt{\sum_{i=1}^{I}q_{I,i}^{2}p_{I,i}^{+}(1-p_{I,i}^{+})}}$. Therefore, in expectation, pair $i$ contributes $q_{I,i}(p_{I,i}-p_{I,i}^{+})$, where $p_{I,i}=P(D_{i}>0\mid \mathcal{F},\mathcal{Z})$ under the dose-response alternative and $p_{I,i}^{+}=\Gamma_{i}/(1+\Gamma_{i})$. A test is therefore generally more powerful when larger scores $q_{I,i}$ are assigned to pairs with larger values of $p_{I,i}-p_{I,i}^{+}$. Recall that outcomes are generated according to $Y_{ij}(z_{ij}) = f(z_{ij})+ \epsilon_{ij}$ in our simulations, where $f$ is the dose-response curve, and $p_{I, i}=P(D_{i}>0\mid \mathcal{F},\mathcal{Z})$ represents the probability that the individual with the larger treatment dose $z_{i}^{**}$ has a larger outcome than the paired individual with the smaller treatment dose $z_{i}^{*}$. Because the noise terms $\epsilon_{ij}$ have the same distribution across all pairs in our simulations, $p_{I,i}$ is expected to be larger when the dose-response curve $f$ changes more between the two doses in pair $i$, that is, when $f(z_{i}^{**})-f(z_{i}^{*})$ is larger. At the same time, $p_{I,i}^{+}=\Gamma_{i}/(1+\Gamma_{i})$ also increases with the dose difference $z_{i}^{**}-z_{i}^{*}$. Consequently, pairs with larger dose differences have larger values of $p_{I,i}-p_{I,i}^{+}$ only when $p_{I,i}$ increases faster than $p_{I,i}^{+}$. In summary, incorporating dose information improves robustness when the outcome signal increases more rapidly with the dose difference than does the considered unmeasured confounding bias, as quantified by $p_{I,i}^{+}$, but can reduce robustness when the dose-response curve flattens and the additional outcome signal fails to keep pace with the corresponding increase in $p_{I,i}^{+}$.

This dependence on the shape of the dose-response curve also suggests that power of sensitivity analysis may be improved by applying a dose transformation before constructing the test statistic. In Appendix~C.3, we use the proposed generalized design sensitivity and Bahadur-Rosenbaum efficiency to compare and select such transformations for candidate test statistics. The results suggest that transformations accentuating the curvature of the dose-response curve, for example by assigning the largest dose weights to matched pairs in its steepest region, can often increase a test's power of sensitivity analysis (i.e., robustness to unmeasured confounding). Thus, choosing the dose transformation to follow the qualitative shape of the dose-response curve is generally a sound strategy. However, the optimal transformation need not match the curve exactly: for a mildly concave dose-response curve, for instance, a more strongly concave transformation may perform better; see Appendix~C.3 for details.

\begin{table}[!htbp]
\caption{The generalized design sensitivities \texorpdfstring{$\overline{\Gamma}_{*}$}{Gamma*} 
 computed from Theorem~\ref{thm: design sensitivity of the general signed rank test} for the Wilcox, D-Wilcox, U, and D-U tests. Below the row of \texorpdfstring{$\overline{\Gamma}_{*}$}{Gamma*} 
, we report the simulated power under $I = 5000$ and various values of $\overline{\Gamma}$ at $\alpha = 0.01$.}
\centering
\footnotesize
\setlength{\tabcolsep}{5pt}
\begin{tabular}{l | c | c c | c c }
\hline
\textbf{Dose-Response Relationship} & $\overline{\Gamma}$ & Wilcox & D-Wilcox & U & D-U \\
\hline\hline 
\multirow{7}{*}{\textbf{Square}} 
& $\overline{\Gamma}_*$ & 2.32 & 2.44 & 3.32 & 4.16 \\
\hline
& 1.50 & 1.00 & 1.00 & 1.00 & 1.00 \\
& 2.00 & 0.86 & 1.00 & 1.00 & 1.00 \\
& 2.50 & 0.00 & 0.00 & 0.97 & 1.00 \\
& 3.00 & 0.00 & 0.00 & 0.14 & 0.91 \\
& 3.50 & 0.00 & 0.00 & 0.00 & 0.23 \\
& 4.00 & 0.00 & 0.00 & 0.00 & 0.01 \\
\hline\hline 
\multirow{7}{*}{\textbf{Kink}} 
& $\overline{\Gamma}_*$ & 2.26 & 2.43 & 3.17 & 4.28 \\
\hline
& 1.50 & 1.00 & 1.00 & 1.00 & 1.00 \\
& 2.00 & 0.66 & 1.00 & 1.00 & 1.00 \\
& 2.50 & 0.00 & 0.00 & 0.87 & 1.00 \\
& 3.00 & 0.00 & 0.00 & 0.05 & 0.95 \\
& 3.50 & 0.00 & 0.00 & 0.00 & 0.36 \\
& 4.00 & 0.00 & 0.00 & 0.00 & 0.02 \\
\hline\hline 
\multirow{7}{*}{\textbf{Linear}} 
& $\overline{\Gamma}_*$ & 2.68 & 2.68 & 4.16 & 4.43 \\
\hline
& 1.50 & 1.00 & 1.00 & 1.00 & 1.00 \\
& 2.00 & 1.00 & 1.00 & 1.00 & 1.00 \\
& 2.50 & 0.17 & 0.20 & 1.00 & 1.00 \\
& 3.00 & 0.00 & 0.00 & 0.98 & 0.97 \\
& 3.50 & 0.00 & 0.00 & 0.40 & 0.47 \\
& 4.00 & 0.00 & 0.00 & 0.03 & 0.06 \\
\hline\hline 
\multirow{7}{*}{\textbf{Square root}} 
& $\overline{\Gamma}_*$ & 2.45 & 2.38 & 3.67 & 3.50 \\
\hline
& 1.50 & 1.00 & 1.00 & 1.00 & 1.00 \\
& 2.00 & 0.99 & 0.98 & 1.00 & 1.00 \\
& 2.50 & 0.00 & 0.00 & 1.00 & 0.94 \\
& 3.00 & 0.00 & 0.00 & 0.59 & 0.20 \\
& 3.50 & 0.00 & 0.00 & 0.03 & 0.00 \\
& 4.00 & 0.00 & 0.00 & 0.00 & 0.00 \\
\hline\hline 
\multirow{7}{*}{\textbf{Flat}} 
& $\overline{\Gamma}_*$ & 2.60 & 2.45 & 4.02 & 3.37 \\
\hline
& 1.50 & 1.00 & 1.00 & 1.00 & 1.00 \\
& 2.00 & 1.00 & 1.00 & 1.00 & 1.00 \\
& 2.50 & 0.06 & 0.00 & 1.00 & 0.88 \\
& 3.00 & 0.00 & 0.00 & 0.93 & 0.12 \\
& 3.50 & 0.00 & 0.00 & 0.24 & 0.00 \\
& 4.00 & 0.00 & 0.00 & 0.01 & 0.00 \\
\hline\hline 
\multirow{7}{*}{\textbf{Log}} 
& $\overline{\Gamma}_*$ & 2.83 & 2.63 & 4.65 & 3.84 \\
\hline
& 1.50 & 1.00 & 1.00 & 1.00 & 1.00 \\
& 2.00 & 1.00 & 1.00 & 1.00 & 1.00 \\
& 2.50 & 0.58 & 0.10 & 1.00 & 1.00 \\
& 3.00 & 0.00 & 0.00 & 1.00 & 0.62 \\
& 3.50 & 0.00 & 0.00 & 0.85 & 0.05 \\
& 4.00 & 0.00 & 0.00 & 0.26 & 0.00 \\
\hline
\end{tabular}
\label{tb:design_sensitivity_and_simulated_power_1}
\end{table}

Table~\ref{tb:bahadur} reports the generalized Bahadur-Rosenbaum efficiencies for small or moderate $\overline{\Gamma}$. At $\overline{\Gamma}=1.00$ and $1.10$, D-Wilcox has the highest generalized Bahadur-Rosenbaum efficiency under all six dose-response curves, even though it does not generally have the highest generalized design sensitivity and has the lowest under the square root, flat, and log dose-response curves. As $\overline{\Gamma}$ increases, the most efficient test (i.e., the test with the highest Bahadur-Rosenbaum efficiency) varies across dose-response curves, further demonstrating that no single test statistic is uniformly optimal across all settings.

\begin{table}[!htbp]
\caption{Each cell shows the generalized Bahadur-Rosenbaum relative efficiency of the Wilcoxon signed rank statistic (Wilcox) compared to the dose-weighted Wilcoxon signed rank statistic (D-Wilcox), $\Upsilon_{\text{Wilcox}}/\Upsilon_{\text{D-Wilcox}}$, or that of the U-statistic (U) compared to the dose-weighted U-statistic (D-U), $\Upsilon_{\text{U}}/\Upsilon_{\text{D-U}}$. The value in parentheses indicates the simulated ratio of the minimum number of matched pairs required for the dose-weighted Wilcoxon signed rank statistic relative to the original Wilcoxon signed rank statistic, $I_{\text{D-Wilcox}}/I_{\text{Wilcox}}$, or for the dose-weighted U-statistic relative to the original U-statistic, $I_{\text{D-U}}/I_{\text{U}}$, to achieve 95\% power at significance level $\alpha = 0.01$. We further label which statistic will give the highest generalized Bahadur-Rosenbaum exact slope $\Upsilon$ at each $\overline{\Gamma}$. }
\centering  
\small
\setlength{\tabcolsep}{3.2pt}
\begin{tabular}{c | c c c  | c c c }
\hline
\multicolumn{1}{c|}{} 
  & \multicolumn{3}{c|}{Square} 
  & \multicolumn{3}{c}{Kink} \\
\hline
$\overline{\Gamma}$ 
  & Highest $\Upsilon$ & (Wilcox, D-Wilcox) & (U, D-U) 
  & Highest $\Upsilon$ & (Wilcox, D-Wilcox) & (U, D-U)\\
\hline
1.00 & D-Wilcox & 0.77(0.77) & 1.03(1.01) & D-Wilcox & 0.74(0.71)& 0.95(0.93)\\
1.10 & D-Wilcox & 0.76(0.78) & 1.01(1.01) & D-Wilcox & 0.72(0.72)& 0.92(0.87)\\
1.30 & D-Wilcox & 0.73(0.73) & 0.98(0.97) & D-Wilcox & 0.68(0.70)& 0.87(0.87) \\
1.50 & D-U      & 0.70(0.72) & 0.93(0.92) & D-U      & 0.63(0.64) & 0.81(0.78)\\
\hline
\multicolumn{1}{c|}{} 
  & \multicolumn{3}{c|}{Linear} 
  & \multicolumn{3}{c}{Square root} \\
\hline
$\overline{\Gamma}$ 
  & Highest $\Upsilon$ & (Wilcox, D-Wilcox) & (U, D-U) 
  & Highest $\Upsilon$ & (Wilcox, D-Wilcox) & (U, D-U) \\
\hline
1.00 & D-Wilcox & 0.84(0.84) & 1.23(1.35) & D-Wilcox & 0.89(0.89)& 1.39(1.36) \\
1.10 & D-Wilcox & 0.84(0.87) & 1.24(1.22)& D-Wilcox & 0.89(0.88) & 1.40(1.37) \\
1.30 & D-Wilcox & 0.84(0.83)& 1.24(1.23)& U        & 0.91(0.96)& 1.44(1.45)\\
1.50 & U        & 0.84(0.83)& 1.25(1.30) & U        & 0.93(0.93)& 1.48(1.50)\\
\hline
\multicolumn{1}{c|}{} 
  & \multicolumn{3}{c|}{Flat} 
  & \multicolumn{3}{c}{Log} \\
\hline
$\overline{\Gamma}$ 
  & Highest $\Upsilon$ & (Wilcox, D-Wilcox) & (U, D-U) 
  & Highest $\Upsilon$ & (Wilcox, D-Wilcox) & (U, D-U) \\
\hline
1.00 & D-Wilcox & 0.93(0.94)& 1.59(1.60)& D-Wilcox & 0.95(0.92) & 1.57(1.63)\\
1.10 & D-Wilcox & 0.95(0.97)& 1.64(1.55) & D-Wilcox & 0.96(0.96) & 1.61(1.65)\\
1.30 & U        & 0.98(1.00) & 1.73(1.72) & U        & 0.99(1.00)& 1.70(1.74) \\
1.50 & U        & 1.02(1.03)& 1.85(1.93) & U        & 1.04(1.04) & 1.80(1.83)\\
\hline
\end{tabular}%
\label{tb:bahadur}
\end{table}

Overall, our results reveal an important insight: no single test achieves optimal performance across all dose-response curves and all levels of unmeasured confounding bias (measured by sensitivity parameters $\overline{\Gamma}$). More importantly, the absence of a universally optimal test means that the tools that we developed are necessary to ensure that practitioners can pick the proper test statistic for their own specific study designs. For example, if the practitioner believes that the dose-response curve is linear, a good baseline statistic would be D-U from the perspective of generalized design sensitivity. However, the D-Wilcox may still be a good choice, especially for small magnitudes of $\overline{\Gamma}$ (from the perspective of generalized Bahadur-Rosenbaum efficiency). Finer assessments can be made by computing the generalized design sensitivity and Bahadur-Rosenbaum efficiency for specific instances of the linear dose-response model.

For example, Tables~\ref{tb:design_sensitivity_and_simulated_power_1} and~\ref{tb:bahadur} show that D-Wilcox has the lowest design sensitivity among the four statistics for the log, flat, and square root dose-response curves. However, when the sensitivity parameter $\overline{\Gamma}$ is as mild as $1.00$ or $1.10$, it is the most efficient statistic among the four. This absence of a universally optimal statistic means that even practitioners who simply want to select the best-performing test statistic, without requiring the detailed quantitative comparisons that the exact values of generalized design sensitivity or the generalized Bahadur-Rosenbaum efficiency provide, still need these theoretical frameworks to identify which statistic performs best under their specific settings.

\section{An Adaptive Approach for Combining Test Statistics under General Treatment Types}\label{sec: adaptive test}

\subsection{Methodology and Properties} \label{sec: adaptive test method}

Suppose that we have two candidate test statistics, such as Wilcox and D-Wilcox. As shown in Section~\ref{sec:design_sensitivity_simulation_studies}, neither test uniformly dominates the other in terms of robustness in sensitivity analyses. Without knowledge of the true data-generating process, it is therefore difficult to determine \textit{a priori} which test will perform better. More generally, this challenge commonly arises in observational studies, where several tests may be available but no single test dominates all others across all possible settings \citep{rosenbaum2012testing, rosenbaum2020design, heller2009split}.

Several methods have been proposed to address this challenge. \citet{heller2009split} and \citet{zhang2011using} introduced a sample-splitting strategy in which one portion of the data (the planning sample) is used to select a test statistic and the remaining portion (the analysis sample) is used to conduct the test. Although conceptually straightforward, sample splitting excludes the planning sample from the subsequent inference and therefore reduces power, particularly in small or moderately sized samples. To mitigate this loss of efficiency, \citet{rosenbaum2012testing} proposed an adaptive procedure that combines two competing test statistics in matched observational studies without requiring sample splitting or prior knowledge of the data-generating process. However, this adaptive approach was developed specifically for binary treatments. Building on the results and insights from Section~\ref{sec: framework}, we extend the binary-treatment adaptive approach of \citet{rosenbaum2012testing} to general treatment types. The key is to show that the desired adaptivity property described in Theorem~\ref{prop: adaptivity} can be achieved by appropriately accounting for the joint behavior of the two competing test statistics under unmeasured confounding, as quantified by the Rosenbaum sensitivity bounds (\ref{eqn: Rosenbaum bounds}).

Specifically, let $T_{1}=\sum_{i=1}^{I}q_{I,i}\mathbbm{1}\{D_{i}>0\}$ and $T_{2}=\sum_{i=1}^{I}s_{I,i}\mathbbm{1}\{D_{i}>0 \}$ denote two competing test statistics, where $q_{I,i}\geq 0$ and $s_{I,i}\geq 0$ are the corresponding scores contributed by matched pair $i$. The corresponding bounding test statistics $T_{1}^{+}$ and $T_{2}^{+}$ for $T_{1}$ and $T_{2}$ are jointly constructed as $T_{1}^{+}=\sum_{i=1}^{I}q_{I,i}B_{i}^{+}$ and $T_{2}^{+}=\sum_{i=1}^{I}s_{I,i}B_{i}^{+}$, where $B_{i}^{+}$ are Bernoulli random variables satisfying $P(B_{i}^{+}=1)=p_{I, i}^{+}=\Gamma_i /(1+\Gamma_i)$ and $P(B_{i}^{+}=0)=1-p_{I, i}^{+}=1/(1+\Gamma_i)$. Let $\mu^{+}_{1}=E(T_{1}^{+}\mid  H_{0}, \mathcal{F}, \mathcal{Z})=\sum_{i=1}^{I}q_{I,i}p_{I,i}^{+}$ and $\mu^{+}_{2}=E(T_{2}^{+}\mid  H_{0}, \mathcal{F}, \mathcal{Z})=\sum_{i=1}^{I}s_{I,i}p_{I,i}^{+}$, and let $\sigma_{1}^{+}=\sqrt{\text{Var}(T_{1}^{+}\mid  H_{0}, \mathcal{F}, \mathcal{Z})}=\sqrt{\sum_{i=1}^{I} q^{2}_{I,i}p^{+}_{I,i}(1-p^{+}_{I,i})}$ and $\sigma_{2}^{+}=\sqrt{\text{Var}(T_{2}^{+}\mid  H_{0}, \mathcal{F}, \mathcal{Z})}=\sqrt{\sum_{i=1}^{I} s^{2}_{I,i}p^{+}_{I,i}(1-p^{+}_{I,i})}$. Under mild regularity conditions (see Appendix A.5), as $I\rightarrow \infty$, $\Big(\frac{T^{+}_{1}-\mu^{+}_{1}}{\sigma^{+}_{1}},\ \frac{T_{2}-\mu^{+}_{2}}{\sigma^{+}_{2}} \Big)$ is asymptotically distributed as $(X_{1}, X_{2})\sim \mathcal{N}\left(\left(\begin{array}{c}
0\\
0
\end{array}\right),\left(\begin{array}{cc}
1 & \mathbf{\rho}^{*} \\
\mathbf{\rho}^{*} & 1 
\end{array}\right)\right)$,
where
\begin{align*}
  \rho^{*}&=\text{corr}\left(\frac{T^{+}_{1}-\mu^{+}_{1}}{\sigma^{+}_{1}},\ \frac{T_{2}-\mu^{+}_{2}}{\sigma^{+}_{2}} \mid  H_{0}, \mathcal{F},\mathcal{Z} \right)\\
  &=\frac{\sum_{i=1}^{I} q_{I,i}s_{I,i}p^{+}_{I,i}(1-p^{+}_{I,i})}{\sqrt{\sum_{i=1}^{I} q_{I,i}^{2}p^{+}_{I,i}(1-p^{+}_{I,i})}\sqrt{\sum_{i=1}^{I} s_{I,i}^{2}p^{+}_{I,i}(1-p^{+}_{I,i})}}.
\end{align*}
Next, let $Q_{\mathbf{\rho^{*}},\alpha}$ denote the quantile satisfying $P(X_{1}\leq Q_{\mathbf{\rho^{*}},\alpha}, X_{2}\leq Q_{\mathbf{\rho^{*}},\alpha})=1-\alpha$. Using this quantile and extending the adaptive approach of \citet{rosenbaum2012testing} to general treatment types, we propose the following procedure for adaptively combining the two competing test statistics $T_{1}$ and $T_{2}$:
\begin{equation}\label{eqn: adaptive test}
    \text{we reject $H_{0}$ iff $\max\left \{\frac{T_{1}-\sum_{i=1}^{I}q_{I,i}p_{I,i}^{+}}{\sqrt{\sum_{i=1}^{I} q_{I,i}^{2}p_{I,i}^{+}(1-p_{I,i}^{+})}}, \frac{T_{2}-\sum_{i=1}^{I}s_{I,i}p_{I,i}^{+}}{\sqrt{\sum_{i=1}^{I} s_{I,i}^{2}p_{I,i}^{+}(1-p_{I,i}^{+})}}\right \} \geq Q_{\mathbf{\rho}^{*},\alpha}$.}
\end{equation}
To the best of our knowledge, the adaptive procedure (\ref{eqn: adaptive test}) is the first adaptive test for matched observational studies that accommodates treatments beyond the binary case. 

We establish two main results for the adaptive testing procedure \eqref{eqn: adaptive test}. First, Theorem~\ref{prop: combine test valid} shows that \eqref{eqn: adaptive test} asymptotically controls the Type I error rate. Second, Theorem~\ref{prop: adaptivity} shows that the adaptive testing procedure (\ref{eqn: adaptive test}) attains both the larger of the two generalized design sensitivities and the larger of the two generalized Bahadur-Rosenbaum exact slopes of its component tests, without requiring knowledge of the underlying data-generating process. 
\begin{theorem}[Type I Error Rate of Adaptive Testing Procedure] \label{prop: combine test valid}
  Under the regularity conditions in Appendix A.5, for any unknown unmeasured confounders $\mathbf{u}_{0}$ that satisfy the Rosenbaum sensitivity bounds (\ref{eqn: Rosenbaum bounds}), the Type I error rate of the adaptive testing procedure (\ref{eqn: adaptive test}) is asymptotically no greater than the prespecified significance level $\alpha \in (0,1/2)$. 
\end{theorem}
\begin{theorem}[Adaptivity of the Adaptive Testing Procedure]\label{prop: adaptivity}
Let $\overline{\Gamma}_{*,1}$ (or $\Upsilon_{1}$) and $\overline{\Gamma}_{*,2}$ (or $\Upsilon_{2}$) be the two generalized design sensitivities (or the two generalized Bahadur-Rosenbaum exact slopes) of the two tests $T_{1}$ and $T_{2}$, and let $\overline{\Gamma}_{*,1:2}$ (or $\Upsilon_{1:2}$) be the generalized design sensitivity (or the generalized Bahadur-Rosenbaum exact slope) of the adaptive testing procedure (\ref{eqn: adaptive test}) with $T_{1}$ and $T_{2}$ as the two component tests. Then, for any treatment types and any underlying data-generating processes or effect models, we have (i) $\overline{\Gamma}_{*,1: 2}= \max\{ \overline{\Gamma}_{*,1}, \overline{\Gamma}_{*,2}\}$ and (ii) $\Upsilon_{1:2}=\max\{\Upsilon_{1}, \Upsilon_{2}\}$ for $\overline{\Gamma}<\min\{ \overline{\Gamma}_{*,1}, \overline{\Gamma}_{*,2}\}$. 
\end{theorem}

Intuitively, the adaptive testing procedure rejects the null hypothesis whenever either normalized candidate test statistic exceeds the common rejection threshold $Q_{\mathbf{\rho}^{*}, \alpha}$. Because $\mathbf{\rho}^{*}\in (0,1)$, Slepian's lemma implies that $Q_{\mathbf{\rho}^{*}, \alpha}\in \big(\Phi^{-1}(1-\alpha),\, \Phi^{-1}(1-\alpha/2)\big)$. Therefore, there exists some $\alpha^{*}\in [\alpha/2,\, \alpha]$ such that $Q_{\mathbf{\rho}^{*}, \alpha}=\Phi^{-1}(1-\alpha^{*})$. The adaptive procedure can thus be viewed as applying both candidate tests at the common significance level $\alpha^{*}$ and rejecting the null hypothesis whenever either test rejects. As shown in Theorems~\ref{thm: design sensitivity of the general signed rank test} and~\ref{thm: Bahadur slope}, neither the generalized design sensitivity nor the Bahadur-Rosenbaum exact slope depends on the fixed significance level. Consequently, combining the two candidate test statistics changes only their common rejection threshold, through the adjustment from $\alpha$ to $\alpha^{*}$, without changing their generalized design sensitivities or Bahadur-Rosenbaum exact slopes. Because the adaptive testing procedure rejects whenever either candidate test rejects, it attains the larger generalized design sensitivity and the larger Bahadur-Rosenbaum exact slope of the two candidate tests while controlling the Type I error rate; see Appendix A.5 for the detailed proof. Thus, in practice, the adaptive procedure allows researchers to asymptotically attain the performance of the better candidate test, regardless of the underlying data-generating process, and without inflating the Type I error rate. In Appendix D.5, we extend the adaptive procedure to combine more than two candidate test statistics.

\subsection{Simulation Studies: Avoiding Poor Test Selection Using the Proposed Adaptive Testing Procedure}
\label{sec:adaptive_simulation_studies}

Table~\ref{tb:design_sensitivity_and_simulated_power} reports the simulated powers of the two adaptive tests, one combining Wilcox and D-Wilcox and one combining U and D-U. It confirms Theorem~\ref{prop: adaptivity} and demonstrates that the proposed adaptive testing approach can effectively address the power loss that occurs when a suboptimal test is selected (e.g., when researchers inappropriately incorporate or discard treatment dose information in their test statistics). For example, consider the performance comparison of U, D-U, and an adaptive test combining U and D-U under the log dose-response curve in Table \ref{tb:design_sensitivity_and_simulated_power}. When $\overline{\Gamma} = 3.00$ and $3.50$, the simulated power of the U test is $1.00$ and $0.85$, respectively, while the D-U test achieves only $0.62$ and $0.05$, respectively. Although the D-U test alone yields substantially lower power, the adaptive test maintains comparably high power ($1.00$ and $0.78$ under $\overline{\Gamma} = 3.00$ and $3.50$) to the U test, effectively mitigating the risk of poor test selection. Similar patterns also hold for other dose-response curves and test combinations (e.g., Wilcox and D-Wilcox), demonstrating that the adaptivity property proved in Theorem~\ref{prop: adaptivity} is agnostic to unknown dose-response relationships. The simulations in Table~\ref{tb:design_sensitivity_and_simulated_power} use a large number of matched pairs ($I=5000$) to highlight the asymptotic behavior predicted by Theorem~\ref{prop: adaptivity}. To examine performance at more moderate sample sizes and assess the finite-sample cost of adaptivity relative to knowing the better component test in advance, Appendix~C.2 of the Supplementary Materials repeats the comparison at $I=100,500,1000,$ and $5000$. It also compares the adaptive test with the sample splitting procedure of \citet{heller2009split}. Two main findings emerge. First, the adaptive test generally attains power close to that of the better-performing component test, suggesting that the finite-sample cost of adaptivity is modest. Second, it typically outperforms sample splitting in the settings considered, in part because it retains all matched pairs for testing rather than reserving a subset for selecting the test statistic. These results complement the asymptotic theory and support the practical usefulness of the adaptive procedure at moderate sample sizes. We refer the reader to Appendix~C.2 for the complete results and further discussion. Code reproducing all numerical results is available at \url{https://github.com/Elainechiu/robust-matched-general-treatments}.

\begin{table}[!htbp]
\caption{The generalized design sensitivities $\overline{\Gamma}_{*}$ computed from Theorem~\ref{thm: design sensitivity of the general signed rank test} and Theorem~\ref{prop: adaptivity} for the Wilcox, D-Wilcox, adaptive test combining Wilcox and D-Wilcox, U, D-U, and adaptive test combining U and D-U. Below the row of $\overline{\Gamma}_{*}$, we report the simulated power under $I = 5000$ and various values of $\overline{\Gamma}$ at $\alpha = 0.01$.}
\centering
\footnotesize
\setlength{\tabcolsep}{3pt}
\begin{tabular}{l | c | c c c | c c c }
\hline
\textbf{Dose-Response Relationship}& $\overline{\Gamma}$ & Wilcox & D-Wilcox & Adaptive & U & D-U & Adaptive \\
\hline\hline 
\multirow{7}{*}{\textbf{Square}} 
& $\overline{\Gamma}_*$ & 2.32 & 2.44 & 2.44 & 3.32 & 4.16 & 4.16 \\
\hline 
& 1.50 & 1.00 & 1.00 & 1.00& 1.00 & 1.00 & 1.00\\
& 2.00 & 0.86 & 1.00 & 0.99& 1.00 & 1.00 & 1.00\\
& 2.50 & 0.00 & 0.00 & 0.00& 0.97 & 1.00 & 1.00\\
& 3.00 & 0.00 & 0.00 & 0.00& 0.14 & 0.91 & 0.85\\
& 3.50 & 0.00 & 0.00 & 0.00& 0.00 & 0.23 & 0.15\\
& 4.00 & 0.00 & 0.00 & 0.00& 0.00 & 0.01 & 0.00\\
\hline\hline 
\multirow{7}{*}{\textbf{Kink}} 
& $\overline{\Gamma}_*$ & 2.26 & 2.43 & 2.43 & 3.17 & 4.28 & 4.28 \\
\hline
& 1.50 & 1.00 & 1.00 & 1.00& 1.00 & 1.00 & 1.00\\
& 2.00 & 0.66 & 1.00 & 0.99& 1.00 & 1.00 & 1.00\\
& 2.50 & 0.00 & 0.00 & 0.00& 0.87 & 1.00 & 1.00\\
& 3.00 & 0.00 & 0.00 & 0.00& 0.05 & 0.95 & 0.92\\
& 3.50 & 0.00 & 0.00 & 0.00& 0.00 & 0.36 & 0.26\\
& 4.00 & 0.00 & 0.00 & 0.00& 0.00 & 0.02 & 0.01\\
\hline\hline 
\multirow{7}{*}{\textbf{Linear}} 
& $\overline{\Gamma}_*$ & 2.68 & 2.68 & 2.68 & 4.16 & 4.43 & 4.43 \\ \hline
& 1.50 & 1.00 & 1.00 & 1.00& 1.00 & 1.00 & 1.00\\
& 2.00 & 1.00 & 1.00 & 1.00& 1.00 & 1.00 & 1.00\\
& 2.50 & 0.17 & 0.20 & 0.20& 1.00 & 1.00 & 1.00\\
& 3.00 & 0.00 & 0.00 & 0.00& 0.98 & 0.97 & 0.99\\
& 3.50 & 0.00 & 0.00 & 0.00& 0.40 & 0.47 & 0.49\\
& 4.00 & 0.00 & 0.00 & 0.00& 0.03 & 0.06 & 0.04\\
\hline\hline 
\multirow{7}{*}{\textbf{Square root}} 
& $\overline{\Gamma}_*$ & 2.45 & 2.38 & 2.45 & 3.67 & 3.50 & 3.67 \\
\hline
& 1.50 & 1.00 & 1.00 & 1.00& 1.00 & 1.00 & 1.00\\
& 2.00 & 0.99 & 0.98 & 0.99& 1.00 & 1.00 & 1.00\\
& 2.50 & 0.00 & 0.00 & 0.00& 1.00 & 0.94 & 1.00\\
& 3.00 & 0.00 & 0.00 & 0.00& 0.59 & 0.20 & 0.51\\
& 3.50 & 0.00 & 0.00 & 0.00& 0.03 & 0.00 & 0.02\\
& 4.00 & 0.00 & 0.00 & 0.00& 0.00 & 0.00 & 0.00\\
\hline\hline 
\multirow{7}{*}{\textbf{Flat}} 
& $\overline{\Gamma}_*$ & 2.60 & 2.45 & 2.60 & 4.02 & 3.37 & 4.02 \\
\hline 
& 1.50 & 1.00 & 1.00 & 1.00& 1.00 & 1.00 & 1.00\\
& 2.00 & 1.00 & 1.00 & 1.00& 1.00 & 1.00 & 1.00\\
& 2.50 & 0.06 & 0.00 & 0.04& 1.00 & 0.88 & 1.00\\
& 3.00 & 0.00 & 0.00 & 0.00& 0.93 & 0.12 & 0.88\\
& 3.50 & 0.00 & 0.00 & 0.00& 0.24 & 0.00 & 0.19\\
& 4.00 & 0.00 & 0.00 & 0.00& 0.01 & 0.00 & 0.01\\
\hline\hline 
\multirow{7}{*}{\textbf{Log}} 
& $\overline{\Gamma}_*$ & 2.83 & 2.63 & 2.83 & 4.65 & 3.84 & 4.65 \\
\hline
& 1.50 & 1.00 & 1.00 & 1.00& 1.00 & 1.00 & 1.00\\
& 2.00 & 1.00 & 1.00 & 1.00& 1.00 & 1.00 & 1.00\\
& 2.50 & 0.58 & 0.10 & 0.52& 1.00 & 1.00 & 1.00\\
& 3.00 & 0.00 & 0.00 & 0.00& 1.00 & 0.62 & 1.00\\
& 3.50 & 0.00 & 0.00 & 0.00& 0.85 & 0.05 & 0.78\\
& 4.00 & 0.00 & 0.00 & 0.00& 0.26 & 0.00 & 0.19\\
\hline
\end{tabular}

\label{tb:design_sensitivity_and_simulated_power}
\end{table}

\section{Real Data Application: Environmental Tobacco Smoke and Lung Function}\label{sec:data_analysis}

In this section, we use the insights obtained from the simulation results to conduct a sensitivity analysis for the effect of tobacco use (i.e., a continuous treatment) on pulmonary function (i.e., a continuous outcome) from the 2011-2012 National Health and Nutrition Examination Survey (NHANES) in the United States. The data are publicly available at \url{https://wwwn.cdc.gov/nchs/nhanes/continuousnhanes/default.aspx?BeginYear=2011}. Environmental tobacco smoke (ETS) is measured by serum cotinine level, a biomarker that provides a precise measure of both active and passive smoking exposures \citep{Benowitz1996}. Pulmonary function is measured by the ratio of the forced expiratory volume in the first second (FEV$_1$) to the forced vital capacity (FVC), the latter being the total volume of air that can be forcibly exhaled after a full inhalation; a lower ratio indicates poorer (more obstructed) lung function. Previous observational studies have found that increased serum cotinine levels are associated with poorer lung function \citep{Gan2005, Flouris2013}. To investigate whether the detected association is due to an actual causal effect or due to confounding bias, we conduct a matched observational study with sensitivity analyses. 

Our study sample consists of $1065$ men who are older than $40$ years old. Because serum cotinine is highly right-skewed, we apply a monotone Box-Cox transformation ($\lambda \approx -0.02$) and use the transformed cotinine as the continuous dose; as the transformation is monotone, it does not alter the direction of the dose-response relationship or of the sensitivity analysis. We then adjust for the following measured confounders using optimal non-bipartite matching (\citealp{lu2001matching, lu2011optimal}): age, race (white or others), income-to-poverty ratio, college education, body mass index (BMI), waist circumference, and self-reported history of asthma. After matching, there are $I=479$ matched pairs. Table~S.13 in Appendix~C.4 reports the post-matching standardized differences in means of the measured confounders between the paired individuals, all of which have absolute values no greater than $0.13$, indicating that all the measured confounders are well balanced based on the commonly used threshold of 0.2 (\citealp{rosenbaum2020design, zhang2023social}).

Before presenting the sensitivity analysis results, we illustrate two ways in which the insights from Section~\ref{sec:design_sensitivity_simulation_studies} can guide the choice of test statistics in practice. If researchers are willing to conduct an outcome-informed exploratory analysis, they may examine the estimated dose-response curve to identify test statistics that are likely to be more robust to unmeasured confounding bias. Figure~\ref{fig:dose_response} displays the estimated dose-response relationship between transformed serum cotinine and $\mathrm{FEV}_1/\mathrm{FVC}$ in the matched sample. The fitted curve is nearly flat at low cotinine levels and declines increasingly steeply at higher levels, so the magnitude of its slope increases with the dose. This corresponds to the steepening dose-response regime considered in Section~\ref{sec:design_sensitivity_simulation_studies}, under which dose-weighted statistics are expected to be more robust to unmeasured confounding than their unweighted counterparts. Therefore, D-Wilcox and D-U appear to be promising test statistics for this data application. Because this assessment uses the observed outcomes, however, it should be regarded as exploratory when the same data are used for the subsequent analysis.

\begin{figure}[ht]
    \centering
       \caption{Dose-response relationship between ETS exposure and pulmonary function. Y-axis: FEV$_1$/FVC ratio. X-axis: serum cotinine after a Box-Cox transformation ($\lambda \approx -0.02$; original units ng/mL). The blue line is the cubic smoothing spline curve \citep[Chapter 2]{GreenSilverman1994} fit of FEV$_1$/FVC on transformed cotinine; its negative slope indicates that higher ETS exposure is associated with lower FEV$_1$/FVC.}
    \label{fig:dose_response}
    \includegraphics[width=0.75\textwidth]{DoseResponseCurveDatawithDots.png} % or DoseResponseCurvewithoutDots.png

\end{figure}

If researchers instead prefer to maintain an outcome-blinded analysis, they should not use an estimated dose-response curve from the same outcome data to select a single test statistic. They may instead prespecify an adaptive test that combines an unweighted statistic with its dose-weighted counterpart, such as Wilcox with D-Wilcox or U with D-U. Because the adaptive procedure calibrates its rejection threshold to account for the simultaneous use of both candidate statistics, it does not require researchers to choose between them after examining the outcomes. It therefore provides an outcome-blinded strategy for protecting against uncertainty about the shape of the dose-response curve.

Table~\ref{tb:sensitivity_value_data_analysis} presents the worst-case (upper-bound) $p$-values reported by the test statistics considered at various values of the sensitivity parameters $\overline{\Gamma}$ and $\gamma$. In addition, we report the largest sensitivity parameter value under which the worst-case $p$-value is still below the $0.05$ significance level, on both the $\overline{\Gamma}$- and $\gamma$-scales. This can be viewed as a generalization of the sensitivity value (\citealp{zhao2019sensitivityvalue}) from the binary treatment case to the continuous treatment case. Each dose-weighted statistic attains a larger sensitivity value than its unweighted counterpart (on the $\overline{\Gamma}$-scale, $1.68$ for D-Wilcox versus $1.43$ for Wilcox, and $1.80$ for D-U versus $1.65$ for U). The data application therefore provides an empirical counterpart to the theoretical and simulation findings of Section~\ref{sec:design_sensitivity_simulation_studies}, and illustrates that those findings can be used prospectively to nominate promising test statistics. Also, Table~\ref{tb:sensitivity_value_data_analysis} shows the advantage of the adaptive test proposed in Section~\ref{sec: adaptive test}. For example, the adaptive test with Wilcox and D-Wilcox as components yields a sensitivity value of $1.65$, compared with $1.43$ for Wilcox alone. This shows that the adaptive test can guard against the risk of selecting a statistic with suboptimal robustness to unmeasured confounding.

\begin{table}[ht]
\caption{Sensitivity analysis of the effect of environmental tobacco smoke on pulmonary function. For each value of the sensitivity parameters $\overline{\Gamma}$ and $\gamma$, we report the worst-case (upper-bound) $p$-value of each component test, together with each test's sensitivity value at the $0.05$ significance level, expressed on the $\overline{\Gamma}$ ($\gamma$) scale in the last row. The fifth (respectively, eighth) column indicates whether the adaptive test combining Wilcox and D-Wilcox (respectively, U and D-U) rejects $H_{0}$ ($\checkmark$) or not ($\times$). See Proposition \ref{prop:sufficiency} for the one-to-one correspondence between $\gamma$ and $\overline{\Gamma}$.}
\centering
\small
\setlength{\tabcolsep}{3.5pt}
\begin{tabular}{ccccc|ccc}
\toprule
$\overline{\Gamma}$ & $\gamma$  & Wilcox & D-Wilcox& Adaptive & U & D-U & Adaptive \\
\midrule
 $1.00$ & 0.000   &  0.000 &  0.000& \checkmark &   0.000 & 0.000 & \checkmark\\
$1.10$ & 0.015   &  0.000 &  0.000& \checkmark &   0.000 &  0.000 & \checkmark\\
 $1.20$ & 0.029  &  0.001 &  0.000 & \checkmark &  0.000 &  0.000 & \checkmark\\
$1.30$ & 0.041  &  0.007 &  0.000 & \checkmark  & 0.001 &  0.001 & \checkmark\\
$1.40$ & 0.052  &  0.034 &  0.001 & \checkmark & 0.004 &  0.002 & \checkmark\\
$1.50$ & 0.062   &  0.109 &  0.004 & \checkmark  & 0.013 &  0.006 & \checkmark\\
 $1.60$ & 0.071  &  0.244 &  0.020 & \checkmark  & 0.034 &  0.014 &  \checkmark\\
$1.70$ & 0.080  &  0.423 &  0.063 & $\times$  &0.071 &  0.029 & \checkmark\\
 $1.80$ & 0.088  &  0.605 &  0.149 & $\times$  & 0.128 &  0.051 &  $\times$ \\
\midrule
\multicolumn{2}{c}{Sensitivity Value}
& 1.43 (0.055) & 1.68 (0.078) & 1.65 (0.075) & 1.65 (0.075) & 1.80 (0.087) & 1.70 (0.080) \\
\bottomrule
\end{tabular}
\label{tb:sensitivity_value_data_analysis}
\end{table}

\section{Discussion: Extensions and Future Directions}

In this work, we focus on sensitivity analysis for pair-matched observational studies with general treatment and outcome types. For binary outcomes, \citet{zhang2024sensitivity} developed a valid and sharp sensitivity analysis for observational studies with continuous treatments under general matching designs. They also derived the associated design sensitivity formula and developed an adaptive testing procedure, but did not investigate Bahadur-Rosenbaum efficiency or characterize the adaptive procedure under this efficiency criterion. One promising direction for future research is therefore to integrate the Bahadur-Rosenbaum efficiency results and techniques developed in our work with the sensitivity analysis framework of \citet{zhang2024sensitivity}. Such an integration may yield a corresponding Bahadur-Rosenbaum efficiency formula for general matched observational studies with continuous treatments and binary outcomes.

For continuous outcomes, \citet{zhang2024sensitivity} showed that obtaining a valid and sharp sensitivity analysis with continuous treatments and outcomes under matching with multiple controls requires solving an NP-hard signomial programming problem. Subsequently, \citet{zhang2025bridging} developed a valid but conservative sensitivity analysis for observational studies with general treatment types under general matching designs. Specifically, for testing Fisher's sharp null hypothesis, \citet{zhang2025bridging} proposed a linear programming approach that yields a conservative upper bound on the sharp worst-case $p$-value in sensitivity analysis. Two main obstacles arise in extending the design sensitivity and Bahadur-Rosenbaum efficiency formulas, as well as the adaptive procedure, to the settings considered by \citet{zhang2025bridging}. First, our arguments and results rely heavily on closed-form expressions for the worst-case $p$-value in sensitivity analysis, while the linear programming approach of \citet{zhang2025bridging} generally does not yield such an expression. Second, the primary source of power and efficiency loss under their approach is the inherent conservativeness of the linear programming bound. Design sensitivity and Bahadur-Rosenbaum efficiency may therefore be most informative once a sharp or nearly sharp sensitivity analysis becomes available, at which point the primary goal shifts to further improving statistical power by selecting the most robust candidate test statistic or adaptively combining multiple candidate test statistics.

Another potential route for extending our work beyond pair matching is to build on the recent work of \citet{fogarty2025tilted}. Specifically, \citet{fogarty2025tilted} proposed a novel tilted sensitivity model that yields a closed-form expression for the sharp worst-case $p$-value in sensitivity analysis under general matching designs. Integrating the approach in \citet{fogarty2025tilted} with our framework may provide a tractable path toward extending the tilted sensitivity model to general treatment types and deriving corresponding robustness measures analogous to generalized design sensitivity and Bahadur-Rosenbaum efficiency.

Our framework may also be extended beyond matched observational studies. The concepts of design sensitivity and Bahadur-Rosenbaum efficiency have recently been extended beyond pair-matched designs to other important observational study frameworks, including observational block designs \citep{rosenbaum2024bahadur, rosenbaum2025conditioning}, weighting-based observational studies \citep{huang2025design}, and controlled pre-post designs \citep{leavitt2025averaged}. These developments, however, remain limited to binary treatments. Our results may offer useful insights for extending these robustness measures to general treatment types in these broader observational study settings.

\section*{Acknowledgments}

This work was supported in part by NIH Grant R21DA060433 and the NYU Research Catalyst Prize. AI tools were used to polish the language, assist with checks of the rigor and presentation of selected theorems and proofs, help explore approaches to the new proof of Theorem~\ref{prop: combine test valid}, and assist in implementing several basic functions in \textsf{R}. The authors take full responsibility for the study design, methodology, theoretical results, analyses, and presentation of all content.

%\section*{Funding}

\section*{Supplementary Materials}

The supplementary materials include technical proofs, additional theoretical and numerical results, more data details, and further discussions.

% ===== SUPPLEMENTARY MATERIAL (INLINE) =====
\clearpage
\appendix

% Print "Supplementary Material" as a header line and add to TOC
\section*{Supplementary Materials}
\addcontentsline{toc}{section}{Supplementary Material}

% Switch visible numbering to S<arabic> for sections/figures/tables/equations
\renewcommand{\thesection}{S\arabic{section}}
\renewcommand{\thefigure}{S\arabic{figure}}
\renewcommand{\thetable}{S\arabic{table}}
\renewcommand{\theequation}{S\arabic{equation}}

% Reset counters
\setcounter{section}{0}
\setcounter{figure}{0}
\setcounter{table}{0}
\setcounter{equation}{0}

% Theorem-like environments: show S<arabic> and reset counters
\setcounter{theorem}{0}
\renewcommand{\thetheorem}{S\arabic{theorem}}
\setcounter{lemma}{0}
\renewcommand{\thelemma}{S\arabic{lemma}}
\renewcommand{\theproposition}{S\arabic{proposition}}
\renewcommand{\thecorollary}{S\arabic{corollary}}
\renewcommand{\thecondition}{S\arabic{condition}}

\section*{Appendix A: Proofs}

\subsection*{A.1: Proof of Theorem 3.1}

The proof of Theorem~3.1 extends the idea from the proof of Theorem 3.5 in \citet{heng2023instrumental} from the instrumental variable setting to the general setting. 

\begin{proof}: By the definition of $\mathcal{L}$, functions $\eta$ and $\zeta$ are strictly positive on the relevant common support and $\zeta\in C^2$, so all ratios, logarithms, and differentiations used below are well defined. For any $P_{Z\mid \mathbf{x}, u}\in \mathcal{L}$, conditional on matching on measured confounders $\mathbf{x}$ (i.e., $\mathbf{x}_{i1}=\mathbf{x}_{i2}$), we have
\begin{align*}
    &\quad \ P(Z_{i 1}=z_{i}^{* *}, Z_{i 2}=z_{i}^{*} \mid \mathcal{F}, \mathcal{Z})\\
    &= \frac{\eta(z_{i}^{**}, \mathbf{x}_{i1}) \zeta (z_{i}^{**},u_{i1})\eta(z_{i}^{*}, \mathbf{x}_{i2}) \zeta (z_{i}^{*},u_{i2})}{\eta(z_{i}^{**}, \mathbf{x}_{i1}) \zeta (z_{i}^{**},u_{i1})\eta(z_{i}^{*}, \mathbf{x}_{i2}) \zeta (z_{i}^{*},u_{i2})+\eta(z_{i}^{*}, \mathbf{x}_{i1}) \zeta (z_{i}^{*},u_{i1})\eta(z_{i}^{**}, \mathbf{x}_{i2}) \zeta (z_{i}^{**},u_{i2})}\\
    &=\frac{\frac{\eta(z_{i}^{**}, \mathbf{x}_{i1}) \zeta (z_{i}^{**},u_{i1})\eta(z_{i}^{*}, \mathbf{x}_{i2}) \zeta (z_{i}^{*},u_{i2})}{\eta(z_{i}^{*}, \mathbf{x}_{i1}) \zeta (z_{i}^{*},u_{i1})\eta(z_{i}^{**}, \mathbf{x}_{i2}) \zeta (z_{i}^{**},u_{i2})}}{\frac{\eta(z_{i}^{**}, \mathbf{x}_{i1}) \zeta (z_{i}^{**},u_{i1})\eta(z_{i}^{*}, \mathbf{x}_{i2}) \zeta (z_{i}^{*},u_{i2})}{\eta(z_{i}^{*}, \mathbf{x}_{i1}) \zeta (z_{i}^{*},u_{i1})\eta(z_{i}^{**}, \mathbf{x}_{i2}) \zeta (z_{i}^{**},u_{i2})}+1}\\
    &=\frac{\frac{ \zeta (z_{i}^{**},u_{i1}) \zeta (z_{i}^{*},u_{i2})}{\zeta (z_{i}^{*},u_{i1})\zeta (z_{i}^{**},u_{i2})}}{\frac{ \zeta (z_{i}^{**},u_{i1}) \zeta (z_{i}^{*},u_{i2})}{\zeta (z_{i}^{*},u_{i1})\zeta (z_{i}^{**},u_{i2})}+1}.
\end{align*}  
This implies that
\begin{equation*}
    \frac{1}{1+\Gamma^{\eta, \zeta}_{i}(z_{i}^{**}, z_{i}^{*})}  \leq   P(Z_{i1}=z_{i}^{**}, Z_{i2}=z^{*}_{i} \mid \mathcal{F}, \mathcal{Z}) \leq \frac{\Gamma^{\eta, \zeta}_{i}(z_{i}^{**}, z_{i}^{*})}{1+\Gamma^{\eta, \zeta}_{i}(z_{i}^{**}, z_{i}^{*})},
\end{equation*}
where
\begin{align*}
    \Gamma^{\eta, \zeta}_{i}(z_{i}^{**}, z_{i}^{*})&=\max_{u_{i1}, u_{i2}\in [0,1] } \frac{ \zeta (z_{i}^{**},u_{i1}) \zeta (z_{i}^{*},u_{i2})}{\zeta (z_{i}^{*},u_{i1})\zeta (z_{i}^{**},u_{i2})}.
\end{align*}
Suppose that the classical Rosenbaum sensitivity bounds are both valid and sharp for any pairs of treatment doses $(z_{i}^{**}, z_{i}^{*})\in \mathcal{S}^{2}$ with $z_{i}^{**}>z_{i}^{*}$, that is, there exists some $\Gamma\geq 1$ such that 
\begin{equation}\label{eqn: Gamma indentity}
    \Gamma^{\eta, \zeta}_{i}(z_{i}^{**}, z_{i}^{*})\equiv \Gamma \text{ for any $(z_{i}^{**}, z_{i}^{*})\in \mathcal{S}^{2}$ with $z_{i}^{**}>z_{i}^{*}$.} 
\end{equation}
For $(a, b, u_{1}, u_{2})\in \mathcal{S}^{2}\times [0,1]^{2}$, define the function
\begin{equation*}
    h(a, b, u_{1}, u_{2})=\frac{ \zeta (b,u_{i1}) \zeta (a,u_{i2})}{\zeta (a,u_{i1})\zeta (b,u_{i2})}.
\end{equation*}
Choose a point $z_{0}\in \text{int}(\mathcal{S})$, where $\text{int}(\mathcal{S})$ is the collection of interior points of $\mathcal{S}$. Because $\mathcal{S}$ is a connected interval and $\text{int}(\mathcal{S})\neq \emptyset$, there exists some $\delta>0$ such that $[z_{0}-\delta, z_{0}+\delta]\subset \mathcal{S}$. Because $\zeta(z,u) \in C^{2}$ over $\mathcal{S}\times [0,1]$, we have $\log \zeta(z,u) \in C^{2}$ over $\mathcal{S}\times [0,1]$. Therefore, there exists a constant $M>0$ such that $|\frac{\partial \log \zeta(z,u)}{\partial z}|\leq M$ for any $(z, u)\in [z_{0}-\delta, z_{0}+\delta]\times [0, 1]$. Therefore, for any $a<b$ with $a, b\in [z_{0}-\delta, z_{0}+\delta]$, we have 
\begin{align*}
    \left|\log h(a, b, u_{1}, u_{2})\right|&=\left|\int_{a}^{b}\left\{ \frac{\partial \log \zeta(z,u_{1})}{\partial z}- \frac{\partial \log \zeta(z,u_{2})}{\partial z}\right\} dz \right|\\
    &\leq 2M(b-a).
\end{align*}
Therefore, for any $z_{i}^{**}, z_{i}^{*}\in [z_{0}-\delta, z_{0}+\delta]$ with $z_{i}^{**}>z_{i}^{*}$, we have 
\begin{equation*}
   1\leq \Gamma \equiv  \Gamma^{\eta, \zeta}_{i}(z_{i}^{**}, z_{i}^{*})=\max_{u_{1}, u_{2}\in [0,1] } h(z_{i}^{*}, z_{i}^{**}, u_{1}, u_{2}) \leq  \exp\left\{2M(z_{i}^{**}-z_{i}^{*}) \right\}. 
\end{equation*}
Setting $z_{i}^{**}\rightarrow z_{i}^{*}$, we have $\Gamma=1$. Therefore, by (\ref{eqn: Gamma indentity}), we have
\begin{equation}\label{eqn: max is 1}
    \Gamma^{\eta, \zeta}_{i}(z_{i}^{**}, z_{i}^{*})=\max_{u_{i1}, u_{i2}\in [0,1] } \frac{ \zeta (z_{i}^{**},u_{i1}) \zeta (z_{i}^{*},u_{i2})}{\zeta (z_{i}^{*},u_{i1})\zeta (z_{i}^{**},u_{i2})} =1 \text{ for any $(z_{i}^{**}, z_{i}^{*})\in \mathcal{S}^{2}$ with $z_{i}^{**}>z_{i}^{*}$.} 
\end{equation}
This implies that 
\begin{equation}\label{eqn: identity of zeta}
    \frac{ \zeta (z_{i}^{**},u_{i1}) \zeta (z_{i}^{*},u_{i2})}{\zeta (z_{i}^{*},u_{i1})\zeta (z_{i}^{**},u_{i2})}=1 \text{ for any $(z_{i}^{**}, z_{i}^{*}, u_{i1}, u_{i2})\in \mathcal{S}^{2}\times [0,1]^{2}$ with $z_{i}^{**}>z_{i}^{*}$.} 
\end{equation}
Otherwise, if (\ref{eqn: identity of zeta}) does not hold, then there are two possibilities: 
\begin{enumerate}
    \item We have $\frac{ \zeta (z_{i}^{**},u_{i1}) \zeta (z_{i}^{*},u_{i2})}{\zeta (z_{i}^{*},u_{i1})\zeta (z_{i}^{**},u_{i2})}>1$ for some $(z_{i}^{**}, z_{i}^{*}, u_{i1}, u_{i2})\in \mathcal{S}^{2}\times [0,1]^{2}$ with $z_{i}^{**}>z_{i}^{*}$. However, this violates (\ref{eqn: max is 1}) and, therefore, cannot hold.

    \item  We have $\frac{ \zeta (z_{i}^{**},u_{i1}) \zeta (z_{i}^{*},u_{i2})}{\zeta (z_{i}^{*},u_{i1})\zeta (z_{i}^{**},u_{i2})}<1$ for some $(z_{i}^{**}, z_{i}^{*}, u_{i1}, u_{i2})\in \mathcal{S}^{2}\times [0,1]^{2}$ with $z_{i}^{**}>z_{i}^{*}$. This implies that $\frac{ \zeta (z_{i}^{*},u_{i1}) \zeta (z_{i}^{**},u_{i2})}{\zeta (z_{i}^{**},u_{i1})\zeta (z_{i}^{*},u_{i2})}>1$, which violates (\ref{eqn: max is 1}) and, therefore, cannot hold. 

\end{enumerate}
Therefore, (\ref{eqn: identity of zeta}) holds. That is, for any $(z_{i}^{**}, z_{i}^{*}, u_{i1}, u_{i2})\in \mathcal{S}^{2}\times [0,1]^{2}$ with $z_{i}^{**}>z_{i}^{*}$, we have
\begin{equation*}
    \log \zeta (z_{i}^{**},u_{i1})-\log \zeta (z_{i}^{**},u_{i2})=  \log \zeta (z_{i}^{*},u_{i1})-\log \zeta (z_{i}^{*},u_{i2}).
\end{equation*}
Therefore, for any $z\in \text{int}(\mathcal{S})$ and any distinct $u_{i1}, u_{i2}\in [0,1]$, we have
\begin{equation*}
    \frac{\partial}{\partial z}\Big (\frac{\log \zeta(z, u_{i1})-\log \zeta(z, u_{i2})}{u_{i1}-u_{i2}}\Big)\equiv 0.
\end{equation*}
Therefore, for any $(z, u)\in \text{int}(\mathcal{S})\times (0,1)$, because $\zeta$ is a $C^{2}$ function (i.e., has continuous second-order derivatives), we have
\begin{align}\label{eqn: change the order of limit}
    &\frac{\partial^{2} \log \zeta(z, u)}{\partial z \partial u}= \frac{\partial}{\partial z}\Big( \lim_{u^{\prime} \rightarrow u}\frac{\log \zeta(z, u^{\prime})-\log \zeta(z, u)}{u^{\prime}-u}\Big)\nonumber\\
    &=\lim_{u^{\prime} \rightarrow u}\frac{\partial}{\partial z}\Big (\frac{\log \zeta(z, u^{\prime})-\log \zeta(z, u)}{u^{\prime}-u}\Big)  \equiv 0.
\end{align}
This implies that, for all $(z, u)\in \text{int}(\mathcal{S})\times (0,1)$, we have
\begin{align}\label{equa: pde}
    \frac{\partial^{2} \vartheta(z,u)}{\partial z \partial u} \equiv 0  \quad (\text{where $\vartheta(z,u)=\log \zeta(z, u)$}).
\end{align}
A basic and well-known result in the partial differential equation (PDE) literature states that if $\vartheta(z,u)=\log \zeta(z, u)$ is a $C^{2}$ solution to the second order linear PDE (\ref{equa: pde}), we have $\vartheta(z,u)=\vartheta_{1}(z)+\vartheta_{2}(u)$ for some functions $\vartheta_{1}$ and $\vartheta_{2}$ \citep{strauss2007partial}. Thus, we have $\zeta(z, u)=\exp\{\vartheta_{1}(z)\}\exp\{\vartheta_{2}(u)\}$. This implies that, for any $(z_{i}^{**}, z_{i}^{*}, u_{i1}, u_{i2})\in \text{int}(\mathcal{S})\times \text{int}(\mathcal{S})\times (0,1)^{2}$ with $z_{i}^{**}>z_{i}^{*}$, we have
\begin{align*}
    P(Z_{i 1}=z_{i}^{* *}, Z_{i 2}=z_{i}^{*} \mid \mathcal{F}, \mathcal{Z})
    &=\frac{\frac{ \zeta (z_{i}^{**},u_{i1}) \zeta (z_{i}^{*},u_{i2})}{\zeta (z_{i}^{*},u_{i1})\zeta (z_{i}^{**},u_{i2})}}{\frac{ \zeta (z_{i}^{**},u_{i1}) \zeta (z_{i}^{*},u_{i2})}{\zeta (z_{i}^{*},u_{i1})\zeta (z_{i}^{**},u_{i2})}+1}\\
    &=\frac{\frac{\exp\{\vartheta_{1}(z^{**}_{i})\}\exp\{\vartheta_{2}(u_{i1})\}\exp\{\vartheta_{1}(z_{i}^{*})\}\exp\{\vartheta_{2}(u_{i2})\}}{\exp\{\vartheta_{1}(z_{i}^{*})\}\exp\{\vartheta_{2}(u_{i1})\}\exp\{\vartheta_{1}(z_{i}^{**})\}\exp\{\vartheta_{2}(u_{i2})\}} }{\frac{\exp\{\vartheta_{1}(z^{**}_{i})\}\exp\{\vartheta_{2}(u_{i1})\}\exp\{\vartheta_{1}(z_{i}^{*})\}\exp\{\vartheta_{2}(u_{i2})\}}{\exp\{\vartheta_{1}(z_{i}^{*})\}\exp\{\vartheta_{2}(u_{i1})\}\exp\{\vartheta_{1}(z_{i}^{**})\}\exp\{\vartheta_{2}(u_{i2})\}} +1}   \\
    &=\frac{1}{2}.
\end{align*}    
Because of the continuity of $\zeta$, for any $(z_{i}^{**}, z_{i}^{*}, u_{i1}, u_{i2})\in \mathcal{S}^{2}\times [0,1]^{2}$ with $z_{i}^{**}>z_{i}^{*}$, we have
\begin{align*}
    P(Z_{i 1}=z_{i}^{* *}, Z_{i 2}=z_{i}^{*} \mid \mathcal{F}, \mathcal{Z})
    =\frac{1}{2},
\end{align*}   
which corresponds to the setting of no unmeasured confounding, violating the settings of sensitivity analysis for unmeasured confounding. 

\end{proof}

\begin{remark}[Remark on the Smoothness Condition of $\zeta$]
In the proof of Theorem~3.1, we impose a smoothness condition on $\zeta$ that allows us to exchange the order of the limit and the partial derivative in (\ref{eqn: change the order of limit}) and thereby derive the second-order PDE in (\ref{equa: pde}). The solution to this PDE implies that the post-matching dose assignment probabilities
$P(Z_{i1}=z_i^{**},Z_{i2}=z_i^*\mid\mathcal{F},\mathcal{Z})$
equal $1/2$ for all matched pairs. This corresponds to the absence of unmeasured confounding and thus contradicts the sensitivity analysis setting considered here. This contradiction establishes that the classical Rosenbaum sensitivity bounds, which impose a common sensitivity bound $\Gamma$ across all pairs, cannot be simultaneously valid and sharp uniformly over all matched pairs.

Without the smoothness condition on $\zeta$, however, this argument need not hold. In fact, there exist nonsmooth choices of $\zeta$ under which the classical Rosenbaum sensitivity bounds with a common sensitivity parameter $\Gamma$ are simultaneously valid and sharp uniformly over all matched pairs. To see this, let the support of the treatment dose be $\mathcal{S}=[0,1]$, fix $\Gamma_0>1$, and consider
\[
\zeta(z,u)=\Gamma_0^{\mathbf{1}\{z\geq u\}},
\quad (z,u)\in[0,1]^2.
\]
Recall that, conditional on matching on the measured confounders $\mathbf{x}$, so that $\mathbf{x}_{i1}=\mathbf{x}_{i2}$, we have
\begin{align*}
P(Z_{i1}=z_i^{**},Z_{i2}=z_i^*\mid\mathcal{F},\mathcal{Z})
=
\frac{
\frac{\zeta(z_i^{**},u_{i1})\zeta(z_i^*,u_{i2})}
{\zeta(z_i^*,u_{i1})\zeta(z_i^{**},u_{i2})}
}{
\frac{\zeta(z_i^{**},u_{i1})\zeta(z_i^*,u_{i2})}
{\zeta(z_i^*,u_{i1})\zeta(z_i^{**},u_{i2})}
+1
}.
\end{align*}
This implies that
\begin{equation*}
\frac{1}{1+\Gamma_i^{\eta,\zeta}(z_i^{**},z_i^*)}
\leq
P(Z_{i1}=z_i^{**},Z_{i2}=z_i^*\mid\mathcal{F},\mathcal{Z})
\leq
\frac{\Gamma_i^{\eta,\zeta}(z_i^{**},z_i^*)}
{1+\Gamma_i^{\eta,\zeta}(z_i^{**},z_i^*)},
\end{equation*}
where
\begin{align*}
\Gamma_i^{\eta,\zeta}(z_i^{**},z_i^*)
=
\max_{u_{i1},u_{i2}\in[0,1]}
\frac{\zeta(z_i^{**},u_{i1})\zeta(z_i^*,u_{i2})}
{\zeta(z_i^*,u_{i1})\zeta(z_i^{**},u_{i2})}.
\end{align*}
For any two distinct treatment doses $0\leq z_i^*<z_i^{**}\leq1$, we have
\[
\frac{\zeta(z_i^{**},u)}{\zeta(z_i^*,u)}
=
\begin{cases}
\Gamma_0, & z_i^*<u\leq z_i^{**},\\
1, & \text{otherwise}.
\end{cases}
\]
It follows that, for any $0\leq z_i^*<z_i^{**}\leq1$,
\begin{align*}
\Gamma_i^{\eta,\zeta}(z_i^{**},z_i^*)
=
\max_{u_{i1},u_{i2}\in[0,1]}
\frac{\zeta(z_i^{**},u_{i1})\zeta(z_i^*,u_{i2})}
{\zeta(z_i^*,u_{i1})\zeta(z_i^{**},u_{i2})}=\Gamma_0.
\end{align*}
Therefore, for every pair of distinct treatment doses
$0\leq z_i^*<z_i^{**}\leq1$, the classical Rosenbaum sensitivity bounds
\[
\frac{1}{1+\Gamma_0}
\leq
P(Z_{i1}=z_i^{**},Z_{i2}=z_i^*\mid\mathcal{F},\mathcal{Z})
\leq
\frac{\Gamma_0}{1+\Gamma_0}
\]
are simultaneously valid and sharp with the same $\Gamma_0>1$. Thus, without an appropriate smoothness condition on $\zeta$, the conclusion of Theorem~3.1 need not hold.
\end{remark}

\begin{remark}
    Theorem~3.1 builds directly on the exclusion-principle argument developed by \citet{heng2023instrumental} for matched studies with continuous instrumental variables. Their result shows that, under a broad class of data-generating processes, a biased post-matching assignment mechanism for a continuous instrumental variable cannot be independent of the paired encouragement doses. Theorem~3.1 adapts the same underlying separability argument to continuous treatments and provides a more precise implication for sensitivity bounds by distinguishing validity from sharpness. Specifically, although sensitivity bounds based on a common $\Gamma$ may remain valid if chosen conservatively, under the conditions of Theorem~3.1 they cannot be simultaneously valid and sharp uniformly across all matched pairs unless there is no unmeasured confounding. Therefore, the contribution of Theorem~3.1 is not a new proof technique; rather, it establishes a critical issue in the continuous treatment setting and motivates the generalized, pair-specific sensitivity bounds and the subsequent developments of generalized design sensitivity and Bahadur-Rosenbaum efficiency.
\end{remark}

\begin{remark}
In Theorem~3.1, we restrict attention to sensitivity models in the family $\mathcal{L}$, which excludes $X$-$U$ interactions. This restriction is not intended to suggest that the conclusion of Theorem~3.1 fails for sensitivity models outside $\mathcal{L}$. When $X$-$U$ interactions are present, sensitivity bounds that are simultaneously valid and sharp generally need to depend not only on the paired treatment doses but also on the pair-specific values of the measured confounders that interact with the unmeasured confounder. Consequently, classical Rosenbaum sensitivity analyses, which use a common sensitivity parameter $\Gamma$ across matched pairs without accounting for such pair-specific heterogeneity, are even less likely to yield sensitivity bounds that are simultaneously valid and sharp. Indeed, \citet{heng2021sharpening} show that, even in the binary-treatment setting, sharp sensitivity bounds generally depend on the pair-specific values of the measured confounders involved in $X$-$U$ interactions. We focus on the family $\mathcal{L}$ in Theorem~3.1 to isolate the central message that a simultaneously valid and sharp sensitivity bound for a continuous treatment must incorporate treatment dose information, while keeping the notation and proof as simple and transparent as possible.
\end{remark}

\begin{remark}
As mentioned in the main text, the classical Rosenbaum sensitivity bounds (4) impose a uniform upper bound across matched pairs, indexed by a single sensitivity parameter $\Gamma$. Specifically, if 
    \begin{equation*}
        \Gamma \geq \max_{i=1,\ldots,I}\max\left\{\frac{P(Z_{i1}=1,Z_{i2}=0\mid \mathcal{F},\mathcal{Z})}{P(Z_{i1}=0,Z_{i2}=1\mid \mathcal{F},\mathcal{Z})}, \frac{P(Z_{i1}=0,Z_{i2}=1\mid \mathcal{F},\mathcal{Z})}{P(Z_{i1}=1,Z_{i2}=0\mid \mathcal{F},\mathcal{Z})}\right\},
    \end{equation*}
   then (4) holds for all $I$ matched pairs. If the true treatment assignment probabilities exhibit pair-level heterogeneity not captured by (2), then the common binary-treatment Rosenbaum sensitivity bounds may be conservative because they impose a single upper bound across all matched pairs. This point is consistent with recent work on heterogeneous unmeasured confounding biases across matched pairs in the binary treatment case \citep{Rosenbaum1987, hasegawa2017sensitivity, fogarty2019extended, heng2021sharpening, wu2025sensitivity}. 
    
    A similar issue arises in the non-binary treatment case. Since the generalized Rosenbaum sensitivity bounds in (3) allow pair-specific values $\Gamma_i=\exp\{\gamma|\phi(z_i^{**})-\phi(z_i^*)|\}$, one can always obtain a valid common-bound version by ignoring pair-specific treatment dose information and setting a uniform upper bound 
    \begin{equation*}
            \Gamma=\max_{i=1,\ldots,I}\Gamma_i=\max_{i=1,\ldots,I}\exp\{\gamma|\phi(z_i^{**})-\phi(z_i^*)|\}.
    \end{equation*}
 This reduces the generalized Rosenbaum sensitivity bounds (3) to a classical Rosenbaum sensitivity bound with a common sensitivity parameter $\Gamma$, but it may be overly conservative because it ignores heterogeneity in dose differences across matched pairs. Therefore, a key technical contribution of our work is to provide a framework for evaluating and improving power of sensitivity analysis while preserving pair-specific treatment dose information.
\end{remark}

\subsection*{A.2: Proof of Proposition 3.2}

\begin{proposition}[An Extended Version of Proposition~3.2] \label{prop:sufficiency}
Under the Rosenbaum sensitivity model (i.e., equation (1) in the main text), $\gamma$, $\overline{\Gamma}$, and the induced bias vector $(\Gamma_{1},\dots,\Gamma_{I})$ constitute equivalent parameterizations: each uniquely determines the other two. In other words, there are pairwise bijections among $\gamma$, $\overline{\Gamma}$, and $(\Gamma_{1},\dots,\Gamma_{I})$. Explicit constructions of these bijections are provided below.
\end{proposition}

\begin{proof}: Throughout the proof, $(\Gamma_1,\ldots,\Gamma_I)$ is understood to range over the one-parameter family of vectors $(\exp\{\gamma|\phi(z^{**}_{1})-\phi(z_{1}^{*})|\}, \dots, \exp\{\gamma|\phi(z^{**}_{I})-\phi(z_{I}^{*})|\})$ induced by $\gamma\geq 0$, rather than over the unrestricted space $[1,\infty)^I$.

Under the Rosenbaum sensitivity model, after matching on measured confounders, we have:
\begin{align*}
    &\quad \ P(Z_{i1}=z_{i}^{**}, Z_{i2}=z^{*}_{i} \mid \mathcal{F}, \mathcal{Z}) \\
   &=P(Z_{i1}=z_{i}^{**}, Z_{i2}=z^{*}_{i}\mid \mathbf{x}_{i1}, u_{i1}, \mathbf{x}_{i2}, u_{i2}, Z_{i1}\wedge Z_{i2}=z_{i}^{*}, Z_{i1}\vee Z_{i2}=z_{i}^{**})\\
   &= \frac{P(Z_{i1}=z_{i}^{**}, Z_{i2}=z_{i}^{*} \mid \mathbf{x}_{i1}, u_{i1}, \mathbf{x}_{i2}, u_{i2})}{P(Z_{i1}=z_{i}^{**}, Z_{i2}=z_{i}^{*} \mid \mathbf{x}_{i1}, u_{i1}, \mathbf{x}_{i2}, u_{i2})+P(Z_{i1}=z_{i}^{*}, Z_{i2}=z_{i}^{**} \mid \mathbf{x}_{i1}, u_{i1}, \mathbf{x}_{i2}, u_{i2})} \\
   &= \frac{P(Z_{i1}=z_{i}^{**}\mid \mathbf{x}_{i1}, u_{i1})P(Z_{i2}=z_{i}^{*}\mid \mathbf{x}_{i2}, u_{i2})}{P(Z_{i1}=z_{i}^{**}\mid \mathbf{x}_{i1}, u_{i1})P(Z_{i2}=z_{i}^{*}\mid \mathbf{x}_{i2}, u_{i2})+P(Z_{i1}=z_{i}^{*}\mid \mathbf{x}_{i1}, u_{i1})P(Z_{i2}=z_{i}^{**}\mid \mathbf{x}_{i2}, u_{i2})} \\
   &=\frac{\exp\{\gamma \phi(z_{i}^{**})u_{i1}\}\exp\{\gamma \phi(z_{i}^{*})u_{i2}\}  }{\exp\{\gamma \phi(z_{i}^{**})u_{i1}\}\exp\{\gamma \phi(z_{i}^{*})u_{i2}\} +\exp\{\gamma \phi(z_{i}^{*})u_{i1}\}\exp\{\gamma \phi(z_{i}^{**})u_{i2}\}}\\
   &=\frac{\exp\{\gamma(\phi(z_{i}^{**})-\phi(z_{i}^{*}))(u_{i1}-u_{i2})\} }{1+\exp\{\gamma(\phi(z_{i}^{**})-\phi(z_{i}^{*}))(u_{i1}-u_{i2})\}}.
\end{align*}
Each $u_{ij}$ has been normalized to $[0,1]$ to make the sensitivity parameter $\gamma\geq 0$ more interpretable. Therefore, we have: 
\begin{equation*}
\frac{1}{1+\Gamma_{i}} \leq P(Z_{i 1}=z_{i}^{* *}, Z_{i 2}=z_{i}^{*} \mid \mathcal{F}, \mathcal{Z}) \leq \frac{\Gamma_{i}}{1+\Gamma_{i}}, \ \text{where $\Gamma_{i} = \exp\{\gamma|\phi(z_{i}^{**})-\phi(z_{i}^{*})|\}\geq 1$.}
\end{equation*} 
 Recall that $\overline{\Gamma}=I^{-1}\sum_{i=1}^{I}\Gamma_{i}=I^{-1}\sum_{i=1}^{I}\exp\{\gamma|\phi(Z_{i1})-\phi(Z_{i2})|\}=I^{-1}\sum_{i=1}^{I}\exp\{\gamma|\phi(z_{i}^{**})-\phi(z_{i}^{*})|\}$, where $\gamma\geq 0$. Since $\phi(Z_{i1})\neq \phi(Z_{i2})$ for each $i$, it is clear that the map $\gamma \mapsto \overline{\Gamma}$ is bijective because $\overline{\Gamma}=I^{-1}\sum_{i=1}^{I}\exp\{\gamma |\phi(z^{**}_{i})-\phi(z_{i}^{*})|\}$ is strictly monotonically increasing with respect to $\gamma$. Similarly, it is clear that the map $\gamma \mapsto (\Gamma_{1}, \dots, \Gamma_{I})=(\exp\{\gamma|\phi(z^{**}_{1})-\phi(z_{1}^{*})|\}, \dots, \exp\{\gamma|\phi(z^{**}_{I})-\phi(z_{I}^{*})|\})$ is bijective. Therefore, the map $(\Gamma_{1}, \dots, \Gamma_{I})\mapsto \overline{\Gamma}$ is bijective. Also, given each fixed value of $\overline{\Gamma}\geq 1$, we can use the bisection method to find the corresponding $\widetilde{\gamma}$ such that $\overline{\Gamma}=I^{-1}\sum_{i=1}^{I}\exp\{\widetilde{\gamma}|\phi(Z_{i1})-\phi(Z_{i2})|\}$, as well as the corresponding vector of within-pair unmeasured confounding biases $(\Gamma_{1},\dots, \Gamma_{I})=(\exp\{\widetilde{\gamma}|\phi(Z_{11})-\phi(Z_{12})|\}, \dots, \exp\{\widetilde{\gamma}|\phi(Z_{I1})-\phi(Z_{I2})|\})$. 
\end{proof}

\subsection*{A.3: Proof of Theorem 3.4}

For notational simplicity, throughout Appendices~A.3--A.5, without loss of generality, we index the sensitivity analysis and state the resulting asymptotic conclusions primarily in terms of the post-matching sensitivity parameter $\overline{\Gamma}$. By the one-to-one correspondence established in Proposition~\ref{prop:sufficiency} and its asymptotic counterpart in Lemma~\ref{lem: inverting Gamma} below, each fixed $\overline{\Gamma}$ uniquely determines the corresponding population-level sensitivity parameter $\gamma$. Because this correspondence is strictly increasing, comparisons of $\overline{\Gamma}$ with $\overline{\Gamma}_{*}$ are equivalent to the corresponding comparisons of $\gamma$ with $\gamma_{*}$. Consequently, all arguments and conclusions below can be equivalently formulated on the $\gamma$-scale.

In Appendices A.3--A.5, we assume that $(Z_{i1}, Z_{i2}, Y_{i1}, Y_{i2})$ are independent and identically distributed realizations from some multivariate distribution $F$, in which $(Z_{i1}, Z_{i2})$ are continuous random variables. We state the following regularity conditions:

\begin{condition}[Bounded First Moment and Bounded Difference of Treatment Doses]\label{condition: dose moment}
Treatment doses $Z_{ij}$ are continuous treatment variables such that $E|\phi(Z_{ij})|<\infty$ and $|\phi(Z_{i1})-\phi(Z_{i2})|$ is bounded. 
\end{condition}
\begin{condition}[Overlap of Supports] \label{condi: overlap of supports}
    There exist $A_{1}, A_{2}, B_{1}, B_{2}, C_{1}, C_{2} >0$ such that $F_{Z}(B_{1})>0$, $F_{Y}(C_{1})>0$, and $P(\mathcal{A})>0$, where the set $\mathcal{A}=\{(Z_{i1}, Z_{i2}, Y_{i1}, Y_{i2}): |\phi(Z_{i1})-\phi(Z_{i2})| \in [A_{1}, A_{2}], |Z_{i1}-Z_{i2}|\in [B_{1}, B_{2}], |Y_{i1}-Y_{i2}|\in [C_{1}, C_{2}]\}$.
\end{condition}
\begin{condition}[Regularity of the Score Function $\psi$]\label{condition: varphi}
    $\psi(z, y)$ is a non-negative differentiable function defined over $[0,1]^{2}$ such that: (i) $0<\psi(z, y)$ for any $(z, y)\in (0,1]^{2}$; (ii) There exists some constant $C_{\psi}<\infty$ such that for any $(z, y)\in [0,1]^{2}$, we have $\norm{\bigtriangledown\psi(z, y)}\leq C_{\psi}$.
\end{condition}
\begin{condition}[Non-Zero and Non-Extreme Treatment Effects]\label{condition: prob of D}
   Let $P_{\psi_i^*}$ denote the distribution of $\psi_i^*$, where $\psi_i^*=\psi\bigl(F_Z(|Z_{i1}-Z_{i2}|),F_Y(|Y_{i1}-Y_{i2}|)\bigr)$. For almost every $s\in \operatorname{supp}(P_{\psi_i^*})\cap(0,\infty)$ over the distribution $P_{\psi_i^*}$, we have $P(D_i>0\mid\psi_i^*=s)\in(1/2,1)$. In other words, under the alternative (i.e., when the treatment effect exists), the association between the $\text{sign}(Z_{i1}-Z_{i2})$ and $\text{sign}(Y_{i1}-Y_{i2})$ cannot equal either zero (i.e., no treatment effect) or one (i.e., extreme or perfect treatment effect).  
\end{condition}

\begin{remark} The iid assumption used to establish the generalized design sensitivity and Bahadur-Rosenbaum efficiency is imposed directly at the matched-pair level on the observed vectors $(Z_{i1},Z_{i2},Y_{i1},Y_{i2})$ under the corresponding data-generating process for the asymptotic power analysis. Therefore, the stated asymptotic results do not require separate distributional assumptions on the measured confounders $\mathbf{x}_{ij}$ or the unmeasured confounders $u_{ij}$. One sufficient superpopulation formulation is that the pair-level vectors $(\mathbf{x}_{i1},\mathbf{x}_{i2},u_{i1},u_{i2})$ are iid across matched pairs and, conditional on these variables, the treatment doses and outcomes are generated independently across pairs according to a common conditional distribution. Under this formulation, $(Z_{i1},Z_{i2},Y_{i1},Y_{i2})$ are iid across pairs after marginalizing over the measured and unmeasured confounders.

This superpopulation formulation is distinct from the framework used to establish the conditional validity of sensitivity analysis (i.e., validity conditional on any fixed values of potential outcomes, matched measured confounders, and unmeasured confounders). Under the latter framework, the matched sample is treated as fixed, and the sensitivity bounds are required to hold for every feasible allocation of the unmeasured confounders $u_{ij}$. Consequently, no probability distribution for $\mathbf{x}_{ij}$ or $u_{ij}$ is required for the conditional validity of sensitivity analysis. For the asymptotic results, it is sufficient to impose the iid assumption directly on $(Z_{i1},Z_{i2},Y_{i1},Y_{i2})$; a distribution for $\mathbf{x}_{ij}$ or $u_{ij}$ is needed only if one chooses to derive this iid assumption from an underlying superpopulation model that explicitly includes these measured and unmeasured confounders.
\end{remark}

We prove some useful lemmas for proving Theorem 3.4. 

\begin{lemma}\label{lem: inverting Gamma}
  Suppose Conditions~\ref{condition: dose moment} and \ref{condi: overlap of supports} hold. Let $\overline{\Gamma}\geq 1$ be a fixed value of the sensitivity parameter. For each $I$, there exists a unique $\widetilde{\gamma}_{I}$ such that $\overline{\Gamma}=I^{-1}\sum_{i=1}^{I}\exp\{\widetilde{\gamma}_{I}|\phi(Z_{i1})-\phi(Z_{i2})|\}$. Also, there exists a unique $\widetilde{\gamma}$ such that $\overline{\Gamma}=E[\exp\{\widetilde{\gamma}|\phi(Z_{i1})-\phi(Z_{i2})|\}]$. Moreover, we have $\widetilde{\gamma}_{I}\xrightarrow{a.s.} \widetilde{\gamma}$ as $I\rightarrow \infty$.
\end{lemma}

\begin{proof}: Let $J_{I}(\gamma)=I^{-1}\sum_{i=1}^{I}\exp\{\gamma|\phi(Z_{i1})-\phi(Z_{i2})|\}$ and $J(\gamma)=E[\exp\{\gamma|\phi(Z_{i1})-\phi(Z_{i2})|\}]$. Since $Z_{i1}\neq Z_{i2}$, $J_{I}(\gamma)$ is a strictly monotonically increasing and continuous function for each $I$. Also, we have $\lim_{\gamma\rightarrow 0^{+}}J_{I}(\gamma)=1$ and $\lim_{\gamma\rightarrow +\infty}J_{I}(\gamma)=+\infty$. Therefore, for each fixed $\overline{\Gamma}\geq 1$ and for each $I$, there exists a unique $\widetilde{\gamma}_{I}$ such that $\overline{\Gamma}=I^{-1}\sum_{i=1}^{I}\exp\{\widetilde{\gamma}_{I}|\phi(Z_{i1})-\phi(Z_{i2})|\}$. 

Then, we prove the following properties for $J(\gamma)$:

(i) $J(\gamma)$ is a continuous function over $\gamma\in [0,+\infty)$: For any $0\leq \gamma_{1}<\gamma_{2}$, we have 
\begin{align*}
    &\quad \ |J(\gamma_{2})-J(\gamma_{1})|\\
    &= E\big[\exp\{\gamma_{2}|\phi(Z_{i1})-\phi(Z_{i2})|\}-\exp\{\gamma_{1}|\phi(Z_{i1})-\phi(Z_{i2})|\}\big]\\
    &\leq E\big[\sup_{\xi(Z_{i1}, Z_{i2})\in [\gamma_{1}, \gamma_{2}] }(\gamma_{2}-\gamma_{1})|\phi(Z_{i1})-\phi(Z_{i2})|\exp\{\xi(Z_{i1}, Z_{i2})|\phi(Z_{i1})-\phi(Z_{i2})|\}\big]\\
    &= (\gamma_{2}-\gamma_{1})E\big[|\phi(Z_{i1})-\phi(Z_{i2})|\exp\{\gamma_{2}|\phi(Z_{i1})-\phi(Z_{i2})|\}\big].
\end{align*}
Then, the continuity of $J(\gamma)$ follows from the fact that the $|\phi(Z_{i1})-\phi(Z_{i2})|$ is bounded (by Condition~\ref{condition: dose moment}).

(ii) $J(\gamma)$ is a strictly monotonically increasing function over $\gamma\in [0,+\infty)$: For any $0\leq \gamma_{1}<\gamma_{2}$, consider the set $\mathcal{A}$ defined in Condition~\ref{condi: overlap of supports}, we have 
\begin{align*}
  &\quad \  J(\gamma_{2})-J(\gamma_{1})\\
  &=E\big[\exp\{\gamma_{2}|\phi(Z_{i1})-\phi(Z_{i2})|\}-\exp\{\gamma_{1}|\phi(Z_{i1})-\phi(Z_{i2})|\}\big]\\
    &\geq E\big[\inf_{\xi(Z_{i1}, Z_{i2})\in [\gamma_{1}, \gamma_{2}] }(\gamma_{2}-\gamma_{1})|\phi(Z_{i1})-\phi(Z_{i2})|\exp\{\xi(Z_{i1}, Z_{i2})|\phi(Z_{i1})-\phi(Z_{i2})|\}\big]\\
    &=(\gamma_{2}-\gamma_{1})E\big[|\phi(Z_{i1})-\phi(Z_{i2})|\exp\{\gamma_{1}|\phi(Z_{i1})-\phi(Z_{i2})|\}\big]\\
    &=(\gamma_{2}-\gamma_{1})P(\mathcal{A})E\big[|\phi(Z_{i1})-\phi(Z_{i2})|\exp\{\gamma_{1}|\phi(Z_{i1})-\phi(Z_{i2})|\}\mid \mathcal{A}\big]\\
    &\quad \quad + (\gamma_{2}-\gamma_{1})P(\mathcal{A}^{c})E\big[|\phi(Z_{i1})-\phi(Z_{i2})|\exp\{\gamma_{1}|\phi(Z_{i1})-\phi(Z_{i2})|\}\mid \mathcal{A}^{c}\big]\\
    &\geq (\gamma_{2}-\gamma_{1})P(\mathcal{A})E\big[|\phi(Z_{i1})-\phi(Z_{i2})|\exp\{\gamma_{1}|\phi(Z_{i1})-\phi(Z_{i2})|\}\mid \mathcal{A}\big] \\
    &\geq (\gamma_{2}-\gamma_{1})P(\mathcal{A}) A_{1}\exp(\gamma_{1}A_{1})>0.
\end{align*}
Therefore, $J(\gamma)$ is strictly monotonically increasing over $\gamma\in [0,+\infty)$.

(iii) We have $\lim_{\gamma\rightarrow 0^{+}}J(\gamma)=1$ and $\lim_{\gamma\rightarrow +\infty}J(\gamma)=+\infty$: Note that $\exp\{\gamma|\phi(Z_{i1})-\phi(Z_{i2})|\}$ is bounded around $\gamma=0$. By the bounded convergence theorem, we have  $\lim_{\gamma\rightarrow 0^{+}}J(\gamma)=E[\lim_{\gamma\rightarrow 0^{+}}\exp\{\gamma|\phi(Z_{i1})-\phi(Z_{i2})|\}]=1$. Also, note that $J(\gamma)= E[\exp\{\gamma|\phi(Z_{i1})-\phi(Z_{i2})|\}]\geq P(\mathcal{A})E\big[\exp\{\gamma|\phi(Z_{i1})-\phi(Z_{i2})|\}\mid \mathcal{A}\big]\geq P(\mathcal{A})\exp(\gamma A_{1})$. Therefore, we have $\lim_{\gamma\rightarrow +\infty}J(\gamma)=+\infty$. 

By the proved facts in (i) -- (iii), we have proven that there exists a unique $\widetilde{\gamma}$ such that $\overline{\Gamma}=E[\exp\{\widetilde{\gamma}|\phi(Z_{i1})-\phi(Z_{i2})|\}]$, where $\overline{\Gamma}\geq 1$ is some fixed value of the sensitivity parameter.

Finally, we show that $\widetilde{\gamma}_{I}\xrightarrow{a.s.}
\widetilde{\gamma}$. If $\overline{\Gamma}=1$, then
$\widetilde{\gamma}_{I}=\widetilde{\gamma}=0$ for every $I$, and the
conclusion follows immediately. Now suppose that
$\overline{\Gamma}>1$, so that $\widetilde{\gamma}>0$. Choose an
integer $m_{0}$ such that $m_{0}^{-1}<\widetilde{\gamma}$. For every
integer $m\geq m_{0}$, the strict monotonicity of $J$ gives $J\left(\widetilde{\gamma}-\frac{1}{m}\right)
<
J(\widetilde{\gamma})
=
\overline{\Gamma}
<
J\left(\widetilde{\gamma}+\frac{1}{m}\right)$. For each fixed $m\geq m_{0}$, the strong law of large numbers gives
$J_{I}\left(\widetilde{\gamma}-\frac{1}{m}\right)
\xrightarrow{a.s.}
J\left(\widetilde{\gamma}-\frac{1}{m}\right)
\quad\text{and}\quad
J_{I}\left(\widetilde{\gamma}+\frac{1}{m}\right)
\xrightarrow{a.s.}
J\left(\widetilde{\gamma}+\frac{1}{m}\right)$. Because the collection of integers $m\geq m_{0}$ is countable, these
convergences hold simultaneously for all $m\geq m_{0}$ on an event
with probability one. Therefore, on this event, for every fixed
$m\geq m_{0}$ and all sufficiently large $I$, $J_{I}\left(\widetilde{\gamma}-\frac{1}{m}\right)
<
\overline{\Gamma}
<
J_{I}\left(\widetilde{\gamma}+\frac{1}{m}\right)$. Since $J_{I}$ is strictly increasing and
$J_{I}(\widetilde{\gamma}_{I})=\overline{\Gamma}$, it follows that $
\widetilde{\gamma}-\frac{1}{m}
<
\widetilde{\gamma}_{I}
<
\widetilde{\gamma}+\frac{1}{m}$ for all sufficiently large $I$. Because this holds for every $m\geq m_{0}$, we conclude that $\widetilde{\gamma}_{I}\xrightarrow{a.s.}\widetilde{\gamma}$. \end{proof}

Under Lemma~\ref{lem: inverting Gamma}, as in the main text, we let $p_{I, i}^{+}=\expit\{\widetilde{\gamma}_{I}|\phi(Z_{i1})-\phi(Z_{i2})|\}$ and $\widetilde{p}_{i}^{+}=\expit\{\widetilde{\gamma}|\phi(Z_{i1})-\phi(Z_{i2})|\}$, where the function $\expit(x)=\exp(x)/(1+\exp(x))$.

%Note that $\widetilde{\gamma}_{I}$ and $\widetilde{\gamma}$ are well-defined (following a similar argument to that of the existence and uniqueness of equation (\ref{eqn: design sensitivity equation})). Because both $I^{-1}\sum_{i=1}^{I}\exp\{\widetilde{\gamma}_{I}|\phi(Z_{i1})-\phi(Z_{i2})|\}$ and $E[\exp\{\widetilde{\gamma}|\phi(Z_{i1})-\phi(Z_{i2})|\}]$ are strictly monotonically increasing function of $\gamma$, and $\overline{\Gamma}=I^{-1}\sum_{i=1}^{I}\exp\{\widetilde{\gamma}_{I}|\phi(Z_{i1})-\phi(Z_{i2})|\}=E[\exp\{\widetilde{\gamma}|\phi(Z_{i1})-\phi(Z_{i2})|\}]$, we have $\lim_{I\rightarrow\infty}\widetilde{\gamma}_{I}=\widetilde{\gamma}$.

\begin{lemma}\label{lemma: convergence}
Let $F_{Z}$ and $F_{Y}$ denote the marginal distributions of $|Z_{i1}-Z_{i2}|$ and $|Y_{i1}-Y_{i2}|$, respectively. Then, under Conditions~\ref{condition: dose moment} and \ref{condition: varphi}, as $I\rightarrow \infty$, we have
\begin{align*}
   &I^{-1} \sum_{i=1}^{I}\psi(r_{I, i}^{z}, r_{I, i}^{y}) \mathbbm{1}\{D_{i}>0\}\xrightarrow{a.s.} E\big [\psi^{*}_{i}  \mathbbm{1}\{D_{i}>0\}\big],\\
    &I^{-1}\sum_{i=1}^{I}\psi(r_{I, i}^{z}, r_{I, i}^{y})p_{I, i}^{+}\xrightarrow{a.s.} E\Big[\psi^{*}_{i}  \expit(\widetilde{\gamma}|\phi(Z_{i1})-\phi(Z_{i2})|) \Big]=E(\psi^{*}_{i}\widetilde{p}_{i}^{+}),
\end{align*}
where $\psi^{*}_{i}=\psi \big( F_{Z}(|Z_{i1}-Z_{i2}|), F_{Y}(|Y_{i1}-Y_{i2}|)\big)$.
\end{lemma}
\begin{proof}: By the Glivenko-Cantelli theorem, as $I\rightarrow \infty$, we have
\begin{align}\label{eqn: uni converge of Z}
    \Big|r^{z}_{I,i}-F_{Z}(|Z_{i1}-Z_{i2}|)\Big|&= \Big|\frac{1}{I}\sum_{i^{\prime}=1}^{I}\mathbbm{1}\{|Z_{i1}-Z_{i2}|\geq |Z_{i^{\prime}1}-Z_{i^{\prime}2}|\}-F_{Z}(|Z_{i1}-Z_{i2}|)\Big|\nonumber\\
    &\leq \sup_{v}\Big|\frac{1}{I}\sum_{i^{\prime}=1}^{I}\mathbbm{1}(v \geq |Z_{i^{\prime}1}-Z_{i^{\prime}2}|)-F_{Z}(v)\Big| \xrightarrow{a.s.} 0,
\end{align}
\begin{align}\label{eqn: uni converge of Y}
    \Big|r^{y}_{I,i}-F_{Y}(|Y_{i1}-Y_{i2}|)\Big|&= \Big|\frac{1}{I}\sum_{i^{\prime}=1}^{I}\mathbbm{1}(|Y_{i1}-Y_{i2}|\geq |Y_{i^{\prime}1}-Y_{i^{\prime}2}|)-F_{Y}(|Y_{i1}-Y_{i2}|)\Big|\nonumber \\
    &\leq \sup_{v}\Big|\frac{1}{I}\sum_{i^{\prime}=1}^{I}\mathbbm{1}(v \geq |Y_{i^{\prime}1}-Y_{i^{\prime}2}|)-F_{Y}(v)\Big| \xrightarrow{a.s.} 0.
\end{align}
By the multivariate mean value theorem, there exists some $\xi_{I, i}\in [0,1]^{2}$ such that
\begin{equation*}
    \psi(r_{I, i}^{z}, r_{I, i}^{y})-\psi^{*}_{i}=\bigtriangledown \psi(\xi_{I, i})\cdot \big((r_{I, i}^{z}, r_{I, i}^{y})-(F_{Z}(|Z_{i1}-Z_{i2}|), F_{Y}(|Y_{i1}-Y_{i2}|))\big).
\end{equation*}
Therefore, by Condition~\ref{condition: varphi} and conclusions (\ref{eqn: uni converge of Z}) and (\ref{eqn: uni converge of Y}), we have 
\begin{align}\label{eqn: var r converge var F}
   \Big|\psi(r_{I, i}^{z}, r_{I, i}^{y})-\psi^{*}_{i}\Big| &=\Big|\bigtriangledown \psi(\xi_{I, i})\cdot \big((r_{I, i}^{z}, r_{I, i}^{y})-(F_{Z}(|Z_{i1}-Z_{i2}|), F_{Y}(|Y_{i1}-Y_{i2}|)\big)\Big|\nonumber\\
   &\leq \norm{\bigtriangledown \psi(\xi_{I, i})}\norm{(r_{I, i}^{z}, r_{I, i}^{y})-(F_{Z}(|Z_{i1}-Z_{i2}|), F_{Y}(|Y_{i1}-Y_{i2}|))}\nonumber \\
   &\leq C_{\psi} \norm{(r_{I, i}^{z}, r_{I, i}^{y})-(F_{Z}(|Z_{i1}-Z_{i2}|), F_{Y}(|Y_{i1}-Y_{i2}|))} \nonumber\\
   &\leq  C_{\psi} \Big(\Big|r^{z}_{I,i}-F_{Z}(|Z_{i1}-Z_{i2}|)\Big|+\Big|r^{y}_{I,i}-F_{Y}(|Y_{i1}-Y_{i2}|)\Big|\Big)\xrightarrow{a.s.} 0.
\end{align}
Note that 
\begin{align}\label{eqn: decomposition of varphi 1D>0}
&\quad \ I^{-1} \sum_{i=1}^{I}  \psi(r_{I, i}^{z}, r_{I, i}^{y}) \mathbbm{1}\{D_{i}>0\}\nonumber\\
&=I^{-1} \sum_{i=1}^{I} \psi^{*}_{i} \mathbbm{1}\{D_{i}>0\}+ I^{-1} \sum_{i=1}^{I}\Big(\psi(r_{I, i}^{z}, r_{I, i}^{y})-\psi^{*}_{i}\Big) \mathbbm{1}\{D_{i}>0\}.
\end{align}
The first term of the right hand side of (\ref{eqn: decomposition of varphi 1D>0}) converges almost surely to $E[\psi^{*}_{i} \mathbbm{1}\{D_{i}>0\}]$. The second term of the right hand side of (\ref{eqn: decomposition of varphi 1D>0}) converges almost surely to zero since (\ref{eqn: var r converge var F}) implies $\Big(\psi(r_{I, i}^{z}, r_{I, i}^{y})-\psi^{*}_{i}\Big) \mathbbm{1}\{D_{i}>0\}\xrightarrow{a.s.}0$. Therefore, we have 
\begin{align*}
   I^{-1} \sum_{i=1}^{I}  \psi(r_{I, i}^{z}, r_{I, i}^{y}) \mathbbm{1}\{D_{i}>0\}\xrightarrow{a.s.} E\big [\psi^{*}_{i}  \mathbbm{1}\{D_{i}>0\}\big].
\end{align*}

Note that
    \begin{align}\label{eqn: double decomposition of varphi gamma}
   &\quad I^{-1}\sum_{i=1}^{I}\psi(r_{I, i}^{z}, r_{I, i}^{y})  \expit(\widetilde{\gamma}_{I} |\phi(Z_{i1})-\phi(Z_{i2})|)\nonumber \\
    &=I^{-1}\sum_{i=1}^{I}\psi^{*}_{i}\expit(\widetilde{\gamma}|\phi(Z_{i1})-\phi(Z_{i2})|)\nonumber\\
    &\quad \quad + \nonumber I^{-1}\sum_{i=1}^{I}\psi^{*}_{i}\big\{\expit(\widetilde{\gamma}_{I} |\phi(Z_{i1})-\phi(Z_{i2})|)-\expit(\widetilde{\gamma} |\phi(Z_{i1})-\phi(Z_{i2})|)\big\}\\
    &\quad \quad +I^{-1}\sum_{i=1}^{I}\Big(\psi(r_{I, i}^{z}, r_{I, i}^{y})-\psi^{*}_{i}\Big)\expit(\widetilde{\gamma}_{I} |\phi(Z_{i1})-\phi(Z_{i2})|).
\end{align}

For the first term on the right-hand side of (\ref{eqn: double decomposition of varphi gamma}), by the law of large numbers, we have 
\begin{equation*}
    I^{-1}\sum_{i=1}^{I}\psi^{*}_{i}\expit(\widetilde{\gamma} |\phi(Z_{i1})-\phi(Z_{i2})|)\xrightarrow{a.s.} E\big\{\psi^{*}_{i} \expit(\widetilde{\gamma} |\phi(Z_{i1})-\phi(Z_{i2})|)\big\}.
\end{equation*}

For the second term, we have 
\begin{align*}
    &\quad \Big|I^{-1}\sum_{i=1}^{I}\psi^{*}_{i}\big\{\expit(\widetilde{\gamma}_{I} |\phi(Z_{i1})-\phi(Z_{i2})|)-\expit(\widetilde{\gamma} |\phi(Z_{i1})-\phi(Z_{i2})|)\big\}\Big|\\
    &\leq I^{-1}\sum_{i=1}^{I}\Big|\psi^{*}_{i}\big\{\expit(\widetilde{\gamma}_{I} |\phi(Z_{i1})-\phi(Z_{i2})|)-\expit(\widetilde{\gamma} |\phi(Z_{i1})-\phi(Z_{i2})|)\big\}\Big| \\
    &\leq I^{-1}\sum_{i=1}^{I}\Big|\psi^{*}_{i}(\widetilde{\gamma}_{I}-\widetilde{\gamma})(\phi(Z_{i1})-\phi(Z_{i2}))\Big|\\
    &\quad \quad \quad \quad \quad \quad \quad \times \Big|\sup_{\xi} \expit(\xi |\phi(Z_{i1})-\phi(Z_{i2})|)\big\{1-\expit(\xi |\phi(Z_{i1})-\phi(Z_{i2})|)\big\}\Big| \\
    &\leq I^{-1}\sum_{i=1}^{I}\frac{1}{4}\big|\psi^{*}_{i}(\widetilde{\gamma}_{I}-\widetilde{\gamma})(\phi(Z_{i1})-\phi(Z_{i2}))\big| 
    \xrightarrow{a.s.} 0,
\end{align*}
in which the last line follows from $\widetilde{\gamma}_{I}\xrightarrow{a.s.} \widetilde{\gamma}$ (as proved in Lemma~\ref{lem: inverting Gamma}) and $\psi^{*}_{i}$ and $\phi(Z_{i1})-\phi(Z_{i2})$ are bounded (by Conditions~\ref{condition: dose moment} and \ref{condition: varphi}).

For the third term, because $\psi(r_{I, i}^{z}, r_{I, i}^{y})-\psi^{*}_{i}\xrightarrow{a.s.} 0$ and $\expit(\widetilde{\gamma}_{I} |\phi(Z_{i1})-\phi(Z_{i2})|)\in [0,1]$ for any $i$, we have 
\begin{equation*}
     I^{-1}\sum_{i=1}^{I}\Big(\psi(r_{I, i}^{z}, r_{I, i}^{y})-\psi^{*}_{i}\Big)\expit(\widetilde{\gamma}_{I} |\phi(Z_{i1})-\phi(Z_{i2})|)\xrightarrow{a.s.} 0.
\end{equation*}

Putting all the above results together, the desired convergence results have been established.
\end{proof}

\begin{lemma}\label{lem: continuity of varphi gamma}
We define $\varphi(\gamma)=\psi^{*}_{i}  \expit(\gamma |\phi(Z_{i1})-\phi(Z_{i2})|)$. Then, under Conditions~\ref{condition: dose moment} and \ref{condition: varphi}, the $E\{\varphi(\gamma)\}$ is a continuous function of $\gamma$.
\end{lemma}

\begin{proof}: For any $0\leq \gamma_{1}<\gamma_{2}$, we have 
    \begin{align*}
       &\quad \ |E\{\varphi(\gamma_{2})\}-E\{\varphi(\gamma_{1})\}|\\
       &\leq E\Big|\psi^{*}_{i}\big(\expit(\gamma_{2} |\phi(Z_{i1})-\phi(Z_{i2})|)- \expit(\gamma_{1} |\phi(Z_{i1})-\phi(Z_{i2})|)\big)\Big|\\
        &\leq \sup \psi^{*}_{i} E\Big(\expit(\gamma_{2} |\phi(Z_{i1})-\phi(Z_{i2})|)- \expit(\gamma_{1} |\phi(Z_{i1})-\phi(Z_{i2})|)\Big)\\
        &\leq \sup \psi^{*}_{i} \big(\sup_{\xi} \expit(\xi)(1-\expit(\xi))\big) E\{(\gamma_{2}-\gamma_{1})|\phi(Z_{i2})-\phi(Z_{i1})|\}\\
        &\leq \sup \psi^{*}_{i} \frac{1}{4}\big(E|\phi(Z_{i1})|+E|\phi(Z_{i2})|\big)(\gamma_{2}-\gamma_{1}).
    \end{align*}
    By Conditions~\ref{condition: dose moment} and \ref{condition: varphi}, there exists some finite constant $H$ such that $|E\{\varphi(\gamma_{2})\}-E\{\varphi(\gamma_{1})\}|\leq H\times (\gamma_{2}-\gamma_{1})$. This implies that $E\{\varphi(\gamma)\}$ is a continuous function of $\gamma$.
\end{proof}

\begin{lemma}\label{lem: monotonicity of varphi gamma}
Under Conditions \ref{condi: overlap of supports} and \ref{condition: varphi}, the $E\{\varphi(\gamma)\}$ is strictly monotonically increasing with respect to $\gamma$ over $\gamma\in [0, +\infty)$.
\end{lemma}

\begin{proof}: Note that for any $(Z_{i1}, Z_{i2}, Y_{i1}, Y_{i2})$, the function $\varphi(\gamma)>0$ is monotonically increasing over $\gamma\in [0, +\infty)$. Under Condition~\ref{condi: overlap of supports}, we let $A_{1}, A_{2}, B_{1}, B_{2}, C_{1}, C_{2}>0$ be six positive constants such that $P(\mathcal{A})>0$ where the event $\mathcal{A}=\{(Z_{i1}, Z_{i2}, Y_{i1}, Y_{i2}): |\phi(Z_{i1})-\phi(Z_{i2})| \in [A_{1}, A_{2}], |Z_{i1}-Z_{i2}|\in [B_{1}, B_{2}], |Y_{i1}-Y_{i2}|\in [C_{1}, C_{2}]\}$. Note that when $|Z_{i1}-Z_{i2}|\in [B_{1}, B_{2}]$ and $|Y_{i1}-Y_{i2}|\in [C_{1}, C_{2}]$, we have $\big( F_{Z}(|Z_{i1}-Z_{i2}|), F_{Y}(|Y_{i1}-Y_{i2}|)\big)\in [F_{Z}(B_{1}), F_{Z}(B_{2})]\times [F_{Y}(C_{1}), F_{Y}(C_{2})]$. Therefore, under Condition~\ref{condition: varphi}, there exist positive constants $L_{1}>0$ and $L_{2}>0$ such that $\psi^{*}_{i}\in [L_{1}, L_{2}]$ for any $(Z_{i1}, Z_{i2}, Y_{i1}, Y_{i2})$ with $|Z_{i1}-Z_{i2}|\in [B_{1}, B_{2}]$ and $|Y_{i1}-Y_{i2}|\in [C_{1}, C_{2}]$. Moreover, given $0\leq \gamma_{1}<\gamma_{2}$, for any $(Z_{i1}, Z_{i2})$ with $|\phi(Z_{i1})-\phi(Z_{i2})|\in [A_{1}, A_{2}]$, we have 
\begin{align*}
   &\quad \expit(\gamma_{2} |\phi(Z_{i1})-\phi(Z_{i2})|)- \expit(\gamma_{1} |\phi(Z_{i1})-\phi(Z_{i2})|)\\
   &\geq \inf_{\xi\in [\gamma_{1} |\phi(Z_{i1})-\phi(Z_{i2})|, \gamma_{2} |\phi(Z_{i1})-\phi(Z_{i2})|] }\expit(\xi)(1-\expit(\xi))(\gamma_{2}-\gamma_{1})|\phi(Z_{i2})-\phi(Z_{i1})|\\
   &\geq A_{1}\times \inf_{\xi\in [\gamma_{1} A_{1}, \gamma_{2} A_{2}] }\expit(\xi)(1-\expit(\xi))(\gamma_{2}-\gamma_{1}):=Q>0.
    \end{align*}

Therefore, for any $0\leq \gamma_{1}<\gamma_{2}$, we have 
\begin{align*}
   &\quad \ E\big\{\varphi(\gamma_{2})\big\}-E\big\{\varphi(\gamma_{1})\big\}\\&=E\Big[\psi^{*}_{i}\times  \Big( \expit(\gamma_{2} |\phi(Z_{i1})-\phi(Z_{i2})|)- \expit(\gamma_{1} |\phi(Z_{i1})-\phi(Z_{i2})|)\Big) \Big]\\
    &= P\big(\mathcal{A} \big)\times E\Big[\psi^{*}_{i}\times  \Big( \expit(\gamma_{2} |\phi(Z_{i1})-\phi(Z_{i2})|)- \expit(\gamma_{1} |\phi(Z_{i1})-\phi(Z_{i2})|)\Big)\mid \mathcal{A} \Big]\\
    & \quad \quad  + P\big(\mathcal{A}^{c} \big)\times E\Big[\psi^{*}_{i}\times   \Big( \expit(\gamma_{2} |\phi(Z_{i1})-\phi(Z_{i2})|)- \expit(\gamma_{1} |\phi(Z_{i1})-\phi(Z_{i2})|)\Big)\mid \mathcal{A}^{c} \Big] \\
    &\geq P\big(\mathcal{A} \big)\times E\Big[\psi^{*}_{i}\times  \Big( \expit(\gamma_{2} |\phi(Z_{i1})-\phi(Z_{i2})|)- \expit(\gamma_{1} |\phi(Z_{i1})-\phi(Z_{i2})|)\Big)\mid \mathcal{A} \Big]\\
    & \geq P\big(\mathcal{A} \big)\times L_{1} \times Q>0.
\end{align*}

This implies that $E\{\varphi(\gamma)\}$ is strictly monotonically increasing over $\gamma\in [0, +\infty)$.

\end{proof}

\begin{lemma}\label{lem: upp and low bounds of varphi}
Under Conditions~\ref{condition: varphi} and \ref{condition: prob of D}, we have: $\lim_{\gamma\rightarrow 0^{+}}E\{\varphi(\gamma)\}<E\big [\psi^{*}_{i}  \mathbbm{1}\{D_{i}>0\}\big]$ and $\lim_{\gamma\rightarrow +\infty}E\{\varphi(\gamma)\}>E\big [\psi^{*}_{i}  \mathbbm{1}\{D_{i}>0\}\big]$.
\end{lemma}
\begin{proof}: Since $\varphi$ is bounded (implied by Condition~\ref{condition: varphi}), by the bounded convergence theorem, we have 
\begin{equation}\label{eqn: converge to 0}
  \lim_{\gamma \rightarrow 0^{+}} E\{\varphi(\gamma)\}
  =E\Big \{ \lim_{\gamma \rightarrow 0^{+}}\varphi(\gamma) \Big\}\\
  =\frac{1}{2}E\big\{ \psi^{*}_{i} \big\},
\end{equation}
\begin{equation}\label{eqn: converge to inf}
  \lim_{\gamma \rightarrow +\infty } E\{\varphi(\gamma)\}
  =E\Big \{ \lim_{\gamma \rightarrow +\infty } \varphi(\gamma) \Big\}=E\big\{ \psi^{*}_{i} \big\}.
\end{equation}
Note that
\begin{align*}
    E\big[\psi^{*}_{i} \mathbbm{1}\{D_{i}>0\}\big]=E\Big[\psi^{*}_{i}  \times E\big\{\mathbbm{1}\{D_{i}>0\}\mid \psi^{*}_{i}  \big\} \Big]=E\big\{\psi^{*}_{i}  \times P(D_{i}>0\mid \psi^{*}_{i}) \big\}.
\end{align*}
Therefore, under Condition~\ref{condition: prob of D}, we have
\begin{equation}\label{eqn: 1/2 less than}
    \frac{1}{2}E(\psi^{*}_{i})<E\big[\psi^{*}_{i} \mathbbm{1}\{D_{i}>0\}\big],
\end{equation}
and 
\begin{equation}\label{eqn: 1/1 greater than}
   E(\psi^{*}_{i})>E\big[\psi^{*}_{i} \mathbbm{1}\{D_{i}>0\}\big].
\end{equation}
By (\ref{eqn: converge to 0}) and (\ref{eqn: 1/2 less than}), we have $\lim_{\gamma\rightarrow 0^{+}}E\{\varphi(\gamma)\}<E\big [\psi^{*}_{i}  \mathbbm{1}\{D_{i}>0\}\big]$. By (\ref{eqn: converge to inf}) and (\ref{eqn: 1/1 greater than}), we have $\lim_{\gamma\rightarrow +\infty}E\{\varphi(\gamma)\}>E\big [\psi^{*}_{i}  \mathbbm{1}\{D_{i}>0\}\big]$.
\end{proof}

\begin{proof}[Proof of Theorem 3.4:]
Lemmas \ref{lem: continuity of varphi gamma}--\ref{lem: upp and low bounds of varphi} immediately imply that the following equation of $\gamma$,
\begin{equation*}\label{eqn: design sensitivity equation}
    E\Big[\psi^{*}_{i}  \expit(\gamma |\phi(Z_{i1})-\phi(Z_{i2})|) \Big]= E\big [\psi^{*}_{i}  \mathbbm{1}\{D_{i}>0\}\big],
\end{equation*}
has a unique solution over $\gamma \in (0, +\infty)$, denoted as $\gamma_{*}$. 

Note that the power of a sensitivity analysis of a rank-based test $T_{\psi}$ is
\begin{align*}
    \Psi_{I, \overline{\Gamma}}&=P\left(\frac{\sum_{i=1}^{I}\psi(r_{I, i}^{z}, r_{I, i}^{y})\mathbbm{1}\{D_{i}>0\}-\sum_{i=1}^{I}\psi(r_{I, i}^{z}, r_{I, i}^{y})p_{I, i}^{+} }{\sqrt{\sum_{i=1}^{I}\psi^{2}(r_{I, i}^{z}, r_{I, i}^{y})p_{I, i}^{+}(1-p_{I, i}^{+})}}\geq \Phi^{-1}(1-\alpha) \right)\\
        &=P\left(\frac{\sqrt{I}\{I^{-1}\sum_{i=1}^{I}\psi(r_{I, i}^{z}, r_{I, i}^{y})\mathbbm{1}\{D_{i}>0\}-I^{-1}\sum_{i=1}^{I}\psi(r_{I, i}^{z}, r_{I, i}^{y})p_{I, i}^{+}\} }{\sqrt{I^{-1}\sum_{i=1}^{I}\psi^{2}(r_{I, i}^{z}, r_{I, i}^{y})p_{I, i}^{+}(1-p_{I, i}^{+})}}\geq \Phi^{-1}(1-\alpha) \right),
\end{align*}
in which the $\sqrt{I^{-1}\sum_{i=1}^{I}\psi^{2}(r_{I, i}^{z}, r_{I, i}^{y})p_{I, i}^{+}(1-p_{I, i}^{+})}$ is bounded above by some constant $C_{\sigma}>0$ for any $I$ because $\psi$ is bounded.

Therefore, if $\overline{\Gamma}<\overline{\Gamma}_{*}$ (or, equivalently, if $\widetilde{\gamma}<\gamma_{*}$), Lemma~\ref{lem: monotonicity of varphi gamma} implies that $E\{\varphi(\widetilde{\gamma})\}<E\{\varphi(\gamma_{*})\}$. Then, by Lemma~\ref{lemma: convergence}, we have
\begin{align*}
  &\quad \ \frac{\sum_{i=1}^{I}\psi(r_{I, i}^{z}, r_{I, i}^{y})\mathbbm{1}\{D_{i}>0\}-\sum_{i=1}^{I}\psi(r_{I, i}^{z}, r_{I, i}^{y})p_{I, i}^{+} }{\sqrt{\sum_{i=1}^{I}\psi^{2}(r_{I, i}^{z}, r_{I, i}^{y})p_{I, i}^{+}(1-p_{I, i}^{+})}}\\
  &\cong \frac{\sqrt{I}\big[E\{\varphi(\gamma_{*} )\}-E\{\varphi(\widetilde{\gamma})\}\big] }{\sqrt{I^{-1}\sum_{i=1}^{I}\psi^{2}(r_{I, i}^{z}, r_{I, i}^{y})p_{I, i}^{+}(1-p_{I, i}^{+})}}\rightarrow +\infty,
\end{align*}
in which ``$\cong$" means ``asymptotically equivalent." Similarly, if $\overline{\Gamma}>\overline{\Gamma}_{*}$ (or, equivalently, if $\widetilde{\gamma}>\gamma_{*}$), Lemma~\ref{lem: monotonicity of varphi gamma} implies that $E\{\varphi(\widetilde{\gamma})\}>E\{\varphi(\gamma_{*})\}$. Similarly, by Lemma~\ref{lemma: convergence}, we have
\begin{align*}
   &\quad \ \frac{\sum_{i=1}^{I}\psi(r_{I, i}^{z}, r_{I, i}^{y})\mathbbm{1}\{D_{i}>0\}-\sum_{i=1}^{I}\psi(r_{I, i}^{z}, r_{I, i}^{y})p_{I, i}^{+} }{\sqrt{\sum_{i=1}^{I}\psi^{2}(r_{I, i}^{z}, r_{I, i}^{y})p_{I, i}^{+}(1-p_{I, i}^{+})}}\\
   &\cong \frac{\sqrt{I}\big[E\{\varphi(\gamma_{*} )\}-E\{\varphi(\widetilde{\gamma})\}\big]}{\sqrt{I^{-1}\sum_{i=1}^{I}\psi^{2}(r_{I, i}^{z}, r_{I, i}^{y})p_{I, i}^{+}(1-p_{I, i}^{+})}}\rightarrow -\infty. 
\end{align*}

That is, the desired conclusion has been established. 
\end{proof}

\subsection*{A.4: Proof of Theorem 3.5}

Recall that, given a general rank-based test $T_{\psi}=\sum_{i=1}^{I}\psi(r_{I, i}^{z}, r_{I, i}^{y}) \mathbbm{1}\{D_{i}>0 \}$, we let $T^{+}_{\psi}=\sum_{i=1}^{I}\psi(r_{I, i}^{z}, r_{I, i}^{y}) B_{i}^{+}$ be the upper-bounding test statistics of $T_{\psi}$ under the prespecified sensitivity parameter $\overline{\Gamma}$, in which $B^{+}_{i}$ are independent random variables taking the value $1$ with probability $p_{I, i}^{+}=\expit\{\widetilde{\gamma}_{I}|\phi(Z_{i1})-\phi(Z_{i2})|\}$ and the value $0$ with probability $1-p_{I, i}^{+}$. For simplicity of notations, we denote $\psi(r_{I, i}^{z}, r_{I, i}^{y})$ as $\psi_{I,i}$. Then, the moment generating function of $T^{+}_{\psi}$ can be expressed as 
\begin{equation*}
    m_{I}(t)=\prod_{i=1}^{I} \Big \{ p_{I, i}^{+}\exp(\psi_{I,i}t)+(1-p_{I, i}^{+})\Big\}, \quad t\in [0, +\infty).
\end{equation*}
We let $\chi_{I}(t)=\log (m_{I}(t))$ denote the cumulant generating function of $T^{+}_{\psi}$. Then, we have 
\begin{align*}
    & \chi_{I}(t)=\sum_{i=1}^{I} \log \Big \{ p_{I, i}^{+}\exp(\psi_{I,i}t)+(1-p_{I, i}^{+})\Big\}, \\
    & \chi^{\prime}_{I}(t)=\sum_{i=1}^{I}\frac{p_{I, i}^{+}\psi_{I,i}\exp(\psi_{I,i}t)}{p_{I, i}^{+}\exp(\psi_{I,i}t)+(1-p_{I, i}^{+})},\\
    & \chi^{\prime \prime}_{I}(t)=\sum_{i=1}^{I}\frac{p_{I, i}^{+}(1-p_{I, i}^{+})\psi_{I,i}^{2}\exp(\psi_{I,i}t)}{\{p_{I, i}^{+}\exp(\psi_{I,i}t)+(1-p_{I, i}^{+})\}^{2}}, \\
    & \chi^{\prime \prime \prime }_{I}(t)=\sum_{i=1}^{I}\frac{p_{I, i}^{+}(1-p_{I, i}^{+})\{(1-p_{I, i}^{+})-p_{I, i}^{+}\exp(\psi_{I,i}t) \}\psi_{I,i}^{3}\exp(\psi_{I,i}t)}{\{p_{I, i}^{+}\exp(\psi_{I,i}t)+(1-p_{I, i}^{+})\}^{3}}.
\end{align*}

\begin{lemma}\label{lemma: convergence for bahadur}
Under Conditions \ref{condition: dose moment}--\ref{condition: varphi}, as $I\rightarrow \infty$, we have 
\begin{align*}
    & I^{-1}\chi_{I}(t)\xrightarrow{a.s.} E\Big[\log \Big \{ \widetilde{p}_{i}^{+}\exp(\psi^{*}_{i}t)+(1-\widetilde{p}_{i}^{+})\Big\}\Big]:= \omega_{0}(t), \\
    & I^{-1}\chi^{\prime}_{I}(t)\xrightarrow{a.s.} E\Big[\frac{\widetilde{p}_{i}^{+}\psi^{*}_{i}\exp(\psi^{*}_{i}t)}{\widetilde{p}_{i}^{+}\exp(\psi^{*}_{i}t)+(1-\widetilde{p}_{i}^{+})} \Big]:=\omega_{1}(t), \\
    & I^{-1}\chi^{\prime\prime}_{I}(t)\xrightarrow{a.s.} E\Big[\frac{\widetilde{p}_{i}^{+}(1-\widetilde{p}_{i}^{+})\psi_{i}^{*2}\exp(\psi^{*}_{i}t)}{\{\widetilde{p}_{i}^{+}\exp(\psi^{*}_{i}t)+(1-\widetilde{p}_{i}^{+})\}^{2}} \Big]:=\omega_{2}(t).
\end{align*}
\end{lemma}

\begin{proof}: Because $\operatorname{expit}$ function is Lipschitz continuous with Lipschitz constant being $1/4$, Lemma~\ref{lem: inverting Gamma} and the boundedness of $|\phi(Z_{i1})-\phi(Z_{i2})|$ in Condition~\ref{condition: dose moment} imply that 
\begin{equation*}
    \max_{i\leq I}\left|p_{I,i}^{+}-\widetilde p_i^{+}\right|\leq\frac{1}{4}\left|\widetilde\gamma_I-\widetilde\gamma\right|\cdot \max_{i\leq I}\left|\phi(Z_{i1})-\phi(Z_{i2})\right|\xrightarrow{a.s.}0.
\end{equation*}
Therefore, we have 
\begin{equation*}
    I^{-1}\sum_{i=1}^{I}|p_{I,i}^{+}-\widetilde p_i^{+}|\xrightarrow{a.s.}0. 
\end{equation*}
Similarly, the uniform Glivenko-Cantelli bounds used in the proof of Lemma~\ref{lemma: convergence}, together with Condition~\ref{condition: varphi}, imply that $\max_{i\leq I}|\psi_{I,i}-\psi_i^*|\xrightarrow{a.s.}0$, which gives
$$I^{-1}\sum_{i=1}^{I}\left|\psi_{I,i}-\psi_i^*\right|\xrightarrow{a.s.}0.$$

 For $I^{-1}\chi_{I}(t)$, note that
\begin{align}\label{eqn: first converge of Bahadur}
  &\quad \ I^{-1}\sum_{i=1}^{I} \log \Big \{ p_{I, i}^{+}\exp(\psi_{I,i}t)+(1-p_{I, i}^{+})\Big\}\nonumber\\
  &=I^{-1}\sum_{i=1}^{I} \log \Big \{ \widetilde{p}_{i}^{+}\exp(\psi^{*}_{i}t)+(1-\widetilde{p}_{i}^{+})\Big\}\nonumber\\
  &\quad \quad +I^{-1}\sum_{i=1}^{I} \log \Big \{ p_{I, i}^{+}\exp(\psi^{*}_{i}t)+(1-p_{I, i}^{+})\Big\}-I^{-1}\sum_{i=1}^{I} \log \Big \{ \widetilde{p}_{i}^{+}\exp(\psi^{*}_{i}t)+(1-\widetilde{p}_{i}^{+})\Big\}\nonumber\\
  &\quad \quad + I^{-1}\sum_{i=1}^{I} \log \Big \{ p_{I, i}^{+}\exp(\psi_{I,i}t)+(1-p_{I, i}^{+})\Big\}-I^{-1}\sum_{i=1}^{I} \log \Big \{ p_{I, i}^{+}\exp(\psi^{*}_{i}t)+(1-p_{I, i}^{+})\Big\}.
\end{align}

For the first term of the right-hand side of (\ref{eqn: first converge of Bahadur}), by the law of large numbers, we have
\begin{equation*}
    I^{-1}\sum_{i=1}^{I} \log \Big \{ \widetilde{p}_{i}^{+}\exp(\psi^{*}_{i}t)+(1-\widetilde{p}_{i}^{+})\Big\}\xrightarrow{a.s.} E\Big[\log \Big \{ \widetilde{p}_{i}^{+}\exp(\psi^{*}_{i}t)+(1-\widetilde{p}_{i}^{+})\Big\}\Big].
\end{equation*}

For the second term, we have 
\begin{align*}                     
    &\quad \ \Big|I^{-1}\sum_{i=1}^{I} \log \Big \{ p_{I, i}^{+}\exp(\psi^{*}_{i}t)+(1-p_{I, i}^{+})\Big\}-I^{-1}\sum_{i=1}^{I} \log \Big \{ \widetilde{p}_{i}^{+}\exp(\psi^{*}_{i}t)+(1-\widetilde{p}_{i}^{+})\Big\}\Big|\\
       &\leq I^{-1}\sum_{i=1}^{I}\Big|\log \Big \{ p_{I, i}^{+}\exp(\psi^{*}_{i}t)+(1-p_{I, i}^{+})\Big\}- \log \Big \{ \widetilde{p}_{i}^{+}\exp(\psi^{*}_{i}t)+(1-\widetilde{p}_{i}^{+})\Big\} \Big| \\
       &\leq\sup_{\xi \in [0,1],\psi_{i}^{*}\in [0,\sup\psi] }\frac{\exp(\psi_{i}^{*} t)-1 }{\xi \exp(\psi_{i}^{*} t)+(1-\xi)} I^{-1}\sum_{i=1}^{I}|p^{+}_{I,i}-\widetilde{p}_{i}^{+}|\xrightarrow{a.s.} 0,
\end{align*}
in which we used the fact that $\sup_{\xi \in [0,1],\psi_{i}^{*}\in [0,\sup\psi]  }\frac{\exp(\psi_{i}^{*} t)-1 }{\xi \exp(\psi_{i}^{*} t)+(1-\xi)}$ is finite for each fixed $t\geq 0$. 

For the third term, we have
    \begin{align*}
       &\quad \ \Big|I^{-1}\sum_{i=1}^{I} \log \Big \{ p_{I, i}^{+}\exp(\psi_{I,i}t)+(1-p_{I, i}^{+})\Big\}-I^{-1}\sum_{i=1}^{I} \log \Big \{ p_{I, i}^{+}\exp(\psi^{*}_{i}t)+(1-p_{I, i}^{+})\Big\}\Big|\\
       &\leq I^{-1}\sum_{i=1}^{I}\Big|\log \Big \{ p_{I, i}^{+}\exp(\psi_{I,i}t)+(1-p_{I, i}^{+})\Big\}- \log \Big \{ p_{I, i}^{+}\exp(\psi^{*}_{i}t)+(1-p_{I, i}^{+})\Big\} \Big| \\
       &\leq\sup_{\xi\in [0, \sup \psi], p_{I, i}^{+} }\frac{p_{I, i}^{+}t\exp(\xi t) }{p_{I, i}^{+}\exp(\xi t)+(1-p_{I, i}^{+})} I^{-1}\sum_{i=1}^{I}|\psi_{I,i}-\psi^{*}_{i}|\xrightarrow{a.s.} 0,
    \end{align*}
    in which we used the fact that $\sup_{\xi\in [0, \sup \psi], p_{I, i}^{+}}\frac{p_{I, i}^{+}t\exp(\xi t) }{p_{I, i}^{+}\exp(\xi t)+(1-p_{I, i}^{+})}$ is finite for each fixed $t\geq 0$. 

    Therefore, we have shown that 
    \begin{equation*}
    I^{-1}\chi_{I}(t)\xrightarrow{a.s.} E\Big[\log \Big \{ \widetilde{p}_{i}^{+}\exp(\psi^{*}_{i}t)+(1-\widetilde{p}_{i}^{+})\Big\}\Big]:= \omega_{0}(t).
    \end{equation*}
    
    For $I^{-1}\chi_{I}^{\prime}(t)$, note that
    \begin{align}\label{eqn: second converge of Bahadur}
  &\quad \ I^{-1}\sum_{i=1}^{I} \frac{p_{I, i}^{+}\psi_{I,i}\exp(\psi_{I,i}t)}{p_{I, i}^{+}\exp(\psi_{I,i}t)+(1-p_{I, i}^{+})}\nonumber\\
  &=I^{-1}\sum_{i=1}^{I} \frac{\widetilde{p}_{i}^{+}\psi^{*}_{i}\exp(\psi^{*}_{i}t)}{\widetilde{p}_{i}^{+}\exp(\psi^{*}_{i}t)+(1-\widetilde{p}_{i}^{+})}\nonumber\\
  &\quad \quad +I^{-1}\sum_{i=1}^{I} \frac{p_{I, i}^{+}\psi^{*}_{i}\exp(\psi^{*}_{i}t)}{p_{I, i}^{+}\exp(\psi^{*}_{i}t)+(1-p_{I, i}^{+})}-I^{-1}\sum_{i=1}^{I} \frac{\widetilde{p}_{i}^{+}\psi^{*}_{i}\exp(\psi^{*}_{i}t)}{\widetilde{p}_{i}^{+}\exp(\psi^{*}_{i}t)+(1-\widetilde{p}_{i}^{+})}\nonumber\\
  &\quad \quad + I^{-1}\sum_{i=1}^{I} \frac{p_{I, i}^{+}\psi_{I,i}\exp(\psi_{I,i}t)}{p_{I, i}^{+}\exp(\psi_{I,i}t)+(1-p_{I, i}^{+})}-I^{-1}\sum_{i=1}^{I} \frac{p_{I, i}^{+}\psi^{*}_{i}\exp(\psi^{*}_{i}t)}{p_{I, i}^{+}\exp(\psi^{*}_{i}t)+(1-p_{I, i}^{+})}.
\end{align}

For the first term of the right-hand side of (\ref{eqn: second converge of Bahadur}), by the law of large numbers, we have 
\begin{equation*}
     I^{-1}\sum_{i=1}^{I} \frac{\widetilde{p}_{i}^{+}\psi^{*}_{i}\exp(\psi^{*}_{i}t)}{\widetilde{p}_{i}^{+}\exp(\psi^{*}_{i}t)+(1-\widetilde{p}_{i}^{+})}\xrightarrow{a.s.} E\Big[\frac{\widetilde{p}_{i}^{+}\psi^{*}_{i}\exp(\psi^{*}_{i}t)}{\widetilde{p}_{i}^{+}\exp(\psi^{*}_{i}t)+(1-\widetilde{p}_{i}^{+})} \Big].
\end{equation*}

For the second term, we have 
    \begin{align*}
       &\quad \ \Big|I^{-1}\sum_{i=1}^{I} \frac{p_{I, i}^{+}\psi^{*}_{i}\exp(\psi^{*}_{i}t)}{p_{I, i}^{+}\exp(\psi^{*}_{i}t)+(1-p_{I, i}^{+})}-I^{-1}\sum_{i=1}^{I} \frac{\widetilde{p}_{i}^{+}\psi^{*}_{i}\exp(\psi^{*}_{i}t)}{\widetilde{p}_{i}^{+}\exp(\psi^{*}_{i}t)+(1-\widetilde{p}_{i}^{+})}\Big|\\
       &\leq I^{-1}\sum_{i=1}^{I}\Big| \frac{p_{I, i}^{+}\psi^{*}_{i}\exp(\psi^{*}_{i}t)}{p_{I, i}^{+}\exp(\psi^{*}_{i}t)+(1-p_{I, i}^{+})}- \frac{\widetilde{p}_{i}^{+}\psi^{*}_{i}\exp(\psi^{*}_{i}t)}{\widetilde{p}_{i}^{+}\exp(\psi^{*}_{i}t)+(1-\widetilde{p}_{i}^{+})} \Big| \\
       &\leq\sup_{\xi\in [0,1], \psi^{*}_{i}\in [0,\sup\psi]  }\frac{\psi_{i}^{*}\exp(\psi_{i}^{*}t) }{\{\xi\exp(\psi^{*}_{i}t)+(1-\xi)\}^{2}} I^{-1}\sum_{i=1}^{I}|p^{+}_{I,i}-\widetilde{p}_{i}^{+}|\xrightarrow{a.s.} 0,
    \end{align*}
    in which we used the fact that $\sup_{\xi\in [0,1], \psi^{*}_{i}\in [0,\sup\psi]  }\frac{\psi_{i}^{*}\exp(\psi_{i}^{*}t) }{\{\xi\exp(\psi^{*}_{i}t)+(1-\xi)\}^{2}}$ is finite for each fixed $t\geq 0$.
    
   For the third term, we have
     \begin{align*}
       &\quad \ \Big|I^{-1}\sum_{i=1}^{I} \frac{p_{I, i}^{+}\psi_{I,i}\exp(\psi_{I,i}t)}{p_{I, i}^{+}\exp(\psi_{I,i}t)+(1-p_{I, i}^{+})}-I^{-1}\sum_{i=1}^{I} \frac{p_{I, i}^{+}\psi^{*}_{i}\exp(\psi^{*}_{i}t)}{p_{I, i}^{+}\exp(\psi^{*}_{i}t)+(1-p_{I, i}^{+})}\Big|\\
       &\leq I^{-1}\sum_{i=1}^{I}\Big|\frac{p_{I, i}^{+}\psi_{I,i}\exp(\psi_{I,i}t)}{p_{I, i}^{+}\exp(\psi_{I,i}t)+(1-p_{I, i}^{+})}- \frac{p_{I, i}^{+}\psi^{*}_{i}\exp(\psi^{*}_{i}t)}{p_{I, i}^{+}\exp(\psi^{*}_{i}t)+(1-p_{I, i}^{+})} \Big| \\
       &\leq\sup_{\xi\in  [0, \sup \psi ], p_{I, i}^{+} }\Big|\frac{p_{I, i}^{+}(1+\xi t)\exp(\xi t)}{p_{I, i}^{+}\exp(\xi t)+(1-p_{I, i}^{+})}-\frac{p_{I, i}^{+,2}\xi t\exp(2\xi t)}{\{p_{I, i}^{+}\exp(\xi t)+(1-p_{I, i}^{+})\}^{2}}\Big|\\
       &\quad \quad \quad \quad \quad \quad \quad \quad \times I^{-1}\sum_{i=1}^{I}|\psi_{I,i}-\psi^{*}_{i}|\xrightarrow{a.s.} 0,
    \end{align*}
    in which we used the fact that $\sup_{\xi\in  [0, \sup \psi ], p_{I, i}^{+} }\Big|\frac{p_{I, i}^{+}(1+\xi t)\exp(\xi t)}{p_{I, i}^{+}\exp(\xi t)+(1-p_{I, i}^{+})}-\frac{p_{I, i}^{+,2}\xi t\exp(2\xi t)}{\{p_{I, i}^{+}\exp(\xi t)+(1-p_{I, i}^{+})\}^{2}}\Big|$ is finite for each fixed $t\geq 0$.

Therefore, we have shown that
\begin{equation*}
     I^{-1}\chi^{\prime}_{I}(t)\xrightarrow{a.s.} E\Big[\frac{\widetilde{p}_{i}^{+}\psi^{*}_{i}\exp(\psi^{*}_{i}t)}{\widetilde{p}_{i}^{+}\exp(\psi^{*}_{i}t)+(1-\widetilde{p}_{i}^{+})} \Big]:=\omega_{1}(t).
\end{equation*}

    For $I^{-1}\chi_{I}^{\prime\prime}(t)$, note that 
    \begin{align}\label{eqn: third converge of Bahadur}
  &\quad \ I^{-1}\sum_{i=1}^{I} \frac{p_{I, i}^{+}(1-p_{I, i}^{+})\psi_{I,i}^{2}\exp(\psi_{I,i}t)}{\{p_{I, i}^{+}\exp(\psi_{I,i}t)+(1-p_{I, i}^{+})\}^{2}}\nonumber\\
  &=I^{-1}\sum_{i=1}^{I} \frac{\widetilde{p}_{i}^{+}(1-\widetilde{p}_{i}^{+})\psi_{i}^{*2}\exp(\psi^{*}_{i}t)}{\{\widetilde{p}_{i}^{+}\exp(\psi^{*}_{i}t)+(1-\widetilde{p}_{i}^{+})\}^{2}}\nonumber\\
  &\quad \quad +I^{-1}\sum_{i=1}^{I} \frac{p_{I, i}^{+}(1-p_{I, i}^{+})\psi_{i}^{*2}\exp(\psi^{*}_{i}t)}{\{p_{I, i}^{+}\exp(\psi^{*}_{i}t)+(1-p_{I, i}^{+})\}^{2}}-I^{-1}\sum_{i=1}^{I} \frac{\widetilde{p}_{i}^{+}(1-\widetilde{p}_{i}^{+})\psi_{i}^{*2}\exp(\psi^{*}_{i}t)}{\{\widetilde{p}_{i}^{+}\exp(\psi^{*}_{i}t)+(1-\widetilde{p}_{i}^{+})\}^{2}} \nonumber\\
  &\quad \quad + I^{-1}\sum_{i=1}^{I} \frac{p_{I, i}^{+}(1-p_{I, i}^{+})\psi_{I,i}^{2}\exp(\psi_{I,i}t)}{\{p_{I, i}^{+}\exp(\psi_{I,i}t)+(1-p_{I, i}^{+})\}^{2}}-I^{-1}\sum_{i=1}^{I} \frac{p_{I, i}^{+}(1-p_{I, i}^{+})\psi_{i}^{*2}\exp(\psi^{*}_{i}t)}{\{p_{I, i}^{+}\exp(\psi^{*}_{i}t)+(1-p_{I, i}^{+})\}^{2}}.
\end{align}

For the first term, by the law of large numbers, we have 
\begin{equation*}
    I^{-1}\sum_{i=1}^{I} \frac{\widetilde{p}_{i}^{+}(1-\widetilde{p}_{i}^{+})\psi_{i}^{*2}\exp(\psi^{*}_{i}t)}{\{\widetilde{p}_{i}^{+}\exp(\psi^{*}_{i}t)+(1-\widetilde{p}_{i}^{+})\}^{2}}\xrightarrow{a.s.} E\Big\{\frac{\widetilde{p}_{i}^{+}(1-\widetilde{p}_{i}^{+})\psi_{i}^{*2}\exp(\psi^{*}_{i}t)}{\{\widetilde{p}_{i}^{+}\exp(\psi^{*}_{i}t)+(1-\widetilde{p}_{i}^{+})\}^{2}}\Big\}.
\end{equation*}

For the second term, we have
    \begin{align*}
       &\quad \ \Big|I^{-1}\sum_{i=1}^{I} \frac{p_{I, i}^{+}(1-p_{I, i}^{+})\psi_{i}^{*2}\exp(\psi^{*}_{i}t)}{\{p_{I, i}^{+}\exp(\psi^{*}_{i}t)+(1-p_{I, i}^{+})\}^{2}}-I^{-1}\sum_{i=1}^{I} \frac{\widetilde{p}_{i}^{+}(1-\widetilde{p}_{i}^{+})\psi_{i}^{*2}\exp(\psi^{*}_{i}t)}{\{\widetilde{p}_{i}^{+}\exp(\psi^{*}_{i}t)+(1-\widetilde{p}_{i}^{+})\}^{2}}\Big|\\
       &\leq I^{-1}\sum_{i=1}^{I}\Big|\frac{p_{I, i}^{+}(1-p_{I, i}^{+})\psi_{i}^{*2}\exp(\psi^{*}_{i}t)}{\{p_{I, i}^{+}\exp(\psi^{*}_{i}t)+(1-p_{I, i}^{+})\}^{2}}- \frac{\widetilde{p}_{i}^{+}(1-\widetilde{p}_{i}^{+})\psi_{i}^{*2}\exp(\psi^{*}_{i}t)}{\{\widetilde{p}_{i}^{+}\exp(\psi^{*}_{i}t)+(1-\widetilde{p}_{i}^{+})\}^{2}}\Big| \\
       &\leq\sup_{\xi\in [0,1], \psi_{i}^{*} }\Big| \frac{\psi_{i}^{*2}\exp(\psi_{i}^{*}t)\{1-\xi-\xi\exp(\psi_{i}^{*}t)\} } {\{\xi\exp(\psi_{i}^{*}t)+(1-\xi)\}^{3}}\Big|I^{-1}\sum_{i=1}^{I}|p_{I,i}^{+}-\widetilde{p}^{+}_{i}|\xrightarrow{a.s.} 0,
    \end{align*}
    in which we used the fact that $\sup_{\xi\in [0,1], \psi_{i}^{*}\in [0, \sup \psi] }\Big| \frac{\psi_{i}^{*2}\exp(\psi_{i}^{*}t)\{1-\xi-\xi\exp(\psi_{i}^{*}t)\} } {\{\xi\exp(\psi_{i}^{*}t)+(1-\xi)\}^{3}}\Big|$ is finite for each fixed $t\geq 0$. 
    
    For the third term, we have
    \begin{align*}
       &\quad \ \Big|I^{-1}\sum_{i=1}^{I} \frac{p_{I, i}^{+}(1-p_{I, i}^{+})\psi_{I,i}^{2}\exp(\psi_{I,i}t)}{\{p_{I, i}^{+}\exp(\psi_{I,i}t)+(1-p_{I, i}^{+})\}^{2}}-I^{-1}\sum_{i=1}^{I} \frac{p_{I, i}^{+}(1-p_{I, i}^{+})\psi_{i}^{*2}\exp(\psi^{*}_{i}t)}{\{p_{I, i}^{+}\exp(\psi^{*}_{i}t)+(1-p_{I, i}^{+})\}^{2}}\Big|\\
       &\leq I^{-1}\sum_{i=1}^{I}\Big|\frac{p_{I, i}^{+}(1-p_{I, i}^{+})\psi_{I,i}^{2}\exp(\psi_{I,i}t)}{\{p_{I, i}^{+}\exp(\psi_{I,i}t)+(1-p_{I, i}^{+})\}^{2}}- \frac{p_{I, i}^{+}(1-p_{I, i}^{+})\psi_{i}^{*2}\exp(\psi^{*}_{i}t)}{\{p_{I, i}^{+}\exp(\psi^{*}_{i}t)+(1-p_{I, i}^{+})\}^{2}} \Big| \\
       &\leq\sup_{\xi\in  [0, \sup \psi ], p_{I, i}^{+}\in [0,1] } p_{I, i}^{+}(1-p_{I, i}^{+})\Big|\frac{2\xi \exp(\xi t)+\xi^{2}t\exp(\xi t)}{\{p_{I, i}^{+}\exp(\xi t)+(1-p_{I, i}^{+})\}^{2}}-\frac{2p_{I, i}^{+}t\xi^{2}\exp(2\xi t)}{\{p_{I, i}^{+}\exp(\xi t)+(1-p_{I, i}^{+})\}^{3}} \Big|\\
       &\quad \quad \quad \quad \times I^{-1}\sum_{i=1}^{I}|\psi_{I,i}-\psi^{*}_{i}|\\
       &\xrightarrow{a.s.} 0,
    \end{align*}
    in which we used the fact that $\sup_{\xi\in  [0, \sup \psi ], p_{I, i}^{+}\in [0,1] }p_{I, i}^{+}(1-p_{I, i}^{+})\Big|\frac{2\xi \exp(\xi t)+\xi^{2}t\exp(\xi t)}{\{p_{I, i}^{+}\exp(\xi t)+(1-p_{I, i}^{+})\}^{2}}-\frac{2p_{I, i}^{+}t\xi^{2}\exp(2\xi t)}{\{p_{I, i}^{+}\exp(\xi t)+(1-p_{I, i}^{+})\}^{3}} \Big|$ is finite for each fixed $t\geq 0$.

Therefore, we have shown that
\begin{equation*}
    I^{-1}\chi^{\prime\prime}_{I}(t)\xrightarrow{a.s.} E\Big[\frac{\widetilde{p}_{i}^{+}(1-\widetilde{p}_{i}^{+})\psi_{i}^{*2}\exp(\psi^{*}_{i}t)}{\{\widetilde{p}_{i}^{+}\exp(\psi^{*}_{i}t)+(1-\widetilde{p}_{i}^{+})\}^{2}} \Big]:=\omega_{2}(t).
\end{equation*}

    Putting the above results together, we have proved the desired result. 
\end{proof}

\begin{lemma}\label{lemma: large deviation}
Under Conditions \ref{condition: dose moment}--\ref{condition: varphi}, given a sequence of constants $c_{I} \rightarrow c \in \{\omega_{1}(t): t>0 \}$, we have 
\begin{equation*}
    \lim_{I\rightarrow \infty} -I^{-1}\log\big\{P(T^{+}_{\psi}\geq c_{I} I)\big\}=c\ \widetilde{t} -\omega_{0}(\widetilde{t}),
\end{equation*}
where $t=\widetilde{t}$ is the unique solution to $\omega_{1}(t)=c$.

\end{lemma}

\begin{proof}: Note that under Conditions \ref{condition: dose moment}--\ref{condition: varphi}, we have: (i) $m_{I}(t)<\infty \text{ for } t\in [0, \infty)$ (because $\psi_{I,i}$ is bounded under Condition~\ref{condition: varphi}); (ii) $\lim_{I\rightarrow \infty}I^{-1}\chi_{I}^{(k)}(t)=\omega_{k}(t)<\infty, t\in [0,\infty), k=0,1,2$ (according to Lemma~\ref{lemma: convergence for bahadur}); (iii) $\omega_{2}(t)>0$, $t\in [0, \infty)$ (we will prove this later); and (iv) $I^{-1}\chi_{I}^{(3)}(t)$ is locally bounded on $t\in [0, \infty)$. That is, the conditions on $\chi_{I}(t)$, $\chi_{I}^{\prime}(t)$, $\chi_{I}^{\prime \prime}(t)$, and $\chi_{I}^{\prime \prime \prime}(t)$ in the theorem in \citet{plachky1971theorem} are satisfied, which implies the desired result. 

Then, the only remaining component is to show that $\omega_{2}(t)>0$, $t\in [0, \infty)$. To prove this, note that 
\begin{align*}
   &\quad \ \omega_{2}(t)\\
   &=E\Big[\frac{\widetilde{p}_{i}^{+}(1-\widetilde{p}_{i}^{+})\psi_{i}^{*2}\exp(\psi^{*}_{i}t)}{\{\widetilde{p}_{i}^{+}\exp(\psi^{*}_{i}t)+(1-\widetilde{p}_{i}^{+})\}^{2}} \Big]\\
    &=P(\mathcal{A})E\Big[\frac{\widetilde{p}_{i}^{+}(1-\widetilde{p}_{i}^{+})\psi_{i}^{*2}\exp(\psi^{*}_{i}t)}{\{\widetilde{p}_{i}^{+}\exp(\psi^{*}_{i}t)+(1-\widetilde{p}_{i}^{+})\}^{2}}\mid \mathcal{A} \Big]+P(\mathcal{A}^{c})E\Big[\frac{\widetilde{p}_{i}^{+}(1-\widetilde{p}_{i}^{+})\psi_{i}^{*2}\exp(\psi^{*}_{i}t)}{\{\widetilde{p}_{i}^{+}\exp(\psi^{*}_{i}t)+(1-\widetilde{p}_{i}^{+})\}^{2}}\mid \mathcal{A}^{c} \Big]\\
    &\geq P(\mathcal{A})E\Big[\frac{\widetilde{p}_{i}^{+}(1-\widetilde{p}_{i}^{+})\psi_{i}^{*2}\exp(\psi^{*}_{i}t)}{\{\widetilde{p}_{i}^{+}\exp(\psi^{*}_{i}t)+(1-\widetilde{p}_{i}^{+})\}^{2}}\mid \mathcal{A} \Big]\\
    &\geq P(\mathcal{A})\times\frac{p_{\Delta}L_1^2\exp(L_1t)}{\max\{\exp(2L_2t),1\}}\\
    &>0,
\end{align*}
where $L_{1}>0$ and $L_{2}>0$ are defined in the proof of Lemma~\ref{lem: monotonicity of varphi gamma} such that $\psi_{i}^{*}\in [L_{1}, L_{2}]$, and 
\begin{equation*}
    p^{\Delta}=\min \big \{\expit(\widetilde{\gamma}A_{1})\{1-\expit(\widetilde{\gamma}A_{1})\}, \expit(\widetilde{\gamma}A_{2})\{1-\expit(\widetilde{\gamma}A_{2})\}\big\},
\end{equation*}
in which $A_{1}>0$ and $A_{2}>0$ are defined in the proof of Lemma~\ref{lem: monotonicity of varphi gamma} such that $|\phi(Z_{i1})-\phi(Z_{i2})|\in [A_{1}, A_{2}]$. Therefore, the desired conclusion has been established.

\end{proof}

\begin{lemma}\label{lemma: continuity of omega1}
   Under Condition~\ref{condition: varphi}, $\omega_{1}(t)$ is a continuous function over $t\in [0,+\infty)$.
\end{lemma}

\begin{proof}: For any $0\leq t_{1}<t_{2}$, we have 
    \begin{align*}
         |\omega_{1}(t_{2})-\omega_{1}(t_{1})|&=\Big|E\Big[\frac{\widetilde{p}_{i}^{+}\psi^{*}_{i}\exp(\psi^{*}_{i}t_{2})}{\widetilde{p}_{i}^{+}\exp(\psi^{*}_{i}t_{2})+(1-\widetilde{p}_{i}^{+})} \Big]-E\Big[\frac{\widetilde{p}_{i}^{+}\psi^{*}_{i}\exp(\psi^{*}_{i}t_{1})}{\widetilde{p}_{i}^{+}\exp(\psi^{*}_{i}t_{1})+(1-\widetilde{p}_{i}^{+})} \Big]\Big|\\
         &\leq E\Big|\frac{\widetilde{p}_{i}^{+}\psi^{*}_{i}\exp(\psi^{*}_{i}t_{2})}{\widetilde{p}_{i}^{+}\exp(\psi^{*}_{i}t_{2})+(1-\widetilde{p}_{i}^{+})} - \frac{\widetilde{p}_{i}^{+}\psi^{*}_{i}\exp(\psi^{*}_{i}t_{1})}{\widetilde{p}_{i}^{+}\exp(\psi^{*}_{i}t_{1})+(1-\widetilde{p}_{i}^{+})} \Big|\\
         &\leq (t_{2}-t_{1})\times \sup_{\psi^{*}_{i}\in  [0, \sup \psi ], \widetilde{p}_{i}^{+}\in [0,1], t\in [t_{1},t_{2}]}\frac{\widetilde{p}_{i}^{+}(1-\widetilde{p}_{i}^{+})\psi^{*2}_{i}\exp(\psi^{*}_{i}t)}{\{\widetilde{p}_{i}^{+}\exp(\psi^{*}_{i}t)+(1-\widetilde{p}_{i}^{+})\}^{2}}.
    \end{align*}
    Then, the continuity of $\omega_{1}(t)$ over $t\in [0,+\infty)$ follows immediately from the fact that the term $\sup_{\psi^{*}_{i}\in [0, \sup \psi ], \widetilde{p}_{i}^{+}\in [0,1], t\in [t_{1},t_{2}]}\frac{\widetilde{p}_{i}^{+}(1-\widetilde{p}_{i}^{+})\psi^{*2}_{i}\exp(\psi^{*}_{i}t)}{\{\widetilde{p}_{i}^{+}\exp(\psi^{*}_{i}t)+(1-\widetilde{p}_{i}^{+})\}^{2}}$ is finite (because it is the maximal value of a continuous function defined over a compact set). 
\end{proof}

\begin{lemma}\label{lemma: monotonicity of omega1}
Under Conditions \ref{condi: overlap of supports} and \ref{condition: varphi}, the $\omega_{1}(t)$ is strictly monotonically increasing over $t\in [0,+\infty)$.
\end{lemma}

\begin{proof}: Let $A$, $A_1$, $A_2$, $L_1$, and $L_2$ be as defined in the proof of Lemma~\ref{lem: monotonicity of varphi gamma}. For the fixed $\widetilde\gamma$, let $P_1=\operatorname{expit}(\widetilde\gamma A_1)$ and $P_2=\operatorname{expit}(\widetilde\gamma A_2)$. Then $0<P_1\leq P_2<1$, and on $\mathcal{A}$, $\widetilde p_i^+\in[P_1,P_2]$ and $\psi_i^*\in[L_1,L_2]$. For any $0\leq t_1<t_2$, we have 
    \begin{align*}
         \omega_{1}(t_{2})-\omega_{1}(t_{1})&=E\Big[\frac{\widetilde{p}_{i}^{+}\psi^{*}_{i}\exp(\psi^{*}_{i}t_{2})}{\widetilde{p}_{i}^{+}\exp(\psi^{*}_{i}t_{2})+(1-\widetilde{p}_{i}^{+})} \Big]-E\Big[\frac{\widetilde{p}_{i}^{+}\psi^{*}_{i}\exp(\psi^{*}_{i}t_{1})}{\widetilde{p}_{i}^{+}\exp(\psi^{*}_{i}t_{1})+(1-\widetilde{p}_{i}^{+})} \Big]\\
         &= P(\mathcal{A})E\Big[\frac{\widetilde{p}_{i}^{+}\psi^{*}_{i}\exp(\psi^{*}_{i}t_{2})}{\widetilde{p}_{i}^{+}\exp(\psi^{*}_{i}t_{2})+(1-\widetilde{p}_{i}^{+})} - \frac{\widetilde{p}_{i}^{+}\psi^{*}_{i}\exp(\psi^{*}_{i}t_{1})}{\widetilde{p}_{i}^{+}\exp(\psi^{*}_{i}t_{1})+(1-\widetilde{p}_{i}^{+})}\mid \mathcal{A} \Big]\\
         &\quad \quad + P(\mathcal{A}^{c})E\Big[\frac{\widetilde{p}_{i}^{+}\psi^{*}_{i}\exp(\psi^{*}_{i}t_{2})}{\widetilde{p}_{i}^{+}\exp(\psi^{*}_{i}t_{2})+(1-\widetilde{p}_{i}^{+})} - \frac{\widetilde{p}_{i}^{+}\psi^{*}_{i}\exp(\psi^{*}_{i}t_{1})}{\widetilde{p}_{i}^{+}\exp(\psi^{*}_{i}t_{1})+(1-\widetilde{p}_{i}^{+})}\mid \mathcal{A}^{c} \Big]\\
         &\geq P(\mathcal{A})E\Big[\frac{\widetilde{p}_{i}^{+}\psi^{*}_{i}\exp(\psi^{*}_{i}t_{2})}{\widetilde{p}_{i}^{+}\exp(\psi^{*}_{i}t_{2})+(1-\widetilde{p}_{i}^{+})} - \frac{\widetilde{p}_{i}^{+}\psi^{*}_{i}\exp(\psi^{*}_{i}t_{1})}{\widetilde{p}_{i}^{+}\exp(\psi^{*}_{i}t_{1})+(1-\widetilde{p}_{i}^{+})}\mid \mathcal{A} \Big]\\
         &\geq P(\mathcal{A})  (t_{2}-t_{1})\times \inf_{\psi^{*}_{i}\in [L_{1}, L_{2}], \widetilde{p}_{i}^{+}\in [P_{1},P_{2}], t\in [t_{1},t_{2}]}\frac{\widetilde{p}_{i}^{+}(1-\widetilde{p}_{i}^{+})\psi^{*2}_{i}\exp(\psi^{*}_{i}t)}{\{\widetilde{p}_{i}^{+}\exp(\psi^{*}_{i}t)+(1-\widetilde{p}_{i}^{+})\}^{2}}\\
         &>0.
    \end{align*}

Therefore, the desired result has been established.
\end{proof}

\begin{lemma}\label{lemma: bounds of omega1}
Under Conditions~\ref{condition: dose moment}--\ref{condition: prob of D}, we have: $\lim_{t\rightarrow 0}\omega_{1}(t)<E\big [\psi^{*}_{i}  \mathbbm{1}\{D_{i}>0\}\big]$ for $\overline{\Gamma}<\overline{\Gamma}_{*}$ and $\lim_{t\rightarrow \infty}\omega_{1}(t)>E\big [\psi^{*}_{i}  \mathbbm{1}\{D_{i}>0\}\big]$ for any $\overline{\Gamma}\geq 1$.
\end{lemma}

\begin{proof}: Let $g(t)=\frac{\widetilde{p}_{i}^{+}\psi^{*}_{i}\exp(\psi^{*}_{i}t)}{\widetilde{p}_{i}^{+}\exp(\psi^{*}_{i}t)+(1-\widetilde{p}_{i}^{+})}$. Under Condition~\ref{condition: varphi}, $g(t)$ is bounded over $t\in [0, \infty)$. By the bounded convergence theorem, we have 
\begin{equation}\label{eqn: omega converge to 0}
  \lim_{t\rightarrow 0^{+}}\omega_{1}(t)=\lim_{t\rightarrow 0^{+} }E\{g(t)\}
  =E\Big \{ \lim_{t \rightarrow 0^{+}}g(t) \Big\}\\
  =E(\psi^{*}_{i}\widetilde{p}_{i}^{+}),
\end{equation}
\begin{equation}\label{eqn: omega converge to inf}
  \lim_{t\rightarrow \infty}\omega_{1}(t)=\lim_{t\rightarrow \infty}E\{g(t)\}
  =E\Big \{ \lim_{t \rightarrow \infty}g(t) \Big\}\\
  =E(\psi^{*}_{i}).
\end{equation}
Note that for $\overline{\Gamma}<\overline{\Gamma}_{*}$ (i.e., for $\widetilde{\gamma}<\gamma_{*}$), by the proof of Theorem 3.4, we have $E(\psi^{*}_{i}\widetilde{p}_{i}^{+})<E[\psi^{*}_{i} \mathbbm{1}\{D_{i}>0\}]$. Also, as shown in the proof of Lemma~\ref{lem: upp and low bounds of varphi}, we have $E( \psi^{*}_{i})>E[\psi^{*}_{i} \mathbbm{1}\{D_{i}>0\}]$. Combining these results with (\ref{eqn: omega converge to 0}) and (\ref{eqn: omega converge to inf}), the desired conclusions have been established. 
\end{proof}

Now, we are ready to prove Theorem 3.5.
\begin{proof}[Proof of Theorem 3.5:] By Lemmas~\ref{lemma: continuity of omega1}, \ref{lemma: monotonicity of omega1}, and \ref{lemma: bounds of omega1}, we know that there exists a unique $\widetilde{t}\in (0,\infty)$ such that $\mu=\omega_{1}(\widetilde{t})$. For any prespecified $\beta\in (0,1)$, let the sequence $\{c_{1}, c_{2}, \dots \}$ be the one that satisfies $P(T_{\psi}\geq I c_{I})\geq 1-\beta$ and $P(T_{\psi}\geq I c_{I}+\epsilon)< 1-\beta$ for all $\epsilon>0$. In Lemma~\ref{lemma: convergence}, we have shown that $I^{-1}T_{\psi}\xrightarrow{a.s.} \mu = E[\psi^{*}_{i} \mathbbm{1}\{D_{i}>0\}]$, which implies that $\lim_{I\rightarrow \infty}c_{I}=\mu$. Applying Lemma~\ref{lemma: large deviation}, we have
   \begin{equation}\label{eqn: convergence of pvalues for bahadur}
       \lim_{I\rightarrow \infty} -I^{-1}\log\big\{P(T^{+}_{\psi}\geq c_{I} I)\big\}=\lim_{I\rightarrow \infty} -I^{-1}\log\big\{p_{\overline{\Gamma}, I}(c_{I}I)\big\}=\mu \widetilde{t} -\omega_{0}(\widetilde{t}),
   \end{equation}
   in which $p_{\overline{\Gamma}, I}(c_{I}I)$ is the worst-case (upper bound) $p$-value given the observed test statistic value $c_{I}I$, and $\mu=\omega_{1}(\widetilde{t})=E[\psi^{*}_{i} \mathbbm{1}\{D_{i}>0\}]$. By Theorem 3.5 of \citet{raghavachari1970theorem}, equation (\ref{eqn: convergence of pvalues for bahadur}) implies the desired result stated in Theorem 3.5 of the main text. 
\end{proof}

\subsection*{A.5: Proofs of Theorems 4.1 and 4.2}

We first give sufficient conditions under which the joint distribution of $(T^{+}_{1}, T^{+}_{2})$ is asymptotically bivariate normal. For $i=1,\dots,I$, we let $\overline{T}^{+}_{1,i}=(q_{I,i}B_{i}^{+}-q_{I,i}p^{+}_{I,i})/\sigma^{+}_{1}$ and $\overline{T}_{2,i}=(s_{I,i}B_{i}^{+}-s_{I,i}p^{+}_{I,i})/\sigma^{+}_{2}$. That is, we have $\frac{T^{+}_{1}-\mu^{+}_{1}}{\sigma^{+}_{1}}=\sum_{i=1}^{I}\overline{T}^{+}_{1,i}$ and $\frac{T^{+}_{2}-\mu^{+}_{2}}{\sigma^{+}_{2}}=\sum_{i=1}^{I}\overline{T}^{+}_{2,i}$. Let $T^{+}_{1:2,i}=(\overline{T}^{+}_{1,i}, \overline{T}^{+}_{2,i})^{T}$ for $i=1,\dots,I$. Then, we have the following Lemma~\ref{prop: regularity assumption}, which gives sufficient conditions for the desired asymptotic bivariate normality (the proof is similar to that of Proposition 3 in Appendix C of \citet{heng2021increasing}).

\begin{lemma}\label{prop: regularity assumption}
Let $\Sigma_{*}=\left(\begin{array}{cc}
1 & \rho^{*} \\
\rho^{*} & 1 
\end{array}\right)$. Suppose that the following two regularity conditions hold: (i) As $I\rightarrow \infty$, $\Sigma_{*}$ has a positive definite limit $\widetilde{\Sigma}$; (ii) For any fixed nonzero vector $\lambda=(\lambda_{1}, \lambda_{2})^{T}$, there exists some $\delta>0$, such that Lyapunov's condition
\begin{equation*}
    \lim_{I\rightarrow \infty}\frac{1}{(\lambda^{T}\Sigma_{*}\lambda)^{1+\delta/2}}\sum_{i=1}^{I}\mathbb{E}\big\{|\lambda^{T}T^{+}_{1:2,i}|^{2+\delta} \big\}=0
\end{equation*}
is satisfied. Then, as $I\rightarrow \infty$, $(T^{+}_{1}, T^{+}_{2})$ is asymptotically bivariate normal: 
\begin{equation*}
\Sigma_{*}^{-1/2}\Big(\frac{T^{+}_{1}-\mu^{+}_{1}}{\sigma^{+}_{1}},\ \frac{T^{+}_{2}-\mu^{+}_{2}}{\sigma^{+}_{2}} \Big)^{T} \xrightarrow{\mathcal{L}} \mathcal{N}(\mathbf{0}, I_{2\times 2}).
\end{equation*}
\end{lemma}

\begin{proof}: We first show that $\widetilde{\Sigma}^{-1/2}\Big(\frac{T^{+}_{1}-\mu^{+}_{1}}{\sigma^{+}_{1}},\ \frac{T^{+}_{2}-\mu^{+}_{2}}{\sigma^{+}_{2}}\Big)^{T}\xrightarrow{\mathcal{L}} \mathcal{N}(\mathbf{0}, I_{2\times 2})$. Invoking the Cram\'er-Wold device (\citealp{Billingsley1995measure}; Theorem 29.4), we just need to ensure that for any nonzero vector $\widetilde{\lambda}=(\widetilde{\lambda}_{1}, \widetilde{\lambda}_{2})^{T}$, we have:
\begin{equation}\label{asymptotic normal of standard deviation}
    \frac{\widetilde{\lambda}^{T} \widetilde{\Sigma}^{-1/2}\Big(\frac{T^{+}_{1}-\mu^{+}_{1}}{\sigma^{+}_{1}},\ \frac{T^{+}_{2}-\mu^{+}_{2}}{\sigma^{+}_{2}}\Big)^{T} }{\sqrt{\widetilde{\lambda}^{T}\widetilde{\lambda}}}\xrightarrow{\mathcal{L}}\mathcal{N}(0,1) \text{ as $I\rightarrow \infty$}.
\end{equation}
By independence across matched pairs, for each $I=1,2,\dots$, the sequence $A_{I}=\{\widetilde{\lambda}^{T}\widetilde{\Sigma}^{-1/2}T^{+}_{1:2,1},\dots,\widetilde{\lambda}^{T}\widetilde{\Sigma}^{-1/2} T^{+}_{1:2,I}\}$ is a sequence of independent random variables. By setting $\lambda^{T}=\widetilde{\lambda}^{T} \widetilde{\Sigma}^{-1/2}$ in condition (ii) and applying the Lyapunov central limit theorem (\citealp{Billingsley1995measure}; Theorem 27.3) to $\{A_{1}, A_{2},\dots,\}$, we have
\begin{equation*}
     \frac{\widetilde{\lambda}^{T} \widetilde{\Sigma}^{-1/2}\Big(\frac{T^{+}_{1}-\mu^{+}_{1}}{\sigma^{+}_{1}},\ \frac{T^{+}_{2}-\mu^{+}_{2}}{\sigma^{+}_{2}}\Big)^{T} }{\sqrt{\widetilde{\lambda}^{T}\widetilde{\Sigma}^{-1/2}\Sigma_{*}\widetilde{\Sigma}^{-1/2} \widetilde{\lambda}}}\xrightarrow{\mathcal{L}}\mathcal{N}(0,1) \text{ as $I\rightarrow \infty$}.
\end{equation*}
Then, (\ref{asymptotic normal of standard deviation}) follows from: as $I\rightarrow \infty$,
\begin{align*}
    \frac{\widetilde{\lambda}^{T} \widetilde{\Sigma}^{-1/2}\Big(\frac{T^{+}_{1}-\mu^{+}_{1}}{\sigma^{+}_{1}},\ \frac{T^{+}_{2}-\mu^{+}_{2}}{\sigma^{+}_{2}}\Big)^{T} }{\sqrt{\widetilde{\lambda}^{T}\widetilde{\lambda}}}&=\frac{\widetilde{\lambda}^{T} \widetilde{\Sigma}^{-1/2}\Big(\frac{T^{+}_{1}-\mu^{+}_{1}}{\sigma^{+}_{1}},\ \frac{T^{+}_{2}-\mu^{+}_{2}}{\sigma^{+}_{2}}\Big)^{T} }{\sqrt{\widetilde{\lambda}^{T}\widetilde{\Sigma}^{-1/2}\Sigma_{*}\widetilde{\Sigma}^{-1/2} \widetilde{\lambda}}}\frac{\sqrt{\widetilde{\lambda}^{T}\widetilde{\Sigma}^{-1/2}\Sigma_{*}\widetilde{\Sigma}^{-1/2}\widetilde{\lambda}}}{\sqrt{\widetilde{\lambda}^{T}\widetilde{\lambda}}}\\
    &\xrightarrow{\mathcal{L}} \mathcal{N}(0,1)\cdot 1 \quad \text{(by condition (i) and Slutsky's theorem)}\\
    &\sim \mathcal{N}(0,1).
\end{align*}
That is, we have shown that $\widetilde{\Sigma}^{-1/2}\Big(\frac{T^{+}_{1}-\mu^{+}_{1}}{\sigma^{+}_{1}},\ \frac{T^{+}_{2}-\mu^{+}_{2}}{\sigma^{+}_{2}}\Big)^{T}\xrightarrow{\mathcal{L}} \mathcal{N}(\mathbf{0}, I_{2\times 2})$, which implies that: as $I\rightarrow \infty$,
\begin{align*}
 \Sigma_{*}^{-1/2}\Big(\frac{T^{+}_{1}-\mu^{+}_{1}}{\sigma^{+}_{1}},\ \frac{T^{+}_{2}-\mu^{+}_{2}}{\sigma^{+}_{2}}\Big)^{T}&= \Sigma_{*}^{-1/2}\widetilde{\Sigma}^{1/2}\widetilde{\Sigma}^{-1/2} \Big(\frac{T^{+}_{1}-\mu^{+}_{1}}{\sigma^{+}_{1}},\ \frac{T^{+}_{2}-\mu^{+}_{2}}{\sigma^{+}_{2}}\Big)^{T}\\
 &\xrightarrow{\mathcal{L}} I_{2\times 2} \cdot \mathcal{N}(\mathbf{0}, I_{2\times 2})\sim \mathcal{N}(\mathbf{0}, I_{2\times 2}).
\end{align*}
That is, the desired result has been established. 

\end{proof}

Then, we are ready to prove Theorem 4.1.

\begin{proof}: Under $H_{0}$, the test scores $q_{I,i}$ and $s_{I,i}$ are fixed test scores. Let $A_{i}=\mathbbm{1}\{D_i>0\}$, and let $\pi_{i}(\mathbf{u}_{0})
=
P\left(A_{i}=1\mid H_{0}, \mathcal{F},\mathcal{Z},\mathbf{u}_{0}\right)$, where $\mathbf{u}_{0}$ denotes the true values of unmeasured confounders. Under the generalized Rosenbaum sensitivity bounds, we have
\begin{align*}
1-p_{I,i}^{+}
\leq
\pi_{i}(\mathbf{u}_{0})
\leq
p_{I,i}^{+} \quad \text{ for each matched pair $i$.}
\end{align*}
 We adopt a coupling argument to rigorously show that $(T^{+}_{1}, T^{+}_{2})$ stochastically dominates $(T_{1}, T_{2})$. Specifically, let $U_{1},\ldots,U_{I}$ be independent uniform random variables with the support $[0,1]$, and define $\widetilde{A}_{i}
=
\mathbbm{1}\left\{
U_{i}\leq\pi_{i}(\mathbf{u}_{0})
\right\}$ and $B_{i}^{+}=
\mathbbm{1}\big\{
U_{i}\leq p_{I,i}^{+}
\big\}$. Then $\widetilde{A}_{i}$ has the same conditional distribution as $A_{i}$, while $B_{i}^{+}$ is a Bernoulli random variable with success probability $p_{I,i}^{+}$. Moreover, because $\pi_{i}(\mathbf{u}_{0})\leq p_{I,i}^{+}$, we have $\widetilde{A}_{i}\leq B_{i}^{+}$ almost surely for every $i$. Define $\widetilde{T}_{1}
=
\sum_{i=1}^{I}q_{I,i}\widetilde{A}_{i}$,
$\widetilde{T}_{2}
=
\sum_{i=1}^{I}s_{I,i}\widetilde{A}_{i}$, $\widetilde{T}_{1}^{+}=
\sum_{i=1}^{I}q_{I,i}B_{i}^{+}$,
and $\widetilde{T}_{2}^{+}=
\sum_{i=1}^{I}s_{I,i}B_{i}^{+}$. Therefore, $(\widetilde{T}_{1},\widetilde{T}_{2})$ has the same distribution as $(T_{1},T_{2})$, and $(\widetilde{T}_{1}^{+},\widetilde{T}_{2}^{+})$ has the same distribution as $(T_{1}^{+},T_{2}^{+})$. Because $q_{I,i}\geq 0$ and $s_{I,i}\geq 0$, the preceding coupling gives 
\begin{equation}\label{eqn: coupling}
    \text{$\widetilde{T}_{1}\leq\widetilde{T}_{1}^{+}$ and $\widetilde{T}_{2}\leq\widetilde{T}_{2}^{+}$ simultaneously almost surely.}
\end{equation}
Let
\begin{align*}
\mathcal{R}
=
\left\{
(t_{1},t_{2}):
\max\left\{
\frac{t_{1}-\mu_{1}^{+}}{\sigma_{1}^{+}},
\frac{t_{2}-\mu_{2}^{+}}{\sigma_{2}^{+}}
\right\}
\geq Q_{\mathbf{\rho}^{*},\alpha}
\right\}
\end{align*}
denote the rejection region of the adaptive procedure. Because $\sigma_{1}^{+}>0$ and $\sigma_{2}^{+}>0$, $\mathcal{R}$ is coordinatewise nondecreasing. Therefore, we have
\begin{equation*}
P
\left(
(\widetilde{T}_{1},\widetilde{T}_{2})\in\mathcal{R}
\mid H_{0}, \mathcal{F},\mathcal{Z}, \mathbf{u}_{0}
\right)\leq
P
\left(
(\widetilde{T}_{1}^{+},\widetilde{T}_{2}^{+})\in\mathcal{R}
\mid H_{0}, \mathcal{F},\mathcal{Z}
\right),
\end{equation*}
which implies that 
\begin{equation}\label{eqn: joint adaptive upper bound}
P
\left(
(T_{1},T_{2})\in\mathcal{R}
\mid H_{0}, \mathcal{F},\mathcal{Z}, \mathbf{u}_{0}
\right)\leq
P
\left(
(T_{1}^{+},T_{2}^{+})\in\mathcal{R}
\mid H_{0}, \mathcal{F},\mathcal{Z}
\right).
\end{equation}
In other words, we have
\begin{align*}
&\quad \ P\left(\max\left \{\frac{T_{1}-\sum_{i=1}^{I}q_{I,i}p_{I,i}^{+}}{\sqrt{\sum_{i=1}^{I} q_{I,i}^{2}p_{I,i}^{+}(1-p_{I,i}^{+})}}, \frac{T_{2}-\sum_{i=1}^{I}s_{I,i}p_{I,i}^{+}}{\sqrt{\sum_{i=1}^{I} s_{I,i}^{2}p_{I,i}^{+}(1-p_{I,i}^{+})}}\right \} \geq Q_{\mathbf{\rho}^{*}, \alpha}\Big|H_{0}, \mathcal{F}, \mathcal{Z}, \mathbf{u}_{0}\right)\\
&\leq P\left(\max\left \{\frac{T^{+}_{1}-\sum_{i=1}^{I}q_{I,i}p_{I,i}^{+}}{\sqrt{\sum_{i=1}^{I} q_{I,i}^{2}p_{I,i}^{+}(1-p_{I,i}^{+})}}, \frac{T^{+}_{2}-\sum_{i=1}^{I}s_{I,i}p_{I,i}^{+}}{\sqrt{\sum_{i=1}^{I} s_{I,i}^{2}p_{I,i}^{+}(1-p_{I,i}^{+})}}\right \} \geq Q_{\mathbf{\rho}^{*}, \alpha}\Big|H_{0}, \mathcal{F}, \mathcal{Z}\right)\\
  &\rightarrow \alpha, \quad \text{as $I \rightarrow \infty$} \quad (\text{by Lemma~\ref{prop: regularity assumption}}).
\end{align*}
That is, the desired conclusion follows.
\end{proof}

Finally, we are ready to prove Theorem 4.2.

\begin{proof}: By Slepian's lemma, for any $\rho^{*}\in (0,1)$, we have $Q_{\mathbf{\rho}^{*}, \alpha}\in (\Phi^{-1}(1-\alpha), \Phi^{-1}(1-\alpha/2))$. That is, $Q_{\mathbf{\rho}^{*}, \alpha}=\Phi^{-1}(1-\alpha^{*})$ for some $\alpha^{*}\in [\alpha/2, \alpha]$. Therefore, the proposed adaptive testing procedure in Section 4.1, i.e.,
\begin{equation*}
    \text{we reject $H_{0}$ iff $\max\left \{\frac{T_{1}-\sum_{i=1}^{I}q_{I,i}p_{I,i}^{+}}{\sqrt{\sum_{i=1}^{I} q_{I,i}^{2}p_{I,i}^{+}(1-p_{I,i}^{+})}}, \frac{T_{2}-\sum_{i=1}^{I}s_{I,i}p_{I,i}^{+}}{\sqrt{\sum_{i=1}^{I} s_{I,i}^{2}p_{I,i}^{+}(1-p_{I,i}^{+})}}\right \} \geq Q_{\mathbf{\rho}^{*},\alpha}$,}
\end{equation*}
is equivalent to: we reject $H_{0}$ iff
\begin{equation}\label{eqn: sensitivity analysis based on T1}
    \frac{T_{1}-\sum_{i=1}^{I}q_{I,i}p_{I,i}^{+}}{\sqrt{\sum_{i=1}^{I} q_{I,i}^{2}p_{I,i}^{+}(1-p_{I,i}^{+})}} \geq \Phi^{-1}(1-\alpha^{*})
\end{equation}
or
\begin{equation}\label{eqn: sensitivity analysis based on T2}
    \frac{T_{2}-\sum_{i=1}^{I}s_{I,i}p_{I,i}^{+}}{\sqrt{\sum_{i=1}^{I} s_{I,i}^{2}p_{I,i}^{+}(1-p_{I,i}^{+})}} \geq \Phi^{-1}(1-\alpha^{*}). 
\end{equation}
That is, the adaptive testing procedure can reject $H_{0}$ if and only if either (\ref{eqn: sensitivity analysis based on T1}) or (\ref{eqn: sensitivity analysis based on T2}) can reject $H_{0}$. Meanwhile, as shown in Theorems 3.4 and 3.5, the generalized design sensitivity and Bahadur-Rosenbaum exact slope do not change with the significance level (e.g., do not change if the significance level changes from $\alpha$ to $\alpha^{*}$, where $\alpha^{*}\in [\alpha/2, \alpha]$). That is, the generalized design sensitivities (or the generalized Bahadur-Rosenbaum exact slopes) of two individual sensitivity analyses (\ref{eqn: sensitivity analysis based on T1}) and (\ref{eqn: sensitivity analysis based on T2}) equal $\overline{\Gamma}_{*,1}$ and $\overline{\Gamma}_{*,2}$ (or $\Upsilon_{1}$ and $\Upsilon_{2}$), respectively. Putting the above results together, by Theorem 2.8 in \citet{berk1978relatively}, we have $\overline{\Gamma}_{*,1: 2}= \max\{ \overline{\Gamma}_{*,1}, \overline{\Gamma}_{*,2}\}$, and $\Upsilon_{1:2}=\max\{\Upsilon_{1}, \Upsilon_{2}\}$ for $\overline{\Gamma}<\min\{ \overline{\Gamma}_{*,1}, \overline{\Gamma}_{*,2}\}$. 
\end{proof}

\noindent \textit{Remark:} Although $\alpha^{*}$ in the proof of Theorem~4.2 may depend on $I$ through $\rho^{*}$, it always satisfies $\alpha/2\leq\alpha^{*}\leq\alpha$. Because the generalized design sensitivity and Bahadur-Rosenbaum exact slope are unchanged for any possibly $I$-dependent significance level within this interval, the corresponding adaptivity conclusions continue to hold.

\section*{Appendix B: Extensions}

\subsection*{B.1: Generalized Design Sensitivity and Bahadur-Rosenbaum Efficiency for Permutation $t$-Test Statistics}

In the main text, we illustrate how to use the generalized design sensitivity and Bahadur-Rosenbaum efficiency to evaluate the performance of competing rank-based test statistics. In this section, we present the formulas for generalized design sensitivity and Bahadur-Rosenbaum efficiency for sign-score test statistics that do not involve ranks. Because the purpose of Appendix B.1 is to illustrate how the proposed framework extends beyond rank-based statistics, the formulas below are understood to hold under regularity conditions analogous to those imposed in Appendices~A.3 and~A.4, including appropriate integrability and non-degeneracy conditions and the existence and uniqueness of the considered solutions. 

Specifically, consider the family of test statistics $T_{\zeta}=\sum_{i=1}^{I}\zeta(|Z_{i1}-Z_{i2}|, |Y_{i1}-Y_{i2}|)\mathbbm{1}\{D_{i}>0\}$, in which $\zeta$ is some function of $(|Z_{i1}-Z_{i2}|, |Y_{i1}-Y_{i2}|)$. For example, when $\zeta=|Y_{i1}-Y_{i2}|$, the $T_{\zeta}$ is a permutation $t$-test for matched designs. When $\zeta=|Z_{i1}-Z_{i2}|\times|Y_{i1}-Y_{i2}|$, the $T_{\zeta}$ is a dose-weighted permutation $t$-test. Then, following similar arguments as those used in Appendix A.3 (with the only major difference being that we do not need to use the uniform convergence arguments to de-correlate the test scores across matched pairs since $\zeta(|Z_{i1}-Z_{i2}|, |Y_{i1}-Y_{i2}|)$ are independent across matched pairs), we can get the following generalized design sensitivity formula for $T_{\zeta}$: Let $\gamma_{*}\in (0, +\infty)$ be the unique solution to the following equation of $\gamma$:
\begin{align*}
    &\quad \ E\Big[\zeta(|Z_{i1}-Z_{i2}|, |Y_{i1}-Y_{i2}|) \frac{\exp\{\gamma |\phi(Z_{i1})-\phi(Z_{i2})|\}}{1+\exp\{\gamma |\phi(Z_{i1})-\phi(Z_{i2})|\}} \Big]\\
    &= E\big [\zeta(|Z_{i1}-Z_{i2}|, |Y_{i1}-Y_{i2}|) \mathbbm{1}\{D_{i}>0\}\big].
\end{align*}
Then, $\gamma_{*}$ is the generalized design sensitivity of $T_{\zeta}$ on the $\gamma$-scale, and $\overline{\Gamma}_{*}=E\big[\exp\{\gamma_{*}|\phi(Z_{i1})-\phi(Z_{i2})|\}\big]$ is the generalized design sensitivity on the $\overline{\Gamma}$-scale. That is, we just need to replace the $\psi_{*}$ in Theorem 3.4 with $\zeta(|Z_{i1}-Z_{i2}|, |Y_{i1}-Y_{i2}|)$ for $T_{\zeta}$.

Similarly, following similar arguments as those used in Appendix A.4, we can get the formula for the generalized Bahadur-Rosenbaum efficiency for $T_{\zeta}$. Specifically, let $\mu_{\zeta}=E\big [\zeta_{i} \mathbbm{1}\{D_{i}>0\}\big]$ and $$\nu_{0}(t)=E\Big[\log \big \{ \widetilde{p}_{i}^{+}\exp(\zeta_{i}t)+(1-\widetilde{p}_{i}^{+})\big\}\Big],$$ where $\zeta_{i}=\zeta(|Z_{i1}-Z_{i2}|, |Y_{i1}-Y_{i2}|)$ and $\widetilde{p}_{i}^{+}=\expit\{\widetilde{\gamma}|\phi(Z_{i1})-\phi(Z_{i2})|\}$ with $\widetilde{\gamma}$ being the unique value such that $\overline{\Gamma}=E[\exp\{\widetilde{\gamma}|\phi(Z_{i1})-\phi(Z_{i2})|\}]$. Also, let $$\nu_{1}(t)=E\Big[\frac{\widetilde{p}_{i}^{+}\zeta_{i}\exp(\zeta_{i}t)}{\widetilde{p}_{i}^{+}\exp(\zeta_{i}t)+(1-\widetilde{p}_{i}^{+})} \Big],$$ and $t=\widetilde{t}$ is the unique solution to $\mu_{\zeta}=\nu_{1}(t)$. Then, $\Upsilon= 2\{  \mu_{\zeta} \widetilde{t} - \nu_{0}(\widetilde{t})\}$ is called a generalized Bahadur-Rosenbaum exact slope of sensitivity analysis.

\subsection*{B.2: Generalized Design Sensitivity and Bahadur-Rosenbaum Efficiency for Discrete Treatments}

In the main text, we derive generalized design sensitivity and Bahadur-Rosenbaum efficiency for continuous treatments, the most challenging and underexplored setting. Here, we present the corresponding formulas for ordinal (discrete) treatments. These illustrative formulas hold under regularity conditions analogous to those in Appendices~A.3 and~A.4, including finiteness of all relevant expectations and moment-generating functions.

Specifically, we suppose that each $|Z_{i1}-Z_{i2}| \in \{\delta_{1}, \delta_{2}, \dots, \delta_{d}, \dots, \}$, where $\delta_{1}<\delta_{2}<\dots<\delta_{d}<\dots$. Here, the paired difference in ordinal treatments $Z_{i1}-Z_{i2}$ can take either finitely or infinitely many possible treatment doses. Specifically, if there are only finitely many $\delta_{d}$ such that $P(|Z_{i1}-Z_{i2}|=\delta_{d})\neq 0$, then the paired difference in treatment $Z_{i1}-Z_{i2}$ can only take finitely many possible treatment doses. Otherwise, $Z_{i1}-Z_{i2}$ can take infinitely many possible treatment doses. Also, without loss of generality, we consider the cases in which researchers consider an identity dose link function $\phi(z)=z$ and apply the sign-score test $T_{\zeta}=\sum_{i=1}^{I}\zeta(|Z_{i1}-Z_{i2}|, |Y_{i1}-Y_{i2}|)\mathbbm{1}\{D_{i}>0\}$ (in which $\zeta\geq 0$ can be any prespecified function and $Y_{ij}$ are bounded outcomes) instead of using rank-transformed treatment doses; otherwise, we only need to set $\delta_{d}=d$ (i.e., the corresponding rank of each dose value) in the corresponding arguments. 

Using similar arguments as those proposed in Appendix A.3 (the only major difference is that we do not need to use the previous arguments to de-correlate the test scores across matched pairs because of the independence of $\zeta(|Z_{i1}-Z_{i2}|, |Y_{i1}-Y_{i2}|)$ across matched pairs), as well as appropriate regularity conditions, we can get the formula for generalized design sensitivity under ordinal treatments. Specifically, let $\gamma_{*}\in (0, +\infty)$ be the solution to the following equation of $\gamma$:
\begin{align*}
   & \sum_{d=1}^{\infty}P(|Z_{i1}-Z_{i2}|=\delta_{d}) E\Big[\zeta(\delta_{d}, |Y_{i1}-Y_{i2}|) \expit (\gamma \delta_{d})\mid |Z_{i1}-Z_{i2}|=\delta_{d} \Big]\\
   &\quad \quad \quad = \sum_{d=1}^{\infty}P(|Z_{i1}-Z_{i2}|=\delta_{d}) E\big [\zeta(\delta_{d}, |Y_{i1}-Y_{i2}|) \mathbbm{1}\{D_{i}>0\}\mid |Z_{i1}-Z_{i2}|=\delta_{d}\big].
\end{align*}
Then, $\gamma_{*}$ is the generalized design sensitivity of $T_{\zeta}$ on the $\gamma$-scale, and $\overline{\Gamma}_{*}=\sum_{d=1}^{\infty}P(|Z_{i1}-Z_{i2}|=\delta_{d})\times \exp(\gamma_{*}\delta_{d})$ is the generalized design sensitivity on the $\overline{\Gamma}$-scale.

Then, using arguments similar to those used in Appendix A.4, as well as appropriate regularity conditions, we can get the formula for the generalized Bahadur-Rosenbaum efficiency under ordinal treatments. Specifically, let 
\begin{equation*}
    \mu=\sum_{d=1}^{\infty}P(|Z_{i1}-Z_{i2}|=\delta_{d}) E\big [\zeta_{i} \mathbbm{1}\{D_{i}>0\}\mid |Z_{i1}-Z_{i2}|=\delta_{d}\big],
\end{equation*}
where $\zeta_{i}= \zeta(\delta_{d}, |Y_{i1}-Y_{i2}|)$, and 
$$\nu_{0}(t)=\sum_{d=1}^{\infty}P(|Z_{i1}-Z_{i2}|=\delta_{d})E\Big[\log \big \{ \widetilde{p}_{i}^{+}\exp(\zeta_{i}t)+(1-\widetilde{p}_{i}^{+})\big\}\mid |Z_{i1}-Z_{i2}|=\delta_{d} \Big],$$ where $\widetilde{\gamma}$ is the unique value such that $\overline{\Gamma}=\sum_{d=1}^{\infty}P(|Z_{i1}-Z_{i2}|=\delta_{d})\times \exp(\widetilde{\gamma}\delta_{d})$. Then, we define $$\nu_{1}(t)=\sum_{d=1}^{\infty}P(|Z_{i1}-Z_{i2}|=\delta_{d}) E\Big[\frac{\widetilde{p}_{i}^{+}\zeta_{i}\exp(\zeta_{i}t)}{\widetilde{p}_{i}^{+}\exp(\zeta_{i}t)+(1-\widetilde{p}_{i}^{+})}\mid  |Z_{i1}-Z_{i2}|=\delta_{d} \Big].$$ Then, the generalized Bahadur-Rosenbaum exact slope of sensitivity analysis under ordinal treatments is 
\begin{equation*}
    \Upsilon= 2\{  \mu \widetilde{t} - \nu_{0}(\widetilde{t})\}, \text{ where $t=\widetilde{t}$ is the unique solution to $\mu=\nu_{1}(t)$}. 
\end{equation*}

\subsection*{B.3: Extensions to the Dose-Dependent Versions of Rosenbaum's U- and M-Statistics}

In the main text, we consider both the original and dose-weighted versions of several rank-based test statistics widely used in pair-matched observational studies. However, our results and methods are not restricted to those particular statistics. They also extend to dose-dependent versions of the broader classes of Rosenbaum's U-statistics \citep{rosenbaum2011new} and M-statistics \citep{rosenbaum2007sensitivity, rosenbaum2014weighted} under pair-matched designs.

\smallskip

\noindent \textit{B.3.1: Extension to the Dose-Dependent Versions of Rosenbaum's U-Statistics}

\smallskip

\noindent Let $D_{i}=(Z_{i1}-Z_{i2})(Y_{i1}-Y_{i2})$ denote the outcome difference in pair $i$ multiplied by the corresponding treatment dose difference. Let $m$ be a fixed positive integer, and let $\mathcal{K}$ be the collection of all ${I\choose m}$ index sequences $\mathcal{I}=(i_{1},\dots, i_{m})$ consisting of $m$ distinct integers with $1\leq i_{1}<\cdots <i_{m}\leq I$. Also, let $\mathbf{D}_{\mathcal{I}}=(D_{i_{1}}, \dots, D_{i_{m}})$. A U-statistic \citep{hoeffding1948class} based on $\{D_{i}\}_{i=1}^{I}$ takes the following form:
\begin{equation*}
    T_{U}={I \choose m}^{-1}\sum_{\mathcal{I}\in \mathcal{K}}h(\mathbf{D}_{\mathcal{I}}), 
\end{equation*}
where the kernel function $h(\mathbf{D}_{\mathcal{I}})$ is symmetric in $D_{i_{1}}, \dots, D_{i_{m}}$ in the sense that it is invariant to permutations of the coordinates of $\mathbf{D}_{\mathcal{I}}$. Assume that $Z_{ij}$ are continuous and that there are no ties among $\{|D_{i}|\}_{i=1}^{I}$. \citet{rosenbaum2011new} proposed defining $h(\mathbf{D}_{\mathcal{I}})$ as follows. For each $\mathcal{I}=(i_{1},\dots, i_{m})\in \mathcal{K}$, reorder $D_{i_{1}}, \dots, D_{i_{m}}$ as $D_{\mathcal{I}, 1}, \dots, D_{\mathcal{I}, m}$ such that $|D_{\mathcal{I}, 1}|\leq \cdots \leq |D_{\mathcal{I}, m}|$. Fix two integers $\overline{m}$ and $\underline{m}$ such that $1\leq \underline{m}\leq \overline{m}\leq m$. The U-statistics in \citet{rosenbaum2011new} define $h(\mathbf{D}_{\mathcal{I}})$ as the number of positive values among $D_{\mathcal{I},\underline{m}},\dots, D_{\mathcal{I},\overline{m}}$. Following the arguments in \citet{rosenbaum2011new}, Rosenbaum's U-statistics can be expressed as sign-score statistics:
\begin{equation}\label{eqn: u as sign-score}
    T_{U}={I \choose m}^{-1}\sum_{\mathcal{I}\in \mathcal{K}}h(\mathbf{D}_{\mathcal{I}})=\sum_{i=1}^{I}q_{I,i} \mathbbm{1}\{D_{i}>0\},
\end{equation}
where 
\begin{equation*}
    q_{I,i}={I \choose m}^{-1}\sum_{l=\underline{m}}^{\overline{m}}{I\cdot r^{d}_{I,i}-1 \choose l-1}{I-I\cdot r^{d}_{I,i} \choose m-l},
\end{equation*}
in which $I\cdot r_{I,i}^{d}=\sum_{i^{\prime}=1}^{I}\mathbbm{1}\{|D_{i}|\geq |D_{i^{\prime}}|\}$ is the rank of $|D_{i}|$. Because the formula for the worst-case $p$-value in equation (5) of the main text applies to any sign-score test statistic, it can be directly applied to the dose-dependent version of Rosenbaum's U-statistics through the sign-score representation in (\ref{eqn: u as sign-score}). \citet{rosenbaum2011new} also showed that
\begin{equation}\label{eqn: u approximation}
   q_{I,i}\approx \frac{1}{I}\sum_{l=\underline{m}}^{\overline{m}}l{m\choose l}\left(r_{I,i}^{d} \right)^{l-1}\left(1-r_{I,i}^{d} \right)^{m-l} \text{ as $I\rightarrow \infty$}. 
\end{equation}
Based on the asymptotic approximation in (\ref{eqn: u approximation}) and following the proof strategy used in Appendix A.3, we obtain the following formula for the generalized design sensitivity of the dose-dependent version of Rosenbaum's U-statistics:
\begin{theorem}[Generalized Design Sensitivity Formula for the Dose-Dependent Version of Rosenbaum's U-Statistics]\label{thm: design sensitivity of U-statistics}
Suppose that $(Z_{i1}, Z_{i2}, Y_{i1}, Y_{i2})$ are independent and identically distributed (iid) realizations from a multivariate distribution $F$ in which $(Z_{i1}, Z_{i2})$ are continuous. Define the function
\begin{equation*}
    \varrho(t)=\sum_{l=\underline{m}}^{\overline{m}}l{m\choose l}t^{l-1}(1-t)^{m-l}.
\end{equation*}
Let $F_{D}$ denote the marginal distribution function of $|D_{i}|$, and let $\varrho^{*}_{i}=\varrho(F_{D}(|D_{i}|))$ and $\mu=E\big [\varrho^{*}_{i} \mathbbm{1}\{D_{i}>0\}\big]$. Under some mild regularity conditions analogous to those in Appendix A.3, suppose that the following equation for $\gamma$, in which $\expit(x)=\exp(x)/\{1+\exp(x)\}$,
\begin{equation}\label{eqn: formula of the generalized design sensitivity}
    E\big[\varrho^{*}_{i} \expit\{\gamma |\phi(Z_{i1})-\phi(Z_{i2})|\} \big]= \mu, 
\end{equation}
has (i) a unique solution $\gamma_{*}\in (0, +\infty)$ and (ii) a left-hand side that is strictly monotonically increasing in $\gamma$. Then, $\gamma_{*}$ is the generalized design sensitivity of $T_{U}$ on the $\gamma$-scale, and $\overline{\Gamma}_{*}=E\big[\exp\{\gamma_{*}|\phi(Z_{i1})-\phi(Z_{i2})|\}\big]$ is the generalized design sensitivity of $T_{U}$ on the $\overline{\Gamma}$-scale. 
\end{theorem}

Using the proof strategy of Appendix~A.4, we similarly obtain a formula for the generalized Bahadur-Rosenbaum efficiency of the dose-dependent version of Rosenbaum's U-statistics. For notational simplicity, we state the result in terms of the post-matching sensitivity parameter $\overline{\Gamma}$. By Proposition~\ref{prop:sufficiency}, the corresponding statement and proof on the population-level $\gamma$-scale follow directly from the one-to-one correspondence between $\gamma$ and $\overline{\Gamma}$.

\begin{theorem}[Generalized Bahadur-Rosenbaum Efficiency for the Dose-Dependent Version of Rosenbaum's U-Statistics]\label{thm: Bahadur slope}
Recall that $\overline{\Gamma}=I^{-1}\sum_{i=1}^{I}\Gamma_{i} =I^{-1}\sum_{i=1}^{I}\exp\{\gamma|\phi(z_{i}^{**})-\phi(z_{i}^{*})|\}$ denotes the generalized Rosenbaum-type sensitivity parameter for the $I$ matched pairs. Consider the same setting as in Theorem~\ref{thm: design sensitivity of U-statistics}. Let $\omega_{0}(t)=E\Big[\log \big \{ \widetilde{p}_{i}^{+}\exp(\varrho^{*}_{i}t)+(1-\widetilde{p}_{i}^{+})\big\}\Big]$ where $\widetilde{p}_{i}^{+}=\expit\{\widetilde{\gamma}|\phi(Z_{i1})-\phi(Z_{i2})|\}$ and $\widetilde{\gamma}$ is the unique value such that $E[\exp\{\widetilde{\gamma}|\phi(Z_{i1})-\phi(Z_{i2})|\}]=\overline{\Gamma}$. Also, let $t=\widetilde{t}$ denote the unique solution to $\mu=\omega_{1}(t)$ where
\begin{equation*}
    \omega_{1}(t)=E\Big[\frac{\widetilde{p}_{i}^{+}\varrho^{*}_{i}\exp(\varrho^{*}_{i}t)}{\widetilde{p}_{i}^{+}\exp(\varrho^{*}_{i}t)+(1-\widetilde{p}_{i}^{+})} \Big].
\end{equation*}
Then, for any fixed $\overline{\Gamma}$ where $1 \leq \overline{\Gamma} < \overline{\Gamma}_{*}$, the worst-case $p$-value $p_{\overline{\Gamma}, I}$ satisfies 
\begin{equation*}
    -I^{-1}\log (p_{\overline{\Gamma}, I}) \xrightarrow{p}   \mu  \widetilde{t}-\omega_{0}(\widetilde{t}) \ \text{ as $I\rightarrow \infty$.}
\end{equation*}
Also, $\Upsilon= 2\{  \mu \widetilde{t} - \omega_{0}(\widetilde{t})\}$ is the generalized Bahadur-Rosenbaum exact slope (efficiency) of the test statistic $T_U$.
\end{theorem}

Finally, because the adaptive testing procedure in equation (7) of the main text applies to any pair of sign-score test statistics, it can be directly applied when one or both component tests are dose-dependent versions of Rosenbaum's U-statistics. Therefore, the Type I error rate control and adaptivity results in Section 4 continue to hold under the corresponding regularity conditions.

\smallskip

\noindent \textit{B.3.2: Extension to the Dose-Dependent Versions of Rosenbaum's M-Statistics}

\smallskip

\noindent Our approaches can also be extended to the dose-dependent versions of Rosenbaum's M-statistics \citep{rosenbaum2007sensitivity, rosenbaum2014weighted}. Specifically, define $\operatorname{sign}(D_{i})=1$ if $D_{i}>0$, $\operatorname{sign}(D_{i})=0$ if $D_{i}=0$, and $\operatorname{sign}(D_{i})=-1$ if $D_{i}<0$. Let $h_{I}>0$ be a function of $\{|D_{1}|, \dots, |D_{I}|\}$. For example, \citet{rosenbaum2007sensitivity} used the median of $\{|D_{1}|, \dots, |D_{I}|\}$. Let $\chi(\cdot)$ be an odd function satisfying $\chi(t)\geq 0$ for $t\geq 0$ and $\chi(-t)=-\chi(t)$. The resulting M-statistic can be written as
\begin{equation*}
    T_{M}=\sum_{i=1}^{I}\chi\left(\frac{|D_{i}|}{h_{I}}\right)\cdot \operatorname{sign}(D_{i}).
\end{equation*}
In the absence of ties among $\{|D_{i}|\}_{i=1}^{I}$, we have
\begin{equation}\label{eqn: m as sign-score}
   T_{M}=\sum_{i=1}^{I}\chi\left(\frac{|D_{i}|}{h_{I}}\right)\cdot \operatorname{sign}(D_{i})=\sum_{i=1}^{I}q_{I,i} \mathbbm{1}\{D_{i}>0\}-\sum_{i=1}^{I}\chi\left(\frac{|D_{i}|}{h_{I}}\right),
\end{equation}
where $q_{I,i}=2\cdot \chi\left(\frac{|D_{i}|}{h_{I}}\right)$. Under $H_{0}$, both the $q_{I,i}$ and $\sum_{i=1}^{I}\chi\left(\frac{|D_{i}|}{h_{I}}\right)$ are fixed. Consequently, the worst-case $p$-value formula in equation (5) of the main text applies directly to $T_{M}$.

Using arguments analogous to those in Appendix~A.3, we obtain the corresponding generalized design sensitivity for the dose-dependent version of Rosenbaum's M-statistics:
\begin{theorem}[Generalized Design Sensitivity Formula for the Dose-Dependent Version of Rosenbaum's M-statistics]\label{thm: design sensitivity of M-statistics}
Suppose that $(Z_{i1}, Z_{i2}, Y_{i1}, Y_{i2})$ are independent and identically distributed (iid) realizations from a multivariate distribution $F$ in which $(Z_{i1}, Z_{i2})$ are continuous. Suppose that $h_I\xrightarrow{\mathrm{a.s.}}h$ for some constant $h>0$ and that $I^{-1}\sum_{i=1}^{I}
\left|q_{I,i}-2\cdot \chi\left(\frac{|D_i|}{h}\right)\right|
\xrightarrow{\mathrm{a.s.}}0$. Let $\mu=E\left[2\cdot \chi\left(\frac{|D_{i}|}{h}\right)\mathbbm{1}\{D_{i}>0\}\right]$. Under the regularity conditions stated above and conditions analogous to those in Appendix~A.3, suppose that the following equation in $\gamma$,
\begin{equation}\label{eqn: formula of the generalized design sensitivity of M-statistics}
    E\left[2\cdot \chi\left(\frac{|D_{i}|}{h}\right)\expit\left\{\gamma|\phi(Z_{i1})-\phi(Z_{i2})|\right\}\right]=\mu,
\end{equation}
has a unique solution $\gamma_{*}\in(0,+\infty)$ and that its left-hand side is strictly monotonically increasing in $\gamma$. Then, $\gamma_{*}$ is the generalized design sensitivity of $T_{M}$ on the $\gamma$-scale, and $\overline{\Gamma}_{*}=E\big[\exp\{\gamma_{*}|\phi(Z_{i1})-\phi(Z_{i2})|\}\big]$ is the generalized design sensitivity of $T_{M}$ on the $\overline{\Gamma}$-scale.
\end{theorem}

Adapting the proof strategy from Appendix~A.4 yields the generalized Bahadur-Rosenbaum efficiency for the dose-dependent version of Rosenbaum's M-statistics. For notational simplicity, we present the result on the post-matching $\overline{\Gamma}$-scale. Proposition~\ref{prop:sufficiency} then provides an equivalent formulation and proof on the population-level $\gamma$-scale through the one-to-one correspondence between the two parameters.

\begin{theorem}[Generalized Bahadur-Rosenbaum Efficiency for the Dose-Dependent Version of Rosenbaum's M-statistics]\label{thm: Bahadur slope of M-statistics}
Recall that $\overline{\Gamma}=I^{-1}\sum_{i=1}^{I}\Gamma_{i}=I^{-1}\sum_{i=1}^{I}\exp\{\gamma|\phi(z_{i}^{**})-\phi(z_{i}^{*})|\}$ denotes the generalized Rosenbaum-type sensitivity parameter for the $I$ matched pairs. Consider the same setting as in Theorem~\ref{thm: design sensitivity of M-statistics}. Let $\omega_{0}(t)=E\Big[\log\big\{\widetilde{p}_{i}^{+}\exp(q_{i}^{*}t)+(1-\widetilde{p}_{i}^{+})\big\}\Big]$, where $\widetilde{p}_{i}^{+}=\expit\{\widetilde{\gamma}|\phi(Z_{i1})-\phi(Z_{i2})|\}$, $q_{i}^{*}=2\cdot \chi\left(\frac{|D_{i}|}{h}\right)$, and $\widetilde{\gamma}$ is the unique value such that
\begin{equation*}
    E\big[\exp\{\widetilde{\gamma}|\phi(Z_{i1})-\phi(Z_{i2})|\}\big]=\overline{\Gamma}.
\end{equation*}
Also, let $t=\widetilde{t}$ denote the unique solution to $\mu=\omega_{1}(t)$, where
$\omega_{1}(t)=E\Big[\frac{\widetilde{p}_{i}^{+}q_{i}^{*}\exp(q_{i}^{*}t)}
    {\widetilde{p}_{i}^{+}\exp(q_{i}^{*}t)+(1-\widetilde{p}_{i}^{+})}\Big]$. Then, for any fixed $\overline{\Gamma}$ satisfying $1\leq\overline{\Gamma}<\overline{\Gamma}_{*}$, the worst-case $p$-value $p_{\overline{\Gamma},I}$ satisfies
\begin{equation*}
    -I^{-1}\log(p_{\overline{\Gamma},I})\xrightarrow{p}
    \mu\widetilde{t}-\omega_{0}(\widetilde{t})
    \ \text{ as $I\rightarrow\infty$.}
\end{equation*}
Moreover, $\Upsilon=2\{\mu\widetilde{t}-\omega_{0}(\widetilde{t})\}$ is the generalized Bahadur-Rosenbaum exact slope (efficiency) of $T_{M}$.
\end{theorem}

Finally, because $T_{M}$ is a sign-score statistic, the adaptive procedure in equation (7) of the main text can combine a dose-dependent M-statistic with another M-statistic or any other sign-score statistic. Therefore, the Type I error rate control and adaptivity results in Section~4 of the main text continue to hold under the corresponding joint asymptotic normality and regularity conditions. Although \citet{rosenbaum2014weighted} developed weighted M-statistics for matched sets with multiple controls, the extension established here concerns the pair-matched version, consistent with the scope of the present paper. Extending the full theory to general matched sets with multiple controls would require a separate treatment of the post-matching dose-assignment sensitivity bounds and is left for future research.

\section*{Appendix C: Additional Simulation Studies and Data Analysis Details}

For notational consistency, unless otherwise stated, the theoretical arguments and numerical results in Appendices~C and~D are presented primarily in terms of the post-matching (data-specific) sensitivity parameter $\overline{\Gamma}$. By Proposition~\ref{prop:sufficiency}, the corresponding formulations and conclusions on the population-level $\gamma$-scale follow directly from the one-to-one correspondence between $\gamma$ and $\overline{\Gamma}$.

\subsection*{C.1: Simulation Studies in the Binary Outcome Case}

In this section, we conduct simulation studies to verify the effectiveness of the proposed formulas for generalized design sensitivity and Bahadur-Rosenbaum efficiency in the binary outcome case. We consider using the proposed formulas for evaluating McNemar's (Mc) test and dose-weighted McNemar's (D-Mc) test under the treatment effect model: $Y_{ij} = \mathbbm{1}\{\beta Z_{ij} + \delta_i \geq \epsilon_{ij}\}$, with $\beta \in \{3,5\}$. We sample the lower treatment dose within each matched pair as \(z_i^* \simiid \mathcal{N}(-1, 1)\), and generate the dose difference as \(z_i^{**} - z_i^* \simiid \text{Unif}(1, 5)\). The pair-level random effect \(\delta_i\) and individual-level random error \(\epsilon_{ij}\) are generated independently from \(\mathcal{N}(0, 1)\). As in the main text, the generalized design sensitivities are computed using Theorem~3.4. The simulated power is estimated as the proportion of $1{,}000$ Monte Carlo iterations in which the worst-case (upper bound) $p$-value falls below the significance level \(\alpha = 0.01\). Each iteration involves \(I = 5{,}000\) matched pairs. The simulation results in Table~\ref{tb:design_sensitivity_mcnemar_variants} show that the simulated power is high when sensitivity analysis is conducted at \(\overline{\Gamma} \ll \overline{\Gamma}_*\), and the simulated power goes to zero when \(\overline{\Gamma} > \overline{\Gamma}_*\). These findings confirm the applicability of our generalized design sensitivity formula to settings with binary outcomes. Similar to the observations in the main text for continuous outcomes, the adaptive test prevents the loss of power when choosing a suboptimal test. For example, when $\beta = 3$ and $\overline{\Gamma} = 2.50$, simulated power of the adaptive test combining Mc and D-Mc is $1.00$, while the simulated power of Mc is $0.15$.

To evaluate the effectiveness of Theorem~3.5 and Corollary~3.6, Table~\ref{tb:binary_outcome_Bahadur} reports the generalized Bahadur-Rosenbaum relative efficiency (denoted as $\Upsilon_{\text{Mc}}/\Upsilon_{\text{D-Mc}}$) alongside the ratio of the simulated minimum numbers of matched pairs required to achieve power of $0.95$ at significance level \(\alpha = 0.01\), comparing Mc to D-Mc (denoted as $I_{\text{D-Mc}}/I_{\text{Mc}}$). From the simulation results in Table~\ref{tb:binary_outcome_Bahadur}, we find that the generalized Bahadur-Rosenbaum relative efficiency closely approximates the simulated sample size ratio, confirming the effectiveness of the proposed formula in the binary outcome case.

\begin{table}[ht]
\centering
\small
\caption{The generalized design sensitivities $\overline{\Gamma}_{*}$ computed from Theorem~3.4 for McNemar's test (Mc) and the dose-weighted McNemar's test (D-Mc), and from Theorem~4.2 for the adaptive test taking them as components, under the binary-outcome treatment effect model $Y_{ij} = \mathbbm{1}\{\beta Z_{ij} + \delta_{i} \geq \epsilon_{ij}\}$ with effect size $\beta \in \{3, 5\}$. The sensitivity model uses the dose link function $\phi(z) = z$, and the test statistics are constructed from the untransformed doses, i.e., the dose transformation of Appendix~C.3 is the identity, $\kappa(z) = z$. Below the row of $\overline{\Gamma}_{*}$, we report the simulated power under $I = 5000$ and various values of $\overline{\Gamma}$ at $\alpha = 0.01$.}
\begin{tabular}{l | c | c | c c c}
\hline
\textbf{Effect size} & $\kappa(z)$ & $\overline{\Gamma}$ & Mc & D-Mc & Adaptive \\
\hline\hline
\multirow{7}{*}{$\beta = 3$} & \multirow{7}{*}{$z$}
& $\overline{\Gamma}_*$ & 2.62 & 3.11 & 3.11\\
\cline{3-6}
& & 1.00 & 1.00 & 1.00 & 1.00\\
& & 1.50 & 1.00 & 1.00 & 1.00\\
& & 2.00 & 1.00 & 1.00 & 1.00\\
& & 2.50 & 0.15 & 1.00 & 1.00\\
& & 3.00 & 0.00 & 0.09 & 0.07\\
& & 3.50 & 0.00 & 0.00 & 0.00\\
\hline\hline
\multirow{7}{*}{$\beta = 5$} & \multirow{7}{*}{$z$}
& $\overline{\Gamma}_*$ & 3.00 & 3.51 & 3.51\\
\cline{3-6}
& & 1.00 & 1.00 & 1.00 & 1.00\\
& & 1.50 & 1.00 & 1.00 & 1.00\\
& & 2.00 & 1.00 & 1.00 & 1.00\\
& & 2.50 & 0.99 & 1.00 & 1.00\\
& & 3.00 & 0.01 & 0.95 & 0.94\\
& & 3.50 & 0.00 & 0.01 & 0.01\\
\hline
\end{tabular}
\label{tb:design_sensitivity_mcnemar_variants}
\end{table}

\begin{table}[ht]
\caption{The generalized Bahadur-Rosenbaum relative efficiencies $\Upsilon_{\text{Mc}}/\Upsilon_{\text{D-Mc}}$ computed from Theorem~3.5 for McNemar's test (Mc) and the dose-weighted McNemar's test (D-Mc), together with the simulated ratio $I_{\text{D-Mc}}/I_{\text{Mc}}$ of the minimum numbers of matched pairs required to attain $95\%$ power at $\alpha = 0.01$ (i.e., Corollary~3.6), under the binary-outcome treatment effect model $Y_{ij} = \mathbbm{1}\{\beta Z_{ij} + \delta_{i} \geq \epsilon_{ij}\}$ with effect size $\beta \in \{3, 5\}$. The sensitivity model uses the dose link function $\phi(z) = z$, and the test statistics are constructed from the untransformed doses, i.e., the dose transformation of Appendix~C.3 is the identity, $\kappa(z) = z$.}
\centering
\small
\begin{tabular}{l | c | c | c c c c}
\hline
\textbf{Effect size} & $\kappa(z)$ & $\overline{\Gamma}$ & $I_{\text{Mc}}$ & $I_{\text{D-Mc}}$ & $\Upsilon_{\text{Mc}}/\Upsilon_{\text{D-Mc}}$ & $I_{\text{D-Mc}}/I_{\text{Mc}}$ \\
\hline\hline
\multirow{4}{*}{$\beta = 3$} & \multirow{4}{*}{$z$}
& 1.125 & 119& 74& 0.67 & 0.62\\
& & 1.250 & 158& 99& 0.64 & 0.63\\
& & 1.500 & 223& 125& 0.60 & 0.56\\
& & 1.750 & 288& 163 & 0.57 & 0.57\\
\hline\hline
\multirow{4}{*}{$\beta = 5$} & \multirow{4}{*}{$z$}
& 1.125 & 94 & 63 & 0.72 & 0.67\\
& & 1.250 & 118& 81& 0.69 & 0.69\\
& & 1.500 & 155& 106 & 0.67 & 0.68\\
& & 1.750 & 204& 128 & 0.64 & 0.63\\
\hline
\end{tabular}
\label{tb:binary_outcome_Bahadur}
\end{table}

\subsection*{C.2: Finite-Sample Performance: Generalized Design Sensitivity, Adaptive Testing, and Sample Splitting}

This section examines how well generalized design sensitivity describes finite-sample performance and compares the adaptive procedure with the sample splitting strategy of \citet{heller2009split}. Tables~\ref{tb:ds_power_split_merged_100}--\ref{tb:ds_power_split_merged_5000} report simulated power for $I=100,500,1000,$ and $5000$ matched pairs. For sample splitting, either $10\%$ or $20\%$ of the pairs are reserved as planning data for selecting between the two candidate test statistics.

First, the simulations show broad agreement between the ordering induced by generalized design sensitivity and relative finite-sample performance. Even at $I=100$, the general pattern is often visible, although some departures from the asymptotic ordering remain. By $I=500$, this ordering is generally reflected in simulated power when the competing design sensitivities are reasonably well separated. One notable exception is the comparison between U and D-U under the linear dose-response curve, for which the design sensitivities are relatively close ($4.16$ versus $4.43$); in this case, substantially larger samples are needed before the anticipated ordering becomes clearly apparent (e.g., $I=5000$). These results suggest that generalized design sensitivity can provide useful guidance at moderate sample sizes, although how clearly the asymptotic ordering is reflected at a given sample size depends on the data-generating process, the separation between the design sensitivities, and how close $\overline{\Gamma}$ is to the design sensitivity boundary. When the generalized design sensitivities of two competing tests are close, or when $\overline{\Gamma}$ is sufficiently below these boundaries that multiple tests already have power close to one at the sample size and under the alternative considered, generalized Bahadur-Rosenbaum efficiency can complement generalized design sensitivity by providing a finer asymptotic comparison among the test statistics, as illustrated in Table~3 of the main text.

Second, the simulations indicate that the finite-sample cost of adaptivity is generally modest. Because the common critical value lies in $\big(\Phi^{-1}(1-\alpha),\Phi^{-1}(1-\alpha/2)\big)$, each component statistic is effectively assessed at a level $\alpha^{*}\in(\alpha/2,\alpha)$ rather than at $\alpha$. At the same time, the adaptive procedure rejects when either component statistic exceeds this common threshold, which helps offset the multiplicity cost. Consequently, its power generally remains close to that of the better-performing component statistic and well above that of the less favorable component statistic. For example, at $I=5000$ under the kink dose-response curve with $\overline{\Gamma}=3.50$, the adaptive test has power $0.26$, compared with $0.36$ for D-U and $0.00$ for U. Similarly, at $I=1000$ under the log dose-response curve with $\overline{\Gamma}=3.00$, the simulated powers of U, D-U, and the adaptive test are $0.55$, $0.08$, and $0.45$, respectively. Thus, the adaptive procedure gives up some power relative to knowing the better statistic in advance, but provides substantial protection against choosing the less favorable statistic.

Finally, the comparison with sample splitting is also favorable. Although the adaptive and sample splitting procedures have the same design sensitivity (i.e., the larger of the two component design sensitivities), the adaptive procedure typically achieves higher finite-sample power in the settings considered. One likely reason is that it uses all matched pairs for testing, while sample splitting reserves a fraction of the pairs for selecting the statistic. For example, at $I=500$ under the kink dose-response curve with $\overline{\Gamma}=2.00$, the adaptive test combining U and D-U attains power $0.63$, compared with $0.48$ and $0.43$ for the $10\%$ and $20\%$ splits of the same pair of tests. At $I=5000$ under the square dose-response curve with $\overline{\Gamma}=3.00$, the corresponding powers are $0.85$, $0.56$, and $0.65$. The few instances in which sample splitting performs marginally better occur mainly when the procedures have very low or nearly equal power. The adaptive procedure is particularly appealing at $I=100$, since a $10\%$ or $20\%$ split provides only approximately $10$ or $20$ planning pairs from which to select the statistic.

\begin{table}[!htbp]
\caption{\textbf{(Simulations under $I=100$)} The generalized design sensitivities $\overline{\Gamma}_{*}$ computed from Theorem~3.4 for the Wilcox, D-Wilcox, U, and D-U tests, and from Theorem~4.2 for the adaptive test combining Wilcox and D-Wilcox and the adaptive test combining U and D-U, together with the sample splitting procedure of \citet{heller2009split} using $10\%$ and $20\%$ of the data as the planning sample, whose generalized design sensitivity equals the larger of the two component design sensitivities. The sensitivity model uses the dose link function $\phi(z) = z$, and the test statistics are constructed from the untransformed doses, i.e., the dose transformation of Appendix~C.3 is the identity, $\kappa(z) = z$. Below the row of $\overline{\Gamma}_{*}$, we report the simulated power under $I = 100$ and various values of $\overline{\Gamma}$ at $\alpha = 0.01$.}
\centering
\footnotesize
\setlength{\tabcolsep}{3pt}
\resizebox{\textwidth}{!}{
\begin{tabular}{l | c | c | c c c c c | c c c c c}
\hline
\textbf{Dose--response} & $\kappa(z)$ & $\overline{\Gamma}$ & Wilcox & D-Wilcox & Adaptive & Split (0.10) & Split (0.20) & U & D-U & Adaptive & Split (0.10) & Split (0.20) \\
\hline\hline
\multirow{7}{*}{\textbf{Square}} & \multirow{7}{*}{$z$}
& $\overline{\Gamma}_*$ & 2.32 & 2.44 & 2.44 & 2.44 & 2.44 & 3.32 & 4.16 & 4.16 & 4.16 & 4.16\\
\cline{3-13}
& & 1.50 & 0.17 & 0.25 & 0.22 & 0.20 & 0.17 & 0.26 & 0.11 & 0.19 & 0.16 & 0.13\\
& & 2.00 & 0.02 & 0.03 & 0.03 & 0.02 & 0.02 & 0.06 & 0.00 & 0.03 & 0.03 & 0.02\\
& & 2.50 & 0.00 & 0.00 & 0.00 & 0.00 & 0.00 & 0.01 & 0.00 & 0.01 & 0.01 & 0.01\\
& & 3.00 & 0.00 & 0.00 & 0.00 & 0.00 & 0.00 & 0.01 & 0.00 & 0.00 & 0.00 & 0.00\\
& & 3.50 & 0.00 & 0.00 & 0.00 & 0.00 & 0.00 & 0.00 & 0.00 & 0.00 & 0.00 & 0.00\\
& & 4.00 & 0.00 & 0.00 & 0.00 & 0.00 & 0.00 & 0.00 & 0.00 & 0.00 & 0.00 & 0.00\\
\hline\hline
\multirow{7}{*}{\textbf{Kink}} & \multirow{7}{*}{$z$}
& $\overline{\Gamma}_*$ & 2.26 & 2.43 & 2.43 & 2.43 & 2.43 & 3.17 & 4.28 & 4.28 & 4.28 & 4.28\\
\cline{3-13}
& & 1.50 & 0.14 & 0.25 & 0.21 & 0.21 & 0.17 & 0.24 & 0.13 & 0.17 & 0.15 & 0.12\\
& & 2.00 & 0.02 & 0.02 & 0.02 & 0.02 & 0.02 & 0.05 & 0.00 & 0.03 & 0.03 & 0.02\\
& & 2.50 & 0.00 & 0.00 & 0.00 & 0.00 & 0.00 & 0.01 & 0.00 & 0.01 & 0.01 & 0.01\\
& & 3.00 & 0.00 & 0.00 & 0.00 & 0.00 & 0.00 & 0.00 & 0.00 & 0.00 & 0.00 & 0.00\\
& & 3.50 & 0.00 & 0.00 & 0.00 & 0.00 & 0.00 & 0.00 & 0.00 & 0.00 & 0.00 & 0.00\\
& & 4.00 & 0.00 & 0.00 & 0.00 & 0.00 & 0.00 & 0.00 & 0.00 & 0.00 & 0.00 & 0.00\\
\hline\hline
\multirow{7}{*}{\textbf{Linear}} & \multirow{7}{*}{$z$}
& $\overline{\Gamma}_*$ & 2.68 & 2.68 & 2.68 & 2.68 & 2.68 & 4.16 & 4.43 & 4.43 & 4.43 & 4.43\\
\cline{3-13}
& & 1.50 & 0.31 & 0.36 & 0.34 & 0.31 & 0.26 & 0.40 & 0.13 & 0.30 & 0.22 & 0.22\\
& & 2.00 & 0.06 & 0.05 & 0.05 & 0.05 & 0.04 & 0.13 & 0.01 & 0.08 & 0.06 & 0.05\\
& & 2.50 & 0.01 & 0.01 & 0.01 & 0.01 & 0.01 & 0.04 & 0.00 & 0.02 & 0.02 & 0.02\\
& & 3.00 & 0.00 & 0.00 & 0.00 & 0.00 & 0.00 & 0.02 & 0.00 & 0.01 & 0.01 & 0.01\\
& & 3.50 & 0.00 & 0.00 & 0.00 & 0.00 & 0.00 & 0.01 & 0.00 & 0.00 & 0.00 & 0.00\\
& & 4.00 & 0.00 & 0.00 & 0.00 & 0.00 & 0.00 & 0.00 & 0.00 & 0.00 & 0.00 & 0.00\\
\hline\hline
\multirow{7}{*}{\textbf{Square root}} & \multirow{7}{*}{$z$}
& $\overline{\Gamma}_*$ & 2.45 & 2.38 & 2.45 & 2.45 & 2.45 & 3.67 & 3.50 & 3.67 & 3.67 & 3.67\\
\cline{3-13}
& & 1.50 & 0.22 & 0.21 & 0.21 & 0.19 & 0.17 & 0.31 & 0.07 & 0.22 & 0.17 & 0.16\\
& & 2.00 & 0.03 & 0.03 & 0.03 & 0.02 & 0.02 & 0.09 & 0.00 & 0.05 & 0.04 & 0.04\\
& & 2.50 & 0.01 & 0.00 & 0.01 & 0.01 & 0.00 & 0.03 & 0.00 & 0.01 & 0.01 & 0.01\\
& & 3.00 & 0.00 & 0.00 & 0.00 & 0.00 & 0.00 & 0.01 & 0.00 & 0.01 & 0.01 & 0.01\\
& & 3.50 & 0.00 & 0.00 & 0.00 & 0.00 & 0.00 & 0.00 & 0.00 & 0.00 & 0.00 & 0.00\\
& & 4.00 & 0.00 & 0.00 & 0.00 & 0.00 & 0.00 & 0.00 & 0.00 & 0.00 & 0.00 & 0.00\\
\hline\hline
\multirow{7}{*}{\textbf{Flat}} & \multirow{7}{*}{$z$}
& $\overline{\Gamma}_*$ & 2.60 & 2.45 & 2.60 & 2.60 & 2.60 & 4.02 & 3.37 & 4.02 & 4.02 & 4.02\\
\cline{3-13}
& & 1.50 & 0.28 & 0.24 & 0.26 & 0.23 & 0.20 & 0.36 & 0.07 & 0.27 & 0.20 & 0.21\\
& & 2.00 & 0.05 & 0.03 & 0.04 & 0.03 & 0.02 & 0.12 & 0.00 & 0.08 & 0.07 & 0.06\\
& & 2.50 & 0.01 & 0.00 & 0.01 & 0.01 & 0.00 & 0.04 & 0.00 & 0.02 & 0.02 & 0.02\\
& & 3.00 & 0.00 & 0.00 & 0.00 & 0.00 & 0.00 & 0.01 & 0.00 & 0.01 & 0.01 & 0.01\\
& & 3.50 & 0.00 & 0.00 & 0.00 & 0.00 & 0.00 & 0.01 & 0.00 & 0.00 & 0.00 & 0.00\\
& & 4.00 & 0.00 & 0.00 & 0.00 & 0.00 & 0.00 & 0.00 & 0.00 & 0.00 & 0.00 & 0.00\\
\hline\hline
\multirow{7}{*}{\textbf{Log}} & \multirow{7}{*}{$z$}
& $\overline{\Gamma}_*$ & 2.83 & 2.63 & 2.83 & 2.83 & 2.83 & 4.65 & 3.84 & 4.65 & 4.65 & 4.65\\
\cline{3-13}
& & 1.50 & 0.38 & 0.34 & 0.36 & 0.31 & 0.26 & 0.46 & 0.09 & 0.36 & 0.26 & 0.26\\
& & 2.00 & 0.07 & 0.06 & 0.06 & 0.06 & 0.04 & 0.18 & 0.00 & 0.11 & 0.09 & 0.08\\
& & 2.50 & 0.01 & 0.01 & 0.01 & 0.01 & 0.01 & 0.06 & 0.00 & 0.03 & 0.03 & 0.02\\
& & 3.00 & 0.00 & 0.00 & 0.00 & 0.00 & 0.00 & 0.03 & 0.00 & 0.01 & 0.01 & 0.01\\
& & 3.50 & 0.00 & 0.00 & 0.00 & 0.00 & 0.00 & 0.01 & 0.00 & 0.01 & 0.01 & 0.01\\
& & 4.00 & 0.00 & 0.00 & 0.00 & 0.00 & 0.00 & 0.01 & 0.00 & 0.00 & 0.00 & 0.00\\
\hline
\end{tabular}
}
\label{tb:ds_power_split_merged_100}
\end{table}

\begin{table}[!htbp]
\caption{\textbf{(Simulations under $I=500$)} The generalized design sensitivities $\overline{\Gamma}_{*}$ computed from Theorem~3.4 for the Wilcox, D-Wilcox, U, and D-U tests, and from Theorem~4.2 for the adaptive test combining Wilcox and D-Wilcox and the adaptive test combining U and D-U, together with the sample splitting procedure of \citet{heller2009split} using $10\%$ and $20\%$ of the data as the planning sample, whose generalized design sensitivity equals the larger of the two component design sensitivities. The sensitivity model uses the dose link function $\phi(z) = z$, and the test statistics are constructed from the untransformed doses, i.e., the dose transformation of Appendix~C.3 is the identity, $\kappa(z) = z$. Below the row of $\overline{\Gamma}_{*}$, we report the simulated power under $I = 500$ and various values of $\overline{\Gamma}$ at $\alpha = 0.01$.}
\centering
\footnotesize
\setlength{\tabcolsep}{3pt}
\resizebox{\textwidth}{!}{
\begin{tabular}{l | c | c | c c c c c | c c c c c}
\hline
\textbf{Dose--response} & $\kappa(z)$ & $\overline{\Gamma}$ & Wilcox & D-Wilcox & Adaptive & Split (0.10) & Split (0.20) & U & D-U & Adaptive & Split (0.10) & Split (0.20) \\
\hline\hline
\multirow{7}{*}{\textbf{Square}} & \multirow{7}{*}{$z$}
& $\overline{\Gamma}_*$ & 2.32 & 2.44 & 2.44 & 2.44 & 2.44 & 3.32 & 4.16 & 4.16 & 4.16 & 4.16\\
\cline{3-13}
& & 1.50 & 0.87 & 0.97 & 0.96 & 0.91 & 0.89 & 0.97 & 0.98 & 0.99 & 0.96 & 0.92\\
& & 2.00 & 0.08 & 0.17 & 0.13 & 0.12 & 0.11 & 0.55 & 0.64 & 0.62 & 0.52 & 0.46\\
& & 2.50 & 0.00 & 0.00 & 0.00 & 0.00 & 0.00 & 0.12 & 0.17 & 0.14 & 0.12 & 0.11\\
& & 3.00 & 0.00 & 0.00 & 0.00 & 0.00 & 0.00 & 0.02 & 0.02 & 0.02 & 0.02 & 0.01\\
& & 3.50 & 0.00 & 0.00 & 0.00 & 0.00 & 0.00 & 0.00 & 0.00 & 0.00 & 0.00 & 0.00\\
& & 4.00 & 0.00 & 0.00 & 0.00 & 0.00 & 0.00 & 0.00 & 0.00 & 0.00 & 0.00 & 0.00\\
\hline\hline
\multirow{7}{*}{\textbf{Kink}} & \multirow{7}{*}{$z$}
& $\overline{\Gamma}_*$ & 2.26 & 2.43 & 2.43 & 2.43 & 2.43 & 3.17 & 4.28 & 4.28 & 4.28 & 4.28\\
\cline{3-13}
& & 1.50 & 0.83 & 0.97 & 0.95 & 0.90 & 0.86 & 0.95 & 0.98 & 0.99 & 0.96 & 0.91\\
& & 2.00 & 0.05 & 0.15 & 0.13 & 0.11 & 0.10 & 0.47 & 0.69 & 0.63 & 0.48 & 0.43\\
& & 2.50 & 0.00 & 0.00 & 0.00 & 0.00 & 0.00 & 0.09 & 0.21 & 0.15 & 0.12 & 0.10\\
& & 3.00 & 0.00 & 0.00 & 0.00 & 0.00 & 0.00 & 0.01 & 0.03 & 0.01 & 0.02 & 0.01\\
& & 3.50 & 0.00 & 0.00 & 0.00 & 0.00 & 0.00 & 0.00 & 0.01 & 0.00 & 0.00 & 0.00\\
& & 4.00 & 0.00 & 0.00 & 0.00 & 0.00 & 0.00 & 0.00 & 0.00 & 0.00 & 0.00 & 0.00\\
\hline\hline
\multirow{7}{*}{\textbf{Linear}} & \multirow{7}{*}{$z$}
& $\overline{\Gamma}_*$ & 2.68 & 2.68 & 2.68 & 2.68 & 2.68 & 4.16 & 4.43 & 4.43 & 4.43 & 4.43\\
\cline{3-13}
& & 1.50 & 0.98 & 0.99 & 0.99 & 0.98 & 0.97 & 1.00 & 0.98 & 1.00 & 0.99 & 0.98\\
& & 2.00 & 0.38 & 0.45 & 0.42 & 0.38 & 0.32 & 0.85 & 0.72 & 0.85 & 0.76 & 0.70\\
& & 2.50 & 0.02 & 0.03 & 0.02 & 0.02 & 0.01 & 0.43 & 0.24 & 0.39 & 0.33 & 0.29\\
& & 3.00 & 0.00 & 0.00 & 0.00 & 0.00 & 0.00 & 0.13 & 0.04 & 0.09 & 0.09 & 0.08\\
& & 3.50 & 0.00 & 0.00 & 0.00 & 0.00 & 0.00 & 0.04 & 0.01 & 0.02 & 0.03 & 0.03\\
& & 4.00 & 0.00 & 0.00 & 0.00 & 0.00 & 0.00 & 0.01 & 0.00 & 0.00 & 0.00 & 0.01\\
\hline\hline
\multirow{7}{*}{\textbf{Square root}} & \multirow{7}{*}{$z$}
& $\overline{\Gamma}_*$ & 2.45 & 2.38 & 2.45 & 2.45 & 2.45 & 3.67 & 3.50 & 3.67 & 3.67 & 3.67\\
\cline{3-13}
& & 1.50 & 0.93 & 0.95 & 0.94 & 0.92 & 0.87 & 0.99 & 0.91 & 0.98 & 0.94 & 0.93\\
& & 2.00 & 0.18 & 0.14 & 0.16 & 0.13 & 0.12 & 0.68 & 0.35 & 0.63 & 0.54 & 0.49\\
& & 2.50 & 0.00 & 0.00 & 0.00 & 0.00 & 0.00 & 0.23 & 0.05 & 0.18 & 0.16 & 0.14\\
& & 3.00 & 0.00 & 0.00 & 0.00 & 0.00 & 0.00 & 0.05 & 0.01 & 0.03 & 0.03 & 0.03\\
& & 3.50 & 0.00 & 0.00 & 0.00 & 0.00 & 0.00 & 0.01 & 0.00 & 0.00 & 0.01 & 0.01\\
& & 4.00 & 0.00 & 0.00 & 0.00 & 0.00 & 0.00 & 0.00 & 0.00 & 0.00 & 0.00 & 0.00\\
\hline\hline
\multirow{7}{*}{\textbf{Flat}} & \multirow{7}{*}{$z$}
& $\overline{\Gamma}_*$ & 2.60 & 2.45 & 2.60 & 2.60 & 2.60 & 4.02 & 3.37 & 4.02 & 4.02 & 4.02\\
\cline{3-13}
& & 1.50 & 0.97 & 0.97 & 0.97 & 0.95 & 0.92 & 1.00 & 0.89 & 1.00 & 0.96 & 0.95\\
& & 2.00 & 0.31 & 0.20 & 0.26 & 0.21 & 0.18 & 0.81 & 0.31 & 0.74 & 0.64 & 0.62\\
& & 2.50 & 0.01 & 0.00 & 0.01 & 0.01 & 0.00 & 0.37 & 0.04 & 0.28 & 0.26 & 0.23\\
& & 3.00 & 0.00 & 0.00 & 0.00 & 0.00 & 0.00 & 0.10 & 0.01 & 0.06 & 0.07 & 0.07\\
& & 3.50 & 0.00 & 0.00 & 0.00 & 0.00 & 0.00 & 0.03 & 0.00 & 0.02 & 0.02 & 0.02\\
& & 4.00 & 0.00 & 0.00 & 0.00 & 0.00 & 0.00 & 0.01 & 0.00 & 0.00 & 0.01 & 0.01\\
\hline\hline
\multirow{7}{*}{\textbf{Log}} & \multirow{7}{*}{$z$}
& $\overline{\Gamma}_*$ & 2.83 & 2.63 & 2.83 & 2.83 & 2.83 & 4.65 & 3.84 & 4.65 & 4.65 & 4.65\\
\cline{3-13}
& & 1.50 & 1.00 & 0.99 & 0.99 & 0.99 & 0.97 & 1.00 & 0.95 & 1.00 & 0.99 & 0.97\\
& & 2.00 & 0.54 & 0.38 & 0.50 & 0.42 & 0.36 & 0.91 & 0.49 & 0.89 & 0.80 & 0.78\\
& & 2.50 & 0.05 & 0.02 & 0.03 & 0.03 & 0.02 & 0.59 & 0.11 & 0.50 & 0.46 & 0.42\\
& & 3.00 & 0.00 & 0.00 & 0.00 & 0.00 & 0.00 & 0.26 & 0.02 & 0.18 & 0.17 & 0.15\\
& & 3.50 & 0.00 & 0.00 & 0.00 & 0.00 & 0.00 & 0.08 & 0.00 & 0.05 & 0.06 & 0.05\\
& & 4.00 & 0.00 & 0.00 & 0.00 & 0.00 & 0.00 & 0.03 & 0.00 & 0.01 & 0.02 & 0.01\\
\hline
\end{tabular}
}
\label{tb:ds_power_split_merged_500}
\end{table}

\begin{table}[!htbp]
\caption{\textbf{(Simulations under $I=1000$)} The generalized design sensitivities $\overline{\Gamma}_{*}$ computed from Theorem~3.4 for the Wilcox, D-Wilcox, U, and D-U tests, and from Theorem~4.2 for the adaptive test combining Wilcox and D-Wilcox and the adaptive test combining U and D-U, together with the sample splitting procedure of \citet{heller2009split} using $10\%$ and $20\%$ of the data as the planning sample, whose generalized design sensitivity equals the larger of the two component design sensitivities. The sensitivity model uses the dose link function $\phi(z) = z$, and the test statistics are constructed from the untransformed doses, i.e., the dose transformation of Appendix~C.3 is the identity, $\kappa(z) = z$. Below the row of $\overline{\Gamma}_{*}$, we report the simulated power under $I = 1000$ and various values of $\overline{\Gamma}$ at $\alpha = 0.01$.}
\centering
\footnotesize
\setlength{\tabcolsep}{3pt}
\resizebox{\textwidth}{!}{
\begin{tabular}{l | c | c | c c c c c | c c c c c}
\hline
\textbf{Dose--response} & $\kappa(z)$ & $\overline{\Gamma}$ & Wilcox & D-Wilcox & Adaptive & Split (0.10) & Split (0.20) & U & D-U & Adaptive & Split (0.10) & Split (0.20) \\
\hline\hline
\multirow{7}{*}{\textbf{Square}} & \multirow{7}{*}{$z$}
& $\overline{\Gamma}_*$ & 2.32 & 2.44 & 2.44 & 2.44 & 2.44 & 3.32 & 4.16 & 4.16 & 4.16 & 4.16\\
\cline{3-13}
& & 1.50 & 1.00 & 1.00 & 1.00 & 1.00 & 0.99 & 1.00 & 1.00 & 1.00 & 1.00 & 1.00\\
& & 2.00 & 0.20 & 0.42 & 0.36 & 0.31 & 0.28 & 0.88 & 0.96 & 0.97 & 0.87 & 0.86\\
& & 2.50 & 0.00 & 0.00 & 0.00 & 0.00 & 0.00 & 0.30 & 0.56 & 0.49 & 0.34 & 0.31\\
& & 3.00 & 0.00 & 0.00 & 0.00 & 0.00 & 0.00 & 0.03 & 0.13 & 0.08 & 0.04 & 0.05\\
& & 3.50 & 0.00 & 0.00 & 0.00 & 0.00 & 0.00 & 0.00 & 0.02 & 0.01 & 0.00 & 0.00\\
& & 4.00 & 0.00 & 0.00 & 0.00 & 0.00 & 0.00 & 0.00 & 0.00 & 0.00 & 0.00 & 0.00\\
\hline\hline
\multirow{7}{*}{\textbf{Kink}} & \multirow{7}{*}{$z$}
& $\overline{\Gamma}_*$ & 2.26 & 2.43 & 2.43 & 2.43 & 2.43 & 3.17 & 4.28 & 4.28 & 4.28 & 4.28\\
\cline{3-13}
& & 1.50 & 0.99 & 1.00 & 1.00 & 1.00 & 0.99 & 1.00 & 1.00 & 1.00 & 1.00 & 1.00\\
& & 2.00 & 0.12 & 0.39 & 0.33 & 0.29 & 0.28 & 0.82 & 0.98 & 0.97 & 0.85 & 0.84\\
& & 2.50 & 0.00 & 0.00 & 0.00 & 0.00 & 0.00 & 0.20 & 0.62 & 0.52 & 0.33 & 0.33\\
& & 3.00 & 0.00 & 0.00 & 0.00 & 0.00 & 0.00 & 0.01 & 0.18 & 0.10 & 0.06 & 0.06\\
& & 3.50 & 0.00 & 0.00 & 0.00 & 0.00 & 0.00 & 0.00 & 0.02 & 0.01 & 0.00 & 0.01\\
& & 4.00 & 0.00 & 0.00 & 0.00 & 0.00 & 0.00 & 0.00 & 0.00 & 0.00 & 0.00 & 0.00\\
\hline\hline
\multirow{7}{*}{\textbf{Linear}} & \multirow{7}{*}{$z$}
& $\overline{\Gamma}_*$ & 2.68 & 2.68 & 2.68 & 2.68 & 2.68 & 4.16 & 4.43 & 4.43 & 4.43 & 4.43\\
\cline{3-13}
& & 1.50 & 1.00 & 1.00 & 1.00 & 1.00 & 1.00 & 1.00 & 1.00 & 1.00 & 1.00 & 1.00\\
& & 2.00 & 0.73 & 0.83 & 0.81 & 0.74 & 0.68 & 1.00 & 0.98 & 1.00 & 0.98 & 0.97\\
& & 2.50 & 0.04 & 0.04 & 0.04 & 0.03 & 0.03 & 0.79 & 0.66 & 0.80 & 0.70 & 0.64\\
& & 3.00 & 0.00 & 0.00 & 0.00 & 0.00 & 0.00 & 0.31 & 0.22 & 0.28 & 0.26 & 0.22\\
& & 3.50 & 0.00 & 0.00 & 0.00 & 0.00 & 0.00 & 0.06 & 0.04 & 0.05 & 0.05 & 0.05\\
& & 4.00 & 0.00 & 0.00 & 0.00 & 0.00 & 0.00 & 0.01 & 0.01 & 0.01 & 0.01 & 0.01\\
\hline\hline
\multirow{7}{*}{\textbf{Square root}} & \multirow{7}{*}{$z$}
& $\overline{\Gamma}_*$ & 2.45 & 2.38 & 2.45 & 2.45 & 2.45 & 3.67 & 3.50 & 3.67 & 3.67 & 3.67\\
\cline{3-13}
& & 1.50 & 1.00 & 1.00 & 1.00 & 1.00 & 0.99 & 1.00 & 1.00 & 1.00 & 1.00 & 1.00\\
& & 2.00 & 0.39 & 0.32 & 0.38 & 0.31 & 0.27 & 0.96 & 0.77 & 0.95 & 0.87 & 0.85\\
& & 2.50 & 0.00 & 0.00 & 0.00 & 0.00 & 0.00 & 0.51 & 0.22 & 0.46 & 0.40 & 0.35\\
& & 3.00 & 0.00 & 0.00 & 0.00 & 0.00 & 0.00 & 0.11 & 0.02 & 0.07 & 0.07 & 0.07\\
& & 3.50 & 0.00 & 0.00 & 0.00 & 0.00 & 0.00 & 0.01 & 0.00 & 0.01 & 0.01 & 0.01\\
& & 4.00 & 0.00 & 0.00 & 0.00 & 0.00 & 0.00 & 0.00 & 0.00 & 0.00 & 0.00 & 0.00\\
\hline\hline
\multirow{7}{*}{\textbf{Flat}} & \multirow{7}{*}{$z$}
& $\overline{\Gamma}_*$ & 2.60 & 2.45 & 2.60 & 2.60 & 2.60 & 4.02 & 3.37 & 4.02 & 4.02 & 4.02\\
\cline{3-13}
& & 1.50 & 1.00 & 1.00 & 1.00 & 1.00 & 1.00 & 1.00 & 1.00 & 1.00 & 1.00 & 1.00\\
& & 2.00 & 0.62 & 0.45 & 0.57 & 0.48 & 0.44 & 0.99 & 0.72 & 0.98 & 0.92 & 0.91\\
& & 2.50 & 0.01 & 0.00 & 0.01 & 0.01 & 0.01 & 0.73 & 0.18 & 0.64 & 0.57 & 0.54\\
& & 3.00 & 0.00 & 0.00 & 0.00 & 0.00 & 0.00 & 0.25 & 0.02 & 0.18 & 0.18 & 0.17\\
& & 3.50 & 0.00 & 0.00 & 0.00 & 0.00 & 0.00 & 0.05 & 0.00 & 0.02 & 0.03 & 0.03\\
& & 4.00 & 0.00 & 0.00 & 0.00 & 0.00 & 0.00 & 0.01 & 0.00 & 0.01 & 0.00 & 0.01\\
\hline\hline
\multirow{7}{*}{\textbf{Log}} & \multirow{7}{*}{$z$}
& $\overline{\Gamma}_*$ & 2.83 & 2.63 & 2.83 & 2.83 & 2.83 & 4.65 & 3.84 & 4.65 & 4.65 & 4.65\\
\cline{3-13}
& & 1.50 & 1.00 & 1.00 & 1.00 & 1.00 & 1.00 & 1.00 & 1.00 & 1.00 & 1.00 & 1.00\\
& & 2.00 & 0.88 & 0.76 & 0.86 & 0.78 & 0.74 & 1.00 & 0.90 & 1.00 & 0.98 & 0.97\\
& & 2.50 & 0.10 & 0.02 & 0.08 & 0.06 & 0.06 & 0.91 & 0.38 & 0.88 & 0.80 & 0.79\\
& & 3.00 & 0.00 & 0.00 & 0.00 & 0.00 & 0.00 & 0.55 & 0.08 & 0.45 & 0.43 & 0.38\\
& & 3.50 & 0.00 & 0.00 & 0.00 & 0.00 & 0.00 & 0.19 & 0.01 & 0.13 & 0.14 & 0.14\\
& & 4.00 & 0.00 & 0.00 & 0.00 & 0.00 & 0.00 & 0.04 & 0.00 & 0.02 & 0.04 & 0.03\\
\hline
\end{tabular}
}
\label{tb:ds_power_split_merged_1000}
\end{table}

\begin{table}[!htbp]
\caption{\textbf{(Simulations under $I=5000$)} The generalized design sensitivities $\overline{\Gamma}_{*}$ computed from Theorem~3.4 for the Wilcox, D-Wilcox, U, and D-U tests, and from Theorem~4.2 for the adaptive test combining Wilcox and D-Wilcox and the adaptive test combining U and D-U, together with the sample splitting procedure of \citet{heller2009split} using $10\%$ and $20\%$ of the data as the planning sample, whose generalized design sensitivity equals the larger of the two component design sensitivities. The sensitivity model uses the dose link function $\phi(z) = z$, and the test statistics are constructed from the untransformed doses, i.e., the dose transformation of Appendix~C.3 is the identity, $\kappa(z) = z$. Below the row of $\overline{\Gamma}_{*}$, we report the simulated power under $I = 5000$ and various values of $\overline{\Gamma}$ at $\alpha = 0.01$.}
\centering
\footnotesize
\setlength{\tabcolsep}{3pt}
\resizebox{\textwidth}{!}{
\begin{tabular}{l | c | c | c c c c c | c c c c c}
\hline
\textbf{Dose--response} & $\kappa(z)$ & $\overline{\Gamma}$ & Wilcox & D-Wilcox & Adaptive & Split (0.10) & Split (0.20) & U & D-U & Adaptive & Split (0.10) & Split (0.20) \\
\hline\hline
\multirow{7}{*}{\textbf{Square}} & \multirow{7}{*}{$z$}
& $\overline{\Gamma}_*$ & 2.32 & 2.44 & 2.44 & 2.44 & 2.44 & 3.32 & 4.16 & 4.16 & 4.16 & 4.16\\
\cline{3-13}
& & 1.50 & 1.00 & 1.00 & 1.00 & 1.00 & 1.00 & 1.00 & 1.00 & 1.00 & 1.00 & 1.00\\
& & 2.00 & 0.86 & 1.00 & 0.99 & 0.97 & 0.96 & 1.00 & 1.00 & 1.00 & 1.00 & 1.00\\
& & 2.50 & 0.00 & 0.00 & 0.00 & 0.00 & 0.00 & 0.97 & 1.00 & 1.00 & 0.97 & 0.97\\
& & 3.00 & 0.00 & 0.00 & 0.00 & 0.00 & 0.00 & 0.14 & 0.91 & 0.85 & 0.56 & 0.65\\
& & 3.50 & 0.00 & 0.00 & 0.00 & 0.00 & 0.00 & 0.00 & 0.23 & 0.15 & 0.11 & 0.14\\
& & 4.00 & 0.00 & 0.00 & 0.00 & 0.00 & 0.00 & 0.00 & 0.01 & 0.00 & 0.00 & 0.00\\
\hline\hline
\multirow{7}{*}{\textbf{Kink}} & \multirow{7}{*}{$z$}
& $\overline{\Gamma}_*$ & 2.26 & 2.43 & 2.43 & 2.43 & 2.43 & 3.17 & 4.28 & 4.28 & 4.28 & 4.28\\
\cline{3-13}
& & 1.50 & 1.00 & 1.00 & 1.00 & 1.00 & 1.00 & 1.00 & 1.00 & 1.00 & 1.00 & 1.00\\
& & 2.00 & 0.66 & 1.00 & 0.99 & 0.95 & 0.97 & 1.00 & 1.00 & 1.00 & 1.00 & 1.00\\
& & 2.50 & 0.00 & 0.00 & 0.00 & 0.00 & 0.00 & 0.87 & 1.00 & 1.00 & 0.95 & 0.97\\
& & 3.00 & 0.00 & 0.00 & 0.00 & 0.00 & 0.00 & 0.05 & 0.95 & 0.92 & 0.70 & 0.78\\
& & 3.50 & 0.00 & 0.00 & 0.00 & 0.00 & 0.00 & 0.00 & 0.36 & 0.26 & 0.22 & 0.23\\
& & 4.00 & 0.00 & 0.00 & 0.00 & 0.00 & 0.00 & 0.00 & 0.02 & 0.01 & 0.01 & 0.01\\
\hline\hline
\multirow{7}{*}{\textbf{Linear}} & \multirow{7}{*}{$z$}
& $\overline{\Gamma}_*$ & 2.68 & 2.68 & 2.68 & 2.68 & 2.68 & 4.16 & 4.43 & 4.43 & 4.43 & 4.43\\
\cline{3-13}
& & 1.50 & 1.00 & 1.00 & 1.00 & 1.00 & 1.00 & 1.00 & 1.00 & 1.00 & 1.00 & 1.00\\
& & 2.00 & 1.00 & 1.00 & 1.00 & 1.00 & 1.00 & 1.00 & 1.00 & 1.00 & 1.00 & 1.00\\
& & 2.50 & 0.17 & 0.20 & 0.20 & 0.17 & 0.15 & 1.00 & 1.00 & 1.00 & 1.00 & 1.00\\
& & 3.00 & 0.00 & 0.00 & 0.00 & 0.00 & 0.00 & 0.98 & 0.97 & 0.99 & 0.97 & 0.93\\
& & 3.50 & 0.00 & 0.00 & 0.00 & 0.00 & 0.00 & 0.40 & 0.47 & 0.49 & 0.39 & 0.33\\
& & 4.00 & 0.00 & 0.00 & 0.00 & 0.00 & 0.00 & 0.03 & 0.06 & 0.04 & 0.02 & 0.03\\
\hline\hline
\multirow{7}{*}{\textbf{Square root}} & \multirow{7}{*}{$z$}
& $\overline{\Gamma}_*$ & 2.45 & 2.38 & 2.45 & 2.45 & 2.45 & 3.67 & 3.50 & 3.67 & 3.67 & 3.67\\
\cline{3-13}
& & 1.50 & 1.00 & 1.00 & 1.00 & 1.00 & 1.00 & 1.00 & 1.00 & 1.00 & 1.00 & 1.00\\
& & 2.00 & 0.99 & 0.98 & 0.99 & 0.98 & 0.96 & 1.00 & 1.00 & 1.00 & 1.00 & 1.00\\
& & 2.50 & 0.00 & 0.00 & 0.00 & 0.00 & 0.00 & 1.00 & 0.94 & 1.00 & 0.99 & 0.98\\
& & 3.00 & 0.00 & 0.00 & 0.00 & 0.00 & 0.00 & 0.59 & 0.20 & 0.51 & 0.46 & 0.44\\
& & 3.50 & 0.00 & 0.00 & 0.00 & 0.00 & 0.00 & 0.03 & 0.00 & 0.02 & 0.02 & 0.03\\
& & 4.00 & 0.00 & 0.00 & 0.00 & 0.00 & 0.00 & 0.00 & 0.00 & 0.00 & 0.00 & 0.00\\
\hline\hline
\multirow{7}{*}{\textbf{Flat}} & \multirow{7}{*}{$z$}
& $\overline{\Gamma}_*$ & 2.60 & 2.45 & 2.60 & 2.60 & 2.60 & 4.02 & 3.37 & 4.02 & 4.02 & 4.02\\
\cline{3-13}
& & 1.50 & 1.00 & 1.00 & 1.00 & 1.00 & 1.00 & 1.00 & 1.00 & 1.00 & 1.00 & 1.00\\
& & 2.00 & 1.00 & 1.00 & 1.00 & 0.99 & 0.99 & 1.00 & 1.00 & 1.00 & 1.00 & 1.00\\
& & 2.50 & 0.06 & 0.00 & 0.04 & 0.03 & 0.04 & 1.00 & 0.88 & 1.00 & 0.99 & 0.99\\
& & 3.00 & 0.00 & 0.00 & 0.00 & 0.00 & 0.00 & 0.93 & 0.12 & 0.88 & 0.84 & 0.81\\
& & 3.50 & 0.00 & 0.00 & 0.00 & 0.00 & 0.00 & 0.24 & 0.00 & 0.19 & 0.20 & 0.17\\
& & 4.00 & 0.00 & 0.00 & 0.00 & 0.00 & 0.00 & 0.01 & 0.00 & 0.01 & 0.01 & 0.01\\
\hline\hline
\multirow{7}{*}{\textbf{Log}} & \multirow{7}{*}{$z$}
& $\overline{\Gamma}_*$ & 2.83 & 2.63 & 2.83 & 2.83 & 2.83 & 4.65 & 3.84 & 4.65 & 4.65 & 4.65\\
\cline{3-13}
& & 1.50 & 1.00 & 1.00 & 1.00 & 1.00 & 1.00 & 1.00 & 1.00 & 1.00 & 1.00 & 1.00\\
& & 2.00 & 1.00 & 1.00 & 1.00 & 1.00 & 1.00 & 1.00 & 1.00 & 1.00 & 1.00 & 1.00\\
& & 2.50 & 0.58 & 0.10 & 0.52 & 0.42 & 0.42 & 1.00 & 1.00 & 1.00 & 1.00 & 1.00\\
& & 3.00 & 0.00 & 0.00 & 0.00 & 0.00 & 0.00 & 1.00 & 0.62 & 1.00 & 0.98 & 0.98\\
& & 3.50 & 0.00 & 0.00 & 0.00 & 0.00 & 0.00 & 0.85 & 0.05 & 0.78 & 0.75 & 0.70\\
& & 4.00 & 0.00 & 0.00 & 0.00 & 0.00 & 0.00 & 0.26 & 0.00 & 0.19 & 0.21 & 0.18\\
\hline
\end{tabular}
}
\label{tb:ds_power_split_merged_5000}
\end{table}

\clearpage
\newpage

\subsection*{C.3: Generalized Design Sensitivity and Bahadur-Rosenbaum Efficiency Under Dose Transformations}

In the main text, we derived formulas for the generalized design sensitivity and Bahadur-Rosenbaum efficiency and used them to evaluate candidate test statistics, such as the original and dose-weighted versions of rank-based test statistics and U-statistics, in terms of their robustness to unmeasured confounding in sensitivity analyses. A key message from the corresponding theoretical and numerical results is that the dose-response relationship can substantially affect the performance of different test statistics in sensitivity analyses. This raises a natural and important question: beyond appropriately selecting the test statistic, can the power of sensitivity analysis be further improved by applying dose transformations to the observed treatment doses before downstream testing?

To answer this question, we use the proposed generalized design sensitivity and Bahadur-Rosenbaum efficiency to examine the effect of applying a dose transformation $\kappa(z)$ during the test-design phase. Specifically, when constructing the dose-dependent test scores, we replace the observed doses ${Z_{11},\ldots,Z_{I2}}$ with their transformed values ${\kappa(Z_{11}),\ldots,\kappa(Z_{I2})}$. The corresponding formulas can be obtained from Theorems~3.4 and~3.5 by replacing $Z_{ij}$ with $\kappa(Z_{ij})$ only where it enters the test score. The sensitivity model, including the pair-specific sensitivity bounds determined by $|\phi(Z_{i1})-\phi(Z_{i2})|$, remains unchanged.

We first confirm that both formulas remain valid after the dose is transformed. Table~\ref{tb:design_sensitivity_and_simulated_power_with_transformation} reports the generalized design sensitivity and simulated power of sensitivity analysis for the four key test statistics considered in the main text: the Wilcoxon signed rank test (Wilcox), the dose-weighted Wilcoxon signed rank test (D-Wilcox), the U-statistic (U) with $(m,\underline{m},\overline{m}) = (8,7,8)$, and the dose-weighted U-statistic (D-U) with $(m,\underline{m},\overline{m}) = (8,7,8)$, under different combinations of dose-response relationships (the square and log dose-response curves defined in Section 3.4) and dose transformations $\kappa(z)=z^{2}$, $z$ (the original dose), and $\log(z)$. Table~\ref{tb:design_sensitivity_and_simulated_power_with_transformation} shows that the simulated power of the dose-transformed test statistics remains high for $\overline{\Gamma}<\overline{\Gamma}^{*}$ and drops to zero for $\overline{\Gamma}>\overline{\Gamma}^{*}$, matching the generalized design sensitivity formula. In addition, Table~\ref{tb:bahadur_transformation_verification} shows that the generalized Bahadur-Rosenbaum relative efficiency derived from the formula generally aligns with the simulated ratio of the minimum sample sizes required by the tests to achieve a common $95\%$ power level ($\alpha=0.01$).

We then use these formulas to study how the dose transformation $\kappa$ affects a test's performance in sensitivity analyses and which transformation is most favorable. The first observation, evident from substituting the transformed treatment doses $\kappa(Z_{ij})$ into the formulas in Theorems 3.4 and 3.5, is that $\kappa$ affects only the performance of the dose-weighted tests (D-Wilcox and D-U): the transformation enters solely through the dose-rank term of the score and is therefore absent from tests that do not use the dose (Wilcox and U). Tables~\ref{tb:design_sensitivity_kappa_transformations} and~\ref{tb:bahadur_slope_kappa_transformations1}--\ref{tb:bahadur_slope_kappa_transformations3} confirm this pattern: the Wilcox and U entries are identical across all $\kappa$, while the D-Wilcox and D-U entries vary substantially.

For maximizing design sensitivity, the best dose transformation applied before constructing a dose-weighted test is one that accentuates the curvature of the dose-response curve, thereby concentrating the largest dose weights on pairs in its steepest region. The transformation need not match the dose-response curve's functional form, although following the curve's trend is usually a sound choice: in Table~\ref{tb:design_sensitivity_kappa_transformations}, the design sensitivity of the dose-weighted U-statistic (D-U) is largest under the trend-matching dose transformation for the square and log dose-response curves, while for the square root curve, the more strongly concave transformation $\kappa(z)=\log(z)$ outperforms the trend-matching dose transformation $\kappa(z)=\sqrt{z}$.

The same principle governs the generalized Bahadur-Rosenbaum efficiency. Tables~\ref{tb:bahadur_slope_kappa_transformations1}--\ref{tb:bahadur_slope_kappa_transformations3} report the generalized Bahadur-Rosenbaum exact slope $\Upsilon$ of each test under each dose-response curve and each dose-transformation $\kappa$ at $\overline{\Gamma}=1.00, 1.30,$ and $1.50$, where a higher slope indicates a more efficient test that requires fewer matched pairs to attain a desired power at a given $\overline{\Gamma}$ (i.e., Corollary~3.6 in the main text). For a given dose-response curve, the dose transformation that maximizes the slope likewise accentuates the dose-response curve's curvature, although the precise optimum may shift with the sensitivity parameter $\overline{\Gamma}$ at which the analysis is conducted. For the square dose-response curve, for instance, the trend-matching transformation $\kappa(z)=z^2$ maximizes the slope of the dose-weighted U-statistic (D-U) at all three values of $\overline{\Gamma}$ (e.g., at $\overline{\Gamma}=1.30$, $\Upsilon=0.0324$, versus $0.0266$, $0.0085$, and $0.0159$ for $\kappa(z)=z$, $\log(z)$, and $\sqrt{z}$), coinciding with its design-sensitivity-optimal dose transformation. For the square root curve, by contrast, the slope-maximizing transformation of the dose-weighted U-statistic (D-U) is the trend-matching transformation $\kappa(z)=\sqrt{z}$ at $\overline{\Gamma} = 1.0$ and $1.30$ but $\kappa(z)=\log(z)$ at $\overline{\Gamma} = 1.50$, the same as the design-sensitivity-optimal choice.

\begin{table}[!htbp]
\caption{The generalized design sensitivities $\overline{\Gamma}_{*}$ computed from Theorem~3.4 for the Wilcox, D-Wilcox, U, and D-U tests, and from Theorem~4.2 for the adaptive test combining Wilcox and D-Wilcox and the adaptive test combining U and D-U, under the square and log dose-response curves defined in Section~3.4 of the main text and three dose transformations, $\kappa(z) = z^{2}$, $z$, and $\log(z)$. The dose transformation $\kappa$ is applied in the test-design phase only, acting on the doses that enter the dose-rank term of the test score; the sensitivity model is unaffected by $\kappa$ and uses the dose link function $\phi(z) = z$ throughout, so that the pair-specific sensitivity bounds $\Gamma_{i}$ are identical across the three blocks. Accordingly, the Wilcox and U tests, which do not use the treatment doses, are invariant to $\kappa$, and the blocks with $\kappa(z) = z$ coincide with the corresponding entries of Table~S.6 in Appendix~C.2. Below the row of $\overline{\Gamma}_{*}$, we report the simulated power under $I = 5000$ and various values of $\overline{\Gamma}$ at $\alpha = 0.01$.}
\centering
\footnotesize
\setlength{\tabcolsep}{3pt}
\begin{tabular}{l | c | c | c c c | c c c}
\hline
\textbf{Dose--response} & $\kappa(z)$ & $\overline{\Gamma}$ & Wilcox & D-Wilcox & Adaptive & U & D-U & Adaptive \\
\hline\hline
\multirow{21}{*}{\textbf{Square}} & \multirow{7}{*}{$z^2$}
& $\overline{\Gamma}_*$ & 2.32 & 2.57 & 2.57 & 3.32 & 5.30 & 5.30\\
\cline{3-9}
& & 1.50 & 1.00 & 1.00 & 1.00 & 1.00 & 1.00 & 1.00\\
& & 2.00 & 0.86 & 1.00 & 1.00 & 1.00 & 1.00 & 1.00\\
& & 2.50 & 0.00 & 0.03 & 0.02 & 0.97 & 1.00 & 1.00\\
& & 3.00 & 0.00 & 0.00 & 0.00 & 0.14 & 1.00 & 1.00\\
& & 3.50 & 0.00 & 0.00 & 0.00 & 0.00 & 0.95 & 0.91\\
& & 4.00 & 0.00 & 0.00 & 0.00 & 0.00 & 0.53 & 0.41\\
\cline{2-9}
& \multirow{7}{*}{$z$}
& $\overline{\Gamma}_*$ & 2.32 & 2.44 & 2.44 & 3.32 & 4.16 & 4.16\\
\cline{3-9}
& & 1.50 & 1.00 & 1.00 & 1.00 & 1.00 & 1.00 & 1.00\\
& & 2.00 & 0.86 & 1.00 & 0.99 & 1.00 & 1.00 & 1.00\\
& & 2.50 & 0.00 & 0.00 & 0.00 & 0.97 & 1.00 & 1.00\\
& & 3.00 & 0.00 & 0.00 & 0.00 & 0.14 & 0.91 & 0.85\\
& & 3.50 & 0.00 & 0.00 & 0.00 & 0.00 & 0.23 & 0.15\\
& & 4.00 & 0.00 & 0.00 & 0.00 & 0.00 & 0.01 & 0.00\\
\cline{2-9}
& \multirow{7}{*}{$\log(z)$}
& $\overline{\Gamma}_*$ & 2.32 & 2.19 & 2.32 & 3.32 & 2.61 & 3.32\\
\cline{3-9}
& & 1.50 & 1.00 & 1.00 & 1.00 & 1.00 & 1.00 & 1.00\\
& & 2.00 & 0.86 & 0.41 & 0.82 & 1.00 & 0.81 & 1.00\\
& & 2.50 & 0.00 & 0.00 & 0.00 & 0.97 & 0.03 & 0.95\\
& & 3.00 & 0.00 & 0.00 & 0.00 & 0.14 & 0.00 & 0.10\\
& & 3.50 & 0.00 & 0.00 & 0.00 & 0.00 & 0.00 & 0.00\\
& & 4.00 & 0.00 & 0.00 & 0.00 & 0.00 & 0.00 & 0.00\\
\hline\hline
\multirow{21}{*}{\textbf{Log}} & \multirow{7}{*}{$z^2$}
& $\overline{\Gamma}_*$ & 2.83 & 2.42 & 2.83 & 4.65 & 2.94 & 4.65\\
\cline{3-9}
& & 1.50 & 1.00 & 1.00 & 1.00 & 1.00 & 1.00 & 1.00\\
& & 2.00 & 1.00 & 0.99 & 1.00 & 1.00 & 0.99 & 1.00\\
& & 2.50 & 0.58 & 0.00 & 0.52 & 1.00 & 0.28 & 1.00\\
& & 3.00 & 0.00 & 0.00 & 0.00 & 1.00 & 0.01 & 1.00\\
& & 3.50 & 0.00 & 0.00 & 0.00 & 0.85 & 0.00 & 0.78\\
& & 4.00 & 0.00 & 0.00 & 0.00 & 0.26 & 0.00 & 0.19\\
\cline{2-9}
& \multirow{7}{*}{$z$}
& $\overline{\Gamma}_*$ & 2.83 & 2.63 & 2.83 & 4.65 & 3.84 & 4.65\\
\cline{3-9}
& & 1.50 & 1.00 & 1.00 & 1.00 & 1.00 & 1.00 & 1.00\\
& & 2.00 & 1.00 & 1.00 & 1.00 & 1.00 & 1.00 & 1.00\\
& & 2.50 & 0.58 & 0.10 & 0.52 & 1.00 & 1.00 & 1.00\\
& & 3.00 & 0.00 & 0.00 & 0.00 & 1.00 & 0.62 & 1.00\\
& & 3.50 & 0.00 & 0.00 & 0.00 & 0.85 & 0.05 & 0.78\\
& & 4.00 & 0.00 & 0.00 & 0.00 & 0.26 & 0.00 & 0.19\\
\cline{2-9}
& \multirow{7}{*}{$\log(z)$}
& $\overline{\Gamma}_*$ & 2.83 & 3.19 & 3.19 & 4.65 & 9.70 & 9.70\\
\cline{3-9}
& & 1.50 & 1.00 & 1.00 & 1.00 & 1.00 & 1.00 & 1.00\\
& & 2.00 & 1.00 & 1.00 & 1.00 & 1.00 & 1.00 & 1.00\\
& & 2.50 & 0.58 & 1.00 & 1.00 & 1.00 & 1.00 & 1.00\\
& & 3.00 & 0.00 & 0.11 & 0.08 & 1.00 & 1.00 & 1.00\\
& & 3.50 & 0.00 & 0.00 & 0.00 & 0.85 & 1.00 & 1.00\\
& & 4.00 & 0.00 & 0.00 & 0.00 & 0.26 & 1.00 & 1.00\\
\hline
\end{tabular}
\label{tb:design_sensitivity_and_simulated_power_with_transformation}
\end{table}

\begin{table}[!htbp]
\caption{The generalized design sensitivities $\overline{\Gamma}_{*}$ computed from Theorem~3.4 for the Wilcox, D-Wilcox, U, and D-U tests, and from Theorem~4.2 for the adaptive test combining Wilcox and D-Wilcox and the adaptive test combining U and D-U, across the six dose-response curves defined in Section~3.4 of the main text and four dose transformations, $\kappa(z) = z$, $\log(z)$, $z^{2}$, and $\sqrt{z}$. The dose transformation $\kappa$ is applied in the test-design phase only, acting on the doses that enter the dose-rank term of the test score; the sensitivity model is unaffected by $\kappa$ and uses the dose link function $\phi(z) = z$ throughout. Accordingly, the Wilcox and U entries are identical across all four choices of $\kappa$, while the D-Wilcox and D-U entries vary, and the rows with $\kappa(z) = z$ coincide with the $\overline{\Gamma}_{*}$ rows of Tables~S.3--S.6 in Appendix~C.2.}
\centering
\footnotesize
\setlength{\tabcolsep}{3pt}
\begin{tabular}{l | l | c c c | c c c}
\hline
\textbf{Dose-response} & $\kappa(z)$ & Wilcox & D-Wilcox & Adaptive & U & D-U & Adaptive \\
\hline\hline
\multirow{4}{*}{\textbf{Square}}
& $z$        &  2.32&  2.44&  2.44&  3.32&  4.16&  4.16\\
& $\log(z)$  &  2.32&  2.19&  2.32&  3.32&  2.61&  3.32\\
& $z^2$      &  2.32&  2.57&  2.57&  3.32&  5.30&  5.30\\
& $\sqrt{z}$ &  2.32&  2.30&  2.32&  3.32&  3.12&  3.32\\
\hline\hline
\multirow{4}{*}{\textbf{Kink}}
& $z$        &  2.26&  2.43&  2.43&  3.17&  4.28&  4.28\\
& $\log(z)$  &  2.26&  2.01&  2.26&  3.17&  2.03&  3.17\\
& $z^2$      &  2.26&  2.67&  2.67&  3.17&  6.62&  6.62\\
& $\sqrt{z}$ &  2.26&  2.19&  2.26&  3.17&  2.70&  3.17\\
\hline\hline
\multirow{4}{*}{\textbf{Linear}}
& $z$        &  2.68&  2.68&  2.68&  4.16&  4.43&  4.43\\
& $\log(z)$  &  2.68&  2.68&  2.68&  4.16&  4.37&  4.37\\
& $z^2$      &  2.68&  2.68&  2.68&  4.16&  4.45&  4.45\\
& $\sqrt{z}$ &  2.68&  2.68&  2.68&  4.16&  4.38&  4.38\\
\hline\hline
\multirow{4}{*}{\textbf{Square root}}
& $z$        &  2.45&  2.38&  2.45&  3.67&  3.50&  3.67\\
& $\log(z)$  &  2.45&  2.57&  2.57&  3.67&  4.87&  4.87\\
& $z^2$      &  2.45&  2.30&  2.45&  3.67&  3.11&  3.67\\
& $\sqrt{z}$ &  2.45&  2.48&  2.48&  3.67&  4.17&  4.17\\
\hline\hline
\multirow{4}{*}{\textbf{Flat}}
& $z$        &  2.60&  2.45&  2.60&  4.02&  3.37&  4.02\\
& $\log(z)$  &  2.60&  2.70&  2.70&  4.02&  5.38&  5.38\\
& $z^2$      &  2.60&  2.35&  2.60&  4.02&  2.72&  4.02\\
& $\sqrt{z}$ &  2.60&  2.58&  2.60&  4.02&  4.53&  4.53\\
\hline\hline
\multirow{4}{*}{\textbf{Log}}
& $z$        &  2.83&  2.63&  2.83&  4.65&  3.84&  4.65\\
& $\log(z)$  &  2.83&  3.19&  3.19&  4.65&  9.70&  9.70\\
& $z^2$      &  2.83&  2.42&  2.83&  4.65&  2.94&  4.65\\
& $\sqrt{z}$ &  2.83&  2.90&  2.90&  4.65&  6.02&  6.02\\
\hline
\end{tabular}
\label{tb:design_sensitivity_kappa_transformations}
\end{table}

\begin{table}[!htbp]
\caption{The generalized Bahadur-Rosenbaum relative efficiencies $\Upsilon_{\text{Wilcox}}/\Upsilon_{\text{D-Wilcox}}$ and $\Upsilon_{\text{U}}/\Upsilon_{\text{D-U}}$ computed from the transformed-dose version of Theorem~3.5, with the corresponding simulated ratios of the minimum numbers of matched pairs required to attain $95\%$ power at $\alpha = 0.01$, $I_{\text{D-Wilcox}}/I_{\text{Wilcox}}$ and $I_{\text{D-U}}/I_{\text{U}}$, reported in parentheses (i.e., Corollary~3.6). Results are reported under the square and log dose-response curves defined in Section~3.4 of the main text and three dose transformations, $\kappa(z) = z$, $\log(z)$, and $z^{2}$. The dose transformation $\kappa$ is applied in the test-design phase only, acting on the doses that enter the dose-rank term of the test score; the sensitivity model is unaffected by $\kappa$ and uses the dose link function $\phi(z) = z$ throughout. All entries satisfy $\overline{\Gamma} < \min(\overline{\Gamma}_{*,1}, \overline{\Gamma}_{*,2})$, so that both component tests remain consistent at the reported values of $\overline{\Gamma}$.}
\centering
\footnotesize
\setlength{\tabcolsep}{4pt}
\begin{tabular}{l | c | c | c c}
\hline
\textbf{Dose--response} & $\kappa(z)$ & $\overline{\Gamma}$ & (Wilcox, D-Wilcox) & (U, D-U) \\
\hline\hline
\multirow{12}{*}{\textbf{Square}} & \multirow{4}{*}{$z$}
& 1.00 & 0.77(0.77) & 1.03(1.01)\\
& & 1.10 & 0.76(0.78) & 1.01(1.01)\\
& & 1.30 & 0.73(0.73) & 0.98(0.97)\\
& & 1.50 & 0.70(0.72) & 0.93(0.92)\\
\cline{2-5}
& \multirow{4}{*}{$\log(z)$}
& 1.00 & 1.07(1.07) & 2.63(2.70)\\
& & 1.10 & 1.09(1.07) & 2.75(2.92)\\
& & 1.30 & 1.14(1.18) & 3.06(3.19)\\
& & 1.50 & 1.23(1.23) & 3.49(3.58)\\
\cline{2-5}
& \multirow{4}{*}{$z^{2}$}
& 1.00 & 0.73(0.73) & 0.90(0.86)\\
& & 1.10 & 0.71(0.71) & 0.87(0.90)\\
& & 1.30 & 0.67(0.65) & 0.80(0.81)\\
& & 1.50 & 0.61(0.62) & 0.73(0.73)\\
\hline\hline
\multirow{12}{*}{\textbf{Log}} & \multirow{4}{*}{$z$}
& 1.00 & 0.95(0.92) & 1.57(1.63)\\
& & 1.10 & 0.96(0.96) & 1.61(1.65)\\
& & 1.30 & 0.99(1.00) & 1.70(1.74)\\
& & 1.50 & 1.04(1.04) & 1.80(1.83)\\
\cline{2-5}
& \multirow{4}{*}{$\log(z)$}
& 1.00 & 0.83(0.78) & 1.09(1.13)\\
& & 1.10 & 0.81(0.81) & 1.06(1.07)\\
& & 1.30 & 0.78(0.77) & 0.99(0.97)\\
& & 1.50 & 0.73(0.73) & 0.93(0.93)\\
\cline{2-5}
& \multirow{4}{*}{$z^{2}$}
& 1.00 & 1.12(1.10) & 2.39(2.59)\\
& & 1.10 & 1.16(1.20) & 2.54(2.74)\\
& & 1.30 & 1.27(1.36) & 2.90(3.11)\\
& & 1.50 & 1.42(1.45) & 3.38(3.59)\\
\hline
\end{tabular}
\label{tb:bahadur_transformation_verification}
\end{table}

\begin{table}[!htbp]
\caption{The generalized Bahadur-Rosenbaum exact slopes $\Upsilon$ computed from Theorem~3.5 for the Wilcox, D-Wilcox, U, and D-U tests, and from Theorem~4.2 for the adaptive test combining Wilcox and D-Wilcox and the adaptive test combining U and D-U, across the six dose-response curves defined in Section~3.4 of the main text and four dose transformations, $\kappa(z) = z$, $\log(z)$, $z^{2}$, and $\sqrt{z}$, evaluated at $\overline{\Gamma} = 1.00$, $1.30$, and $1.50$. The dose transformation $\kappa$ is applied in the test-design phase only, acting on the doses that enter the dose-rank term of the test score; the sensitivity model is unaffected by $\kappa$ and uses the dose link function $\phi(z) = z$ throughout. Accordingly, the Wilcox and U entries are identical across all four choices of $\kappa$, while the D-Wilcox and D-U entries vary. Part 1 of 3: square and kink dose-response curves.}
\centering
\footnotesize
\setlength{\tabcolsep}{3pt}
\begin{tabular}{l | c | c | c c c | c c c}
\hline
\textbf{Dose--response} & $\kappa(z)$ & $\overline{\Gamma}$ & Wilcox & D-Wilcox & Adaptive & U & D-U & Adaptive \\
\hline\hline
\multirow{12}{*}{\textbf{Square}}
& \multirow{3}{*}{$z$}        & 1.00 & 0.0520 & 0.0672 & 0.0672 & 0.0474 & 0.0460 & 0.0474 \\
&                             & 1.30 & 0.0224 & 0.0305 & 0.0305 & 0.0260 & 0.0266 & 0.0266 \\
&                             & 1.50 & 0.0120 & 0.0173 & 0.0173 & 0.0175 & 0.0189 & 0.0189 \\
\cline{2-9}
& \multirow{3}{*}{$\log(z)$}  & 1.00 & 0.0520 & 0.0486 & 0.0520 & 0.0474 & 0.0180 & 0.0474 \\
&                             & 1.30 & 0.0224 & 0.0196 & 0.0224 & 0.0260 & 0.0085 & 0.0260 \\
&                             & 1.50 & 0.0120 & 0.0098 & 0.0120 & 0.0175 & 0.0050 & 0.0175 \\
\cline{2-9}
& \multirow{3}{*}{$z^{2}$}    & 1.00 & 0.0520 & 0.0707 & 0.0707 & 0.0474 & 0.0527 & 0.0527 \\
&                             & 1.30 & 0.0224 & 0.0335 & 0.0335 & 0.0260 & 0.0324 & 0.0324 \\
&                             & 1.50 & 0.0120 & 0.0198 & 0.0198 & 0.0175 & 0.0241 & 0.0241 \\
\cline{2-9}
& \multirow{3}{*}{$\sqrt{z}$} & 1.00 & 0.0520 & 0.0588 & 0.0588 & 0.0474 & 0.0305 & 0.0474 \\
&                             & 1.30 & 0.0224 & 0.0252 & 0.0252 & 0.0260 & 0.0159 & 0.0260 \\
&                             & 1.50 & 0.0120 & 0.0134 & 0.0134 & 0.0175 & 0.0103 & 0.0175 \\
\hline\hline
\multirow{12}{*}{\textbf{Kink}}
& \multirow{3}{*}{$z$}        & 1.00 & 0.0494 & 0.0667 & 0.0667 & 0.0452 & 0.0478 & 0.0478 \\
&                             & 1.30 & 0.0206 & 0.0302 & 0.0302 & 0.0243 & 0.0279 & 0.0279 \\
&                             & 1.50 & 0.0107 & 0.0170 & 0.0170 & 0.0161 & 0.0200 & 0.0200 \\
\cline{2-9}
& \multirow{3}{*}{$\log(z)$}  & 1.00 & 0.0494 & 0.0403 & 0.0494 & 0.0452 & 0.0107 & 0.0452 \\
&                             & 1.30 & 0.0206 & 0.0143 & 0.0206 & 0.0243 & 0.0038 & 0.0243 \\
&                             & 1.50 & 0.0107 & 0.0061 & 0.0107 & 0.0161 & 0.0017 & 0.0161 \\
\cline{2-9}
& \multirow{3}{*}{$z^{2}$}    & 1.00 & 0.0494 & 0.0760 & 0.0760 & 0.0452 & 0.0627 & 0.0627 \\
&                             & 1.30 & 0.0206 & 0.0373 & 0.0373 & 0.0243 & 0.0400 & 0.0400 \\
&                             & 1.50 & 0.0107 & 0.0228 & 0.0228 & 0.0161 & 0.0306 & 0.0306 \\
\cline{2-9}
& \multirow{3}{*}{$\sqrt{z}$} & 1.00 & 0.0494 & 0.0534 & 0.0534 & 0.0452 & 0.0250 & 0.0452 \\
&                             & 1.30 & 0.0206 & 0.0216 & 0.0216 & 0.0243 & 0.0120 & 0.0243 \\
&                             & 1.50 & 0.0107 & 0.0108 & 0.0108 & 0.0161 & 0.0072 & 0.0161 \\
\hline
\end{tabular}
\label{tb:bahadur_slope_kappa_transformations1}
\end{table}

\begin{table}[!htbp]
\caption{The generalized Bahadur-Rosenbaum exact slopes $\Upsilon$, continued from Table~\ref{tb:bahadur_slope_kappa_transformations1}; see that table for the definitions of the tests, the dose transformations $\kappa$, and the dose link function $\phi(z) = z$. Part 2 of 3: linear and square root dose-response curves.}
\centering
\footnotesize
\setlength{\tabcolsep}{3pt}
\begin{tabular}{l | c | c | c c c | c c c}
\hline
\textbf{Dose--response} & $\kappa(z)$ & $\overline{\Gamma}$ & Wilcox & D-Wilcox & Adaptive & U & D-U & Adaptive \\
\hline\hline
\multirow{12}{*}{\textbf{Linear}}
& \multirow{3}{*}{$z$}        & 1.00 & 0.0667 & 0.0790 & 0.0790 & 0.0600 & 0.0487 & 0.0600 \\
&                             & 1.30 & 0.0325 & 0.0387 & 0.0387 & 0.0357 & 0.0287 & 0.0357 \\
&                             & 1.50 & 0.0197 & 0.0236 & 0.0236 & 0.0257 & 0.0207 & 0.0257 \\
\cline{2-9}
& \multirow{3}{*}{$\log(z)$}  & 1.00 & 0.0667 & 0.0697 & 0.0697 & 0.0600 & 0.0352 & 0.0600 \\
&                             & 1.30 & 0.0325 & 0.0340 & 0.0340 & 0.0357 & 0.0208 & 0.0357 \\
&                             & 1.50 & 0.0197 & 0.0207 & 0.0207 & 0.0257 & 0.0150 & 0.0257 \\
\cline{2-9}
& \multirow{3}{*}{$z^{2}$}    & 1.00 & 0.0667 & 0.0762 & 0.0762 & 0.0600 & 0.0453 & 0.0600 \\
&                             & 1.30 & 0.0325 & 0.0373 & 0.0373 & 0.0357 & 0.0268 & 0.0357 \\
&                             & 1.50 & 0.0197 & 0.0227 & 0.0227 & 0.0257 & 0.0193 & 0.0257 \\
\cline{2-9}
& \multirow{3}{*}{$\sqrt{z}$} & 1.00 & 0.0667 & 0.0764 & 0.0764 & 0.0600 & 0.0441 & 0.0600 \\
&                             & 1.30 & 0.0325 & 0.0373 & 0.0373 & 0.0357 & 0.0259 & 0.0357 \\
&                             & 1.50 & 0.0197 & 0.0227 & 0.0227 & 0.0257 & 0.0186 & 0.0257 \\
\hline\hline
\multirow{12}{*}{\textbf{Square root}}
& \multirow{3}{*}{$z$}        & 1.00 & 0.0570 & 0.0642 & 0.0642 & 0.0516 & 0.0373 & 0.0516 \\
&                             & 1.30 & 0.0259 & 0.0285 & 0.0285 & 0.0294 & 0.0204 & 0.0294 \\
&                             & 1.50 & 0.0147 & 0.0158 & 0.0158 & 0.0206 & 0.0139 & 0.0206 \\
\cline{2-9}
& \multirow{3}{*}{$\log(z)$}  & 1.00 & 0.0570 & 0.0642 & 0.0642 & 0.0516 & 0.0380 & 0.0516 \\
&                             & 1.30 & 0.0259 & 0.0303 & 0.0303 & 0.0294 & 0.0230 & 0.0294 \\
&                             & 1.50 & 0.0147 & 0.0179 & 0.0179 & 0.0206 & 0.0170 & 0.0206 \\
\cline{2-9}
& \multirow{3}{*}{$z^{2}$}    & 1.00 & 0.0570 & 0.0581 & 0.0581 & 0.0516 & 0.0295 & 0.0516 \\
&                             & 1.30 & 0.0259 & 0.0249 & 0.0259 & 0.0294 & 0.0154 & 0.0294 \\
&                             & 1.50 & 0.0147 & 0.0133 & 0.0147 & 0.0206 & 0.0100 & 0.0206 \\
\cline{2-9}
& \multirow{3}{*}{$\sqrt{z}$} & 1.00 & 0.0570 & 0.0663 & 0.0663 & 0.0516 & 0.0407 & 0.0516 \\
&                             & 1.30 & 0.0259 & 0.0305 & 0.0305 & 0.0294 & 0.0236 & 0.0294 \\
&                             & 1.50 & 0.0147 & 0.0175 & 0.0175 & 0.0206 & 0.0168 & 0.0206 \\
\hline
\end{tabular}
\label{tb:bahadur_slope_kappa_transformations2}
\end{table}

\begin{table}[!htbp]
\caption{The generalized Bahadur-Rosenbaum exact slopes $\Upsilon$, continued from Table~\ref{tb:bahadur_slope_kappa_transformations1}; see that table for the definitions of the tests, the dose transformations $\kappa$, and the dose link function $\phi(z) = z$. Part 3 of 3: flat and log dose-response curves.}
\centering
\footnotesize
\setlength{\tabcolsep}{3pt}
\begin{tabular}{l | c | c | c c c | c c c}
\hline
\textbf{Dose--response} & $\kappa(z)$ & $\overline{\Gamma}$ & Wilcox & D-Wilcox & Adaptive & U & D-U & Adaptive \\
\hline\hline
\multirow{12}{*}{\textbf{Flat}}
& \multirow{3}{*}{$z$}        & 1.00 & 0.0630 & 0.0675 & 0.0675 & 0.0569 & 0.0357 & 0.0569 \\
&                             & 1.30 & 0.0300 & 0.0307 & 0.0307 & 0.0335 & 0.0193 & 0.0335 \\
&                             & 1.50 & 0.0179 & 0.0175 & 0.0179 & 0.0240 & 0.0130 & 0.0240 \\
\cline{2-9}
& \multirow{3}{*}{$\log(z)$}  & 1.00 & 0.0630 & 0.0695 & 0.0695 & 0.0569 & 0.0422 & 0.0569 \\
&                             & 1.30 & 0.0300 & 0.0340 & 0.0340 & 0.0335 & 0.0261 & 0.0335 \\
&                             & 1.50 & 0.0179 & 0.0208 & 0.0208 & 0.0240 & 0.0196 & 0.0240 \\
\cline{2-9}
& \multirow{3}{*}{$z^{2}$}    & 1.00 & 0.0630 & 0.0607 & 0.0630 & 0.0569 & 0.0245 & 0.0569 \\
&                             & 1.30 & 0.0300 & 0.0265 & 0.0300 & 0.0335 & 0.0118 & 0.0335 \\
&                             & 1.50 & 0.0179 & 0.0145 & 0.0179 & 0.0240 & 0.0072 & 0.0240 \\
\cline{2-9}
& \multirow{3}{*}{$\sqrt{z}$} & 1.00 & 0.0630 & 0.0708 & 0.0708 & 0.0569 & 0.0441 & 0.0569 \\
&                             & 1.30 & 0.0300 & 0.0336 & 0.0336 & 0.0335 & 0.0262 & 0.0335 \\
&                             & 1.50 & 0.0179 & 0.0199 & 0.0199 & 0.0240 & 0.0190 & 0.0240 \\
\hline\hline
\multirow{12}{*}{\textbf{Log}}
& \multirow{3}{*}{$z$}        & 1.00 & 0.0722 & 0.0763 & 0.0763 & 0.0654 & 0.0417 & 0.0654 \\
&                             & 1.30 & 0.0365 & 0.0368 & 0.0368 & 0.0400 & 0.0236 & 0.0400 \\
&                             & 1.50 & 0.0230 & 0.0221 & 0.0230 & 0.0296 & 0.0164 & 0.0296 \\
\cline{2-9}
& \multirow{3}{*}{$\log(z)$}  & 1.00 & 0.0722 & 0.0873 & 0.0873 & 0.0654 & 0.0602 & 0.0654 \\
&                             & 1.30 & 0.0365 & 0.0470 & 0.0470 & 0.0400 & 0.0403 & 0.0403 \\
&                             & 1.50 & 0.0230 & 0.0313 & 0.0313 & 0.0296 & 0.0319 & 0.0319 \\
\cline{2-9}
& \multirow{3}{*}{$z^{2}$}    & 1.00 & 0.0722 & 0.0642 & 0.0722 & 0.0654 & 0.0273 & 0.0654 \\
&                             & 1.30 & 0.0365 & 0.0289 & 0.0365 & 0.0400 & 0.0138 & 0.0400 \\
&                             & 1.50 & 0.0230 & 0.0162 & 0.0230 & 0.0296 & 0.0088 & 0.0296 \\
\cline{2-9}
& \multirow{3}{*}{$\sqrt{z}$} & 1.00 & 0.0722 & 0.0845 & 0.0845 & 0.0654 & 0.0554 & 0.0654 \\
&                             & 1.30 & 0.0365 & 0.0434 & 0.0434 & 0.0400 & 0.0347 & 0.0400 \\
&                             & 1.50 & 0.0230 & 0.0277 & 0.0277 & 0.0296 & 0.0263 & 0.0296 \\
\hline
\end{tabular}
\label{tb:bahadur_slope_kappa_transformations3}
\end{table}

\clearpage

\newpage
\subsection*{C.4: Additional Remarks on the Simulation Studies and Real Data Analysis}

\begin{remark}
    In practice, dose-response curves can take different shapes, becoming either flatter or steeper as the treatment or exposure dose increases; both patterns arise in real-world studies. As an example of a flattening curve, \citet{Lanphear2017LowLevelToxicity} reports that, for toxic exposures such as lead and benzene, dose-response curves for outcomes such as cognitive impairment and mortality rise steeply at low exposure levels and flatten at higher levels. As an example of a steepening curve, in chemical carcinogenesis, the superposition of multiple mechanisms of action can cause cancer risk to increase more than proportionally with exposure \citep{Lutz1998DoseResponse}. Because the shape of the dose-response curve varies across applications, it is important to understand how the robustness of each test depends on this shape.
\end{remark}

\begin{remark}
    In Table~\ref{tab:matching_quality}, we report the detailed post-matching covariate balance in our real data application. 

\begin{table}[ht]
\caption{The post-matching balance table of measured confounders between the paired high and low dose individuals, in which we report the post-matching means of each measured confounder among the paired high and low dose groups, as well as the corresponding standardized differences in means. }
\centering
\small
\begin{tabular}{l r r r}
\toprule
Measured Confounders &
\multicolumn{1}{c}{Low Dose} &
\multicolumn{1}{c}{High Dose} &
\multicolumn{1}{c}{Standardized Difference} \\
\midrule
Age                     & 56.49  & 56.24& 0.02  \\
White (Proportion)               & 0.35   & 0.39   & -0.07 \\
Income to Poverty Ratio & 2.68   & 2.69   & -0.01 \\
Some College Degree          & 0.57   & 0.51   & 0.13  \\
Body Mass Index (BMI)                   & 29.05  & 28.72  & 0.06  \\
Waist Circumference     & 102.90 & 102.85 & 0.00  \\
Asthma                  & 0.10   & 0.12   & -0.07 \\
\bottomrule
\end{tabular}
\label{tab:matching_quality}
\end{table}
% Reduce the space between columns globally.
\setlength{\tabcolsep}{4pt} % Default is usually 6pt; adjust as needed.

\end{remark}

\section*{Appendix D: Some Additional Remarks}

\subsection*{D.1: Some Remarks on the Nuisance Functions in the Rosenbaum Sensitivity Model}

In the Rosenbaum sensitivity model for a general treatment $Z$ \citep{Rosenbaum1987, rosenbaum1989sensitivity, rosenbaum2002observational, gastwirth1998dual}, i.e.,
    $$P(Z=z \mid \mathbf{x}, u)=\xi(\mathbf{x}, u) \eta (z,\mathbf{x}) \exp\{\gamma \phi(z) u \},$$
    $\xi$ and $\eta$ are unspecified nuisance functions, while $\phi$ is a prespecified dose link function. The nuisance functions $\xi$ and $\eta$ do not need to be modeled or estimated for the sensitivity analysis. Conditional on the matched pair, the two observed dose values, and the measured confounders used in matching, these nuisance functions cancel from the post-matching dose assignment probability. Specifically, when $\mathbf{x}_{i1}=\mathbf{x}_{i2}$,
\begin{align*}
    &\quad \ P(Z_{i1}=z_{i}^{**}, Z_{i2}=z^{*}_{i} \mid \mathcal{F}, \mathcal{Z}) \\
   &=P(Z_{i1}=z_{i}^{**}, Z_{i2}=z^{*}_{i}\mid \mathbf{x}_{i1}, u_{i1}, \mathbf{x}_{i2}, u_{i2}, Z_{i1}\wedge Z_{i2}=z_{i}^{*}, Z_{i1}\vee Z_{i2}=z_{i}^{**})\\
   &= \frac{P(Z_{i1}=z_{i}^{**}, Z_{i2}=z_{i}^{*} \mid \mathbf{x}_{i1}, u_{i1}, \mathbf{x}_{i2}, u_{i2})}{P(Z_{i1}=z_{i}^{**}, Z_{i2}=z_{i}^{*} \mid \mathbf{x}_{i1}, u_{i1}, \mathbf{x}_{i2}, u_{i2})+P(Z_{i1}=z_{i}^{*}, Z_{i2}=z_{i}^{**} \mid \mathbf{x}_{i1}, u_{i1}, \mathbf{x}_{i2}, u_{i2})} \\
   &= \frac{P(Z_{i1}=z_{i}^{**}\mid \mathbf{x}_{i1}, u_{i1})P(Z_{i2}=z_{i}^{*}\mid \mathbf{x}_{i2}, u_{i2})}{P(Z_{i1}=z_{i}^{**}\mid \mathbf{x}_{i1}, u_{i1})P(Z_{i2}=z_{i}^{*}\mid \mathbf{x}_{i2}, u_{i2})+P(Z_{i1}=z_{i}^{*}\mid \mathbf{x}_{i1}, u_{i1})P(Z_{i2}=z_{i}^{**}\mid \mathbf{x}_{i2}, u_{i2})} \\
   &=\frac{\frac{P(Z_{i1}=z_{i}^{**}\mid \mathbf{x}_{i1}, u_{i1})P(Z_{i2}=z_{i}^{*}\mid \mathbf{x}_{i2}, u_{i2})}{P(Z_{i1}=z_{i}^{*}\mid \mathbf{x}_{i1}, u_{i1})P(Z_{i2}=z_{i}^{**}\mid \mathbf{x}_{i2}, u_{i2})}}{\frac{P(Z_{i1}=z_{i}^{**}\mid \mathbf{x}_{i1}, u_{i1})P(Z_{i2}=z_{i}^{*}\mid \mathbf{x}_{i2}, u_{i2})}{P(Z_{i1}=z_{i}^{*}\mid \mathbf{x}_{i1}, u_{i1})P(Z_{i2}=z_{i}^{**}\mid \mathbf{x}_{i2}, u_{i2})}+1} \\
   &=\frac{\frac{\xi(\mathbf{x}_{i1}, u_{i1}) \eta (z_{i}^{**},\mathbf{x}_{i1}) \exp\{\gamma \phi(z_{i}^{**}) u_{i1} \}\xi(\mathbf{x}_{i2}, u_{i2}) \eta (z_{i}^{*},\mathbf{x}_{i2}) \exp\{\gamma \phi(z_{i}^{*}) u_{i2} \}}{\xi(\mathbf{x}_{i1}, u_{i1}) \eta (z_{i}^{*},\mathbf{x}_{i1}) \exp\{\gamma \phi(z_{i}^{*}) u_{i1} \}\xi(\mathbf{x}_{i2}, u_{i2}) \eta (z_{i}^{**},\mathbf{x}_{i2}) \exp\{\gamma \phi(z_{i}^{**}) u_{i2} \}}}{\frac{\xi(\mathbf{x}_{i1}, u_{i1}) \eta (z_{i}^{**},\mathbf{x}_{i1}) \exp\{\gamma \phi(z_{i}^{**}) u_{i1} \}\xi(\mathbf{x}_{i2}, u_{i2}) \eta (z_{i}^{*},\mathbf{x}_{i2}) \exp\{\gamma \phi(z_{i}^{*}) u_{i2} \}}{\xi(\mathbf{x}_{i1}, u_{i1}) \eta (z_{i}^{*},\mathbf{x}_{i1}) \exp\{\gamma \phi(z_{i}^{*}) u_{i1} \}\xi(\mathbf{x}_{i2}, u_{i2}) \eta (z_{i}^{**},\mathbf{x}_{i2}) \exp\{\gamma \phi(z_{i}^{**}) u_{i2} \}}+1}       \\
   &=\frac{\frac{\exp\{\gamma \phi(z_{i}^{**})u_{i1}\}\exp\{\gamma \phi(z_{i}^{*})u_{i2}\}}{\exp\{\gamma \phi(z_{i}^{*})u_{i1}\}\exp\{\gamma \phi(z_{i}^{**})u_{i2}\}}  }{\frac{\exp\{\gamma \phi(z_{i}^{**})u_{i1}\}\exp\{\gamma \phi(z_{i}^{*})u_{i2}\}}{\exp\{\gamma \phi(z_{i}^{*})u_{i1}\}\exp\{\gamma \phi(z_{i}^{**})u_{i2}\}}+1}\\
   &=\frac{\exp\{\gamma(\phi(z_{i}^{**})-\phi(z_{i}^{*}))(u_{i1}-u_{i2})\} }{\exp\{\gamma(\phi(z_{i}^{**})-\phi(z_{i}^{*}))(u_{i1}-u_{i2})\}+1}.
\end{align*}
Therefore, the resulting sensitivity bounds depend on the pair-specific dose contrast through $\phi(z_{i}^{**})-\phi(z_{i}^{*})$, but not on the nuisance functions $\xi$ and $\eta$.

By contrast, $\phi$ remains in the sensitivity model and therefore must be specified by the analyst. In practice, $\phi$ should be chosen based on subject-matter considerations about how an unmeasured confounder is expected to affect treatment assignment across the dose scale. For example, $\phi(z)=z$ is a natural default when hidden bias is believed to act approximately linearly on the original dose scale (\citealp{rosenbaum1989sensitivity, gastwirth1998dual, zhang2024sensitivity}), while nonlinear choices of $\phi$, such as logarithmic, square root, or polynomial transformations, may be more appropriate when the relationship between the dose assignment mechanism and unmeasured confounding is thought to be nonlinear. Although the observed data may help suggest scientifically meaningful dose scales or transformations, $\phi$ is not identifiable from the observed data alone without additional assumptions. We therefore recommend prespecifying $\phi$ using domain knowledge and, when appropriate, reporting sensitivity analyses under several plausible choices of $\phi$; our proposed methods can be applied to any such prespecified $\phi$.

\subsection*{D.2: Extension of the Rosenbaum Sensitivity Model to More General Functional Forms Allowing for $X$-$U$ Interactions}

Recall that the original Rosenbaum sensitivity model \citep{rosenbaum1989sensitivity, rosenbaum2002observational, gastwirth1998dual} takes the form
\begin{equation*}
    P(Z=z \mid \mathbf{x}, u)=\xi(\mathbf{x}, u) \eta (z,\mathbf{x}) \exp\{\gamma \phi(z) u \}, \quad{} u \in [0,1], \quad{} \gamma \geq 0,
\end{equation*}
where $\xi(\mathbf{x}, u)$ is an unknown normalizing constant, $\eta(z,\mathbf{x})$ is an unknown function, and $\phi(z)$ is a prespecified dose link function. As shown in Appendix D.1, this model yields
\begin{align*}
  P(Z_{i1}=z_{i}^{**}, Z_{i2}=z^{*}_{i} \mid \mathcal{F}, \mathcal{Z}) =\frac{\exp\{\gamma(\phi(z_{i}^{**})-\phi(z_{i}^{*}))(u_{i1}-u_{i2})\} }{\exp\{\gamma(\phi(z_{i}^{**})-\phi(z_{i}^{*}))(u_{i1}-u_{i2})\}+1},
\end{align*}
and hence the generalized Rosenbaum sensitivity bounds
\begin{equation}\label{eqn: Rosenbaum bounds}
    \frac{1}{1+\Gamma_{i}} \leq P(Z_{i 1}=z_{i}^{* *}, Z_{i 2}=z_{i}^{*} \mid \mathcal{F}, \mathcal{Z}) \leq \frac{\Gamma_{i}}{1+\Gamma_{i}}, \text{ where } \Gamma_i=\exp\{\gamma|\phi(z_{i}^{**})-\phi(z_{i}^{*})|\} \geq 1,
\end{equation}
which form the basis for subsequent sensitivity analyses, such as calculating the worst-case $p$-value subject to (\ref{eqn: Rosenbaum bounds}).

In principle, the corresponding post-matching sensitivity bounds can be derived for any sensitivity model. For example, the most general functional form of a sensitivity model can be written as
\begin{equation}\label{eqn: general sensitivity model}
   P(Z=z \mid \mathbf{x}, u)=\varsigma(z, \mathbf{x}, u),
\end{equation}
where $\varsigma$ is a positive function of the treatment dose $z$, the measured confounders $\mathbf{x}$, and the normalized unmeasured confounder $u$. It then follows that
\begin{align*}
  P(Z_{i1}=z_{i}^{**}, Z_{i2}=z^{*}_{i} \mid \mathcal{F}, \mathcal{Z}) =\frac{\varsigma(z_{i}^{**}, \mathbf{x}_{i1}, u_{i1})\varsigma(z_{i}^{*}, \mathbf{x}_{i2}, u_{i2}) }{\varsigma(z_{i}^{**}, \mathbf{x}_{i1}, u_{i1})\varsigma(z_{i}^{*}, \mathbf{x}_{i2}, u_{i2}) + \varsigma(z_{i}^{*}, \mathbf{x}_{i1}, u_{i1})\varsigma(z_{i}^{**}, \mathbf{x}_{i2}, u_{i2}) }. 
\end{align*}
Conditional on matching ($\mathbf{x}_{i1}=\mathbf{x}_{i2}$), this yields the following sensitivity bounds:
\begin{equation}\label{eqn: most general sensitivity bounds}
\frac{1}{1+\Gamma^{\varsigma}_{i}}  \leq   P(Z_{i1}=z_{i}^{**}, Z_{i2}=z^{*}_{i} \mid \mathcal{F}, \mathcal{Z}) \leq \frac{\Gamma^{\varsigma}_{i}}{1+\Gamma^{\varsigma}_{i}},
\end{equation}
where
\begin{equation*}
    \Gamma^{\varsigma}_{i}=\sup_{u_{i1}, u_{i2}\in [0,1] } \frac{\varsigma(z_{i}^{**}, \mathbf{x}_{i1}, u_{i1})\varsigma(z_{i}^{*}, \mathbf{x}_{i2}, u_{i2}) }{\varsigma(z_{i}^{*}, \mathbf{x}_{i1}, u_{i1})\varsigma(z_{i}^{**}, \mathbf{x}_{i2}, u_{i2}) }.
\end{equation*}
The advantage of this extension is that it accommodates a broader class of sensitivity models than the Rosenbaum sensitivity model in (\ref{eqn: Rosenbaum bounds}). However, this generality comes at a cost: $\Gamma^{\varsigma}_{i}$ may depend on the measured confounders $\mathbf{x}$ because of potential interaction terms between $\mathbf{x}$ and $u$ (see Remark~\ref{rem: independence of X without interaction} below for details). These terms are also referred to as $X$-$U$ interactions (\citealp{heng2021sharpening}). In this setting, applying the sensitivity bounds in (\ref{eqn: most general sensitivity bounds}) requires careful specification of the structure of the $X$-$U$ interactions. For example, as suggested by \citet{heng2021sharpening}, even if only additive, two-way interaction terms between $\mathbf{x}$ and $u$ are considered, the number of additional sensitivity parameters required is at least equal to the dimension of the measured confounders $\mathbf{x}$. This requirement makes the resulting sensitivity analysis more difficult to conduct and interpret in practice. A more practical way to accommodate $X$-$U$ interactions is to examine only one or two $X$-$U$ interaction terms at a time, introduce one or two corresponding sensitivity parameters, derive the resulting sensitivity bounds, and use these bounds to calculate the corresponding worst-case $p$-values for each set of sensitivity parameters. Within each set, one parameter quantifies the main effect of $u$ on the dose assignment mechanism, while the remaining parameters quantify the effects of the selected $X$-$U$ interaction terms on the dose assignment mechanism; see \citet{heng2021sharpening} for examples.

\begin{remark}\label{rem: independence of X without interaction}

Suppose that the sensitivity model does not allow $X$-$U$ interactions; that is, $P(Z=z \mid \mathbf{x}, u)\propto \eta(z, \mathbf{x}) \zeta (z,u)$ for some functions $\eta(z, \mathbf{x})$ and $\zeta (z,u)$. Conditional on matching on the measured confounders, so that $\mathbf{x}_{i1}=\mathbf{x}_{i2}$, we have
    \begin{align*}
    &\quad \ P(Z_{i 1}=z_{i}^{* *}, Z_{i 2}=z_{i}^{*} \mid \mathcal{F}, \mathcal{Z})\\
    &= \frac{\eta(z_{i}^{**}, \mathbf{x}_{i1}) \zeta (z_{i}^{**},u_{i1})\eta(z_{i}^{*}, \mathbf{x}_{i2}) \zeta (z_{i}^{*},u_{i2})}{\eta(z_{i}^{**}, \mathbf{x}_{i1}) \zeta (z_{i}^{**},u_{i1})\eta(z_{i}^{*}, \mathbf{x}_{i2}) \zeta (z_{i}^{*},u_{i2})+\eta(z_{i}^{*}, \mathbf{x}_{i1}) \zeta (z_{i}^{*},u_{i1})\eta(z_{i}^{**}, \mathbf{x}_{i2}) \zeta (z_{i}^{**},u_{i2})}\\
    &=\frac{\frac{\eta(z_{i}^{**}, \mathbf{x}_{i1}) \zeta (z_{i}^{**},u_{i1})\eta(z_{i}^{*}, \mathbf{x}_{i2}) \zeta (z_{i}^{*},u_{i2})}{\eta(z_{i}^{*}, \mathbf{x}_{i1}) \zeta (z_{i}^{*},u_{i1})\eta(z_{i}^{**}, \mathbf{x}_{i2}) \zeta (z_{i}^{**},u_{i2})}}{\frac{\eta(z_{i}^{**}, \mathbf{x}_{i1}) \zeta (z_{i}^{**},u_{i1})\eta(z_{i}^{*}, \mathbf{x}_{i2}) \zeta (z_{i}^{*},u_{i2})}{\eta(z_{i}^{*}, \mathbf{x}_{i1}) \zeta (z_{i}^{*},u_{i1})\eta(z_{i}^{**}, \mathbf{x}_{i2}) \zeta (z_{i}^{**},u_{i2})}+1}\\
    &=\frac{\frac{ \zeta (z_{i}^{**},u_{i1}) \zeta (z_{i}^{*},u_{i2})}{\zeta (z_{i}^{*},u_{i1})\zeta (z_{i}^{**},u_{i2})}}{\frac{ \zeta (z_{i}^{**},u_{i1}) \zeta (z_{i}^{*},u_{i2})}{\zeta (z_{i}^{*},u_{i1})\zeta (z_{i}^{**},u_{i2})}+1}.
\end{align*}    
Therefore, in the absence of $X$-$U$ interactions, the following sensitivity bounds hold:
\begin{equation}\label{eqn: more general bounds without interactions}
    \frac{1}{1+\Gamma^{\eta, \zeta}_{i}}  \leq   P(Z_{i1}=z_{i}^{**}, Z_{i2}=z^{*}_{i} \mid \mathcal{F}, \mathcal{Z}) \leq \frac{\Gamma^{\eta, \zeta}_{i}}{1+\Gamma^{\eta, \zeta}_{i}},
\end{equation}
where
\begin{equation*}
    \Gamma^{\eta, \zeta}_{i}=\sup_{u_{i1}, u_{i2}\in [0,1] } \frac{ \zeta (z_{i}^{**},u_{i1}) \zeta (z_{i}^{*},u_{i2})}{\zeta (z_{i}^{*},u_{i1})\zeta (z_{i}^{**},u_{i2})}.
\end{equation*}
That is, when there are no $X$-$U$ interactions, the sensitivity bounds in (\ref{eqn: more general bounds without interactions}) do not depend on the measured confounders $\mathbf{x}$. In contrast, under the general sensitivity model (\ref{eqn: general sensitivity model}) associated with the general sensitivity bounds (\ref{eqn: most general sensitivity bounds}), the measured confounders $\mathbf{x}$ generally do not cancel when the sensitivity bounds are calculated.
\end{remark}

\subsection*{D.3: Relaxation of the Bijectivity and Monotonicity Requirements of the Dose Link Function $\phi$}

For notational simplicity, we assume in the main text that the dose link function $\phi$ in the Rosenbaum sensitivity model is bijective and monotonic. However, most of the methods and results developed in the paper require neither bijectivity nor monotonicity of $\phi$, although additional mild regularity conditions on $\phi$ may be needed to rule out certain degenerate cases. We discuss the relevant settings in turn:
\begin{itemize}
    \item For any prespecified dose link function $\phi$, the generalized Rosenbaum sensitivity bounds continue to hold:
    \begin{equation*}
    \frac{1}{1+\Gamma_{i}} \leq P(Z_{i 1}=z_{i}^{* *}, Z_{i 2}=z_{i}^{*} \mid \mathcal{F}, \mathcal{Z}) \leq \frac{\Gamma_{i}}{1+\Gamma_{i}}, 
    \end{equation*}
    where $\Gamma_i=\exp\{\gamma|\phi(z_{i}^{**})-\phi(z_{i}^{*})|\} \geq 1$. Based on these sensitivity bounds, for any prespecified sensitivity parameter $\gamma>0$, the procedures for conducting the sensitivity analysis, such as calculating the worst-case $p$-value, remain unchanged and require neither bijectivity nor monotonicity of the dose link function $\phi$.

    \item To establish a one-to-one correspondence between the original sensitivity parameter $\gamma$ and the new sensitivity parameter $\overline{\Gamma}=I^{-1}\sum_{i=1}^{I}\Gamma_{i}=I^{-1}\sum_{i=1}^{I}\exp\{\gamma|\phi(z_{i}^{**})-\phi(z_{i}^{*})|\}$, neither bijectivity nor monotonicity of the dose link function $\phi$ is required, provided that there exists at least one matched pair $i$ such that $\phi(z_{i}^{**})-\phi(z_{i}^{*})\neq 0$. Under this condition, $\overline{\Gamma}$ is a bijective function of $\gamma$; equivalently, there is a one-to-one correspondence between $\gamma$ and $\overline{\Gamma}$. Similarly, provided that there exists at least one matched pair $i$ such that $\phi(z_{i}^{**})-\phi(z_{i}^{*})\neq 0$, there is a one-to-one correspondence among $\gamma$, $\overline{\Gamma}$, and the bias vector $(\Gamma_{1},\dots, \Gamma_{I})$. Therefore, Proposition~3.2 continues to hold.

    \item By similar reasoning, neither bijectivity nor monotonicity of $\phi$ is required when deriving the asymptotic properties of the individual and adaptive test statistics, including generalized design sensitivity and Bahadur-Rosenbaum efficiency, provided that the corresponding regularity conditions in Appendices~A.3--A.5 continue to hold. In particular, the test-specific non-degeneracy condition $P\big(|\phi(z^{**}_{i})-\phi(z^{*}_{i})|>0,\psi_i^{*}>0\big)>0$ ensures the strict monotonicity required in the generalized design sensitivity argument. The remaining regularity conditions remain unchanged.

\end{itemize}

\subsection*{D.4: The Necessity of Specifying the Functional Form of $\phi$ for Non-Binary Treatments}

When the treatment is binary, there is no need to specify the functional form of $\phi$ because $\phi$ can be incorporated into a sensitivity parameter that constrains the magnitude of unmeasured confounding bias. Specifically, when the treatment variable $Z$ is binary and $\phi(1)\neq \phi(0)$, we have 
\begin{align*}
  P(Z_{i1}=1, Z_{i2}=0 \mid \mathcal{F}, \mathcal{Z}) =\frac{\exp\{\gamma(\phi(1)-\phi(0))(u_{i1}-u_{i2})\} }{\exp\{\gamma(\phi(1)-\phi(0))(u_{i1}-u_{i2})\}+1},
\end{align*}
which implies that 
\begin{equation}\label{eqn: sensitivity bounds for binary phi}
   \frac{1}{1+\exp\{\gamma|\phi(1)-\phi(0)|\}}\leq  P(Z_{i1}=1, Z_{i2}=0 \mid \mathcal{F}, \mathcal{Z})\leq \frac{\exp\{\gamma|\phi(1)-\phi(0)|\} }{\exp\{\gamma|\phi(1)-\phi(0)|\}+1}.
\end{equation}
Thus, the functional form of $\phi$ need not be specified separately. Instead, the composite quantity $\gamma|\phi(1)-\phi(0)|$ can be treated directly as a sensitivity parameter because it uniquely determines the magnitude of unmeasured confounding bias represented by the sensitivity bounds in (\ref{eqn: sensitivity bounds for binary phi}) for all matched pairs. 

In contrast, for more general treatment types, such as continuous treatments, the exact functional form of $\phi$ typically needs to be prespecified as a component of the sensitivity model for $P(Z=z\mid \mathbf{x}, u)$. Specifically, for continuous treatments, recall that 
\begin{align*}
  P(Z_{i1}=z_{i}^{**}, Z_{i2}=z^{*}_{i} \mid \mathcal{F}, \mathcal{Z}) =\frac{\exp\{\gamma(\phi(z_{i}^{**})-\phi(z_{i}^{*}))(u_{i1}-u_{i2})\} }{\exp\{\gamma(\phi(z_{i}^{**})-\phi(z_{i}^{*}))(u_{i1}-u_{i2})\}+1},
\end{align*}
which implies that
\begin{align*}
 \frac{1 }{1+\exp\{\gamma|\phi(z_{i}^{**})-\phi(z_{i}^{*})|\}}\leq  P(Z_{i1}=z_{i}^{**}, Z_{i2}=z^{*}_{i} \mid \mathcal{F}, \mathcal{Z}) \leq \frac{\exp\{\gamma|\phi(z_{i}^{**})-\phi(z_{i}^{*})|\} }{\exp\{\gamma|\phi(z_{i}^{**})-\phi(z_{i}^{*})|\} +1}.
\end{align*}
Because $\phi(z_{i}^{**})-\phi(z_{i}^{*})$ may vary across matched pairs, the full collection of sensitivity bounds generally cannot be summarized by a single parameter involving $\gamma$ and $\phi$. Instead, these sensitivity bounds depend on the exact functional form of $\phi$.

\subsection*{D.5: Extension of the Adaptive Testing Procedure to $K\geq 2$ Candidate Test Statistics}
\label{sec: adaptive K tests}

Although Section~4.1 presents the adaptive procedure for two candidate test statistics, the same idea extends directly to any fixed number $K\geq 2$ of candidate tests. Specifically, consider $K$ candidate test statistics of the form $T_{k}=\sum_{i=1}^{I}q_{k, i}\mathbbm{1}\{D_{i}>0\}, \quad k=1,\ldots,K$, where $q_{k, i}\geq 0$ is the test score contributed by matched pair $i$ to the $k$th test statistic. For each matched pair $i$, let $B_{i}^{+}$ be an independent Bernoulli random variable satisfying $P(B_{i}^{+}=1)=p_{I,i}^{+}=\Gamma_i/(1+\Gamma_i)$ and define the bounding statistics $T_{k}^{+}=\sum_{i=1}^{I}q_{k, i}B_{i}^{+}$, $k=1,\ldots,K$. Importantly, the same Bernoulli variable $B_{i}^{+}$ is used to construct the $i$th summand of every bounding statistic $T_{k}^{+}$. For $k=1,\ldots,K$, we have
\begin{align*}
&\mu_{k}^{+}=E(T_{k}^{+}\mid H_{0}, \mathcal{F}, \mathcal{Z}) =\sum_{i=1}^{I}q_{k, i}p_{I,i}^{+},\\
&\sigma_{k}^{+}=\sqrt{\text{Var}(T_{k}^{+}\mid H_{0}, \mathcal{F}, \mathcal{Z})}=\sqrt{\sum_{i=1}^{I}q_{k, i}^{2}p_{I,i}^{+}(1-p_{I,i}^{+})}.
\end{align*}
In addition, for $k_{1}, k_{2}\in \{1,\dots, K\}$, we have
\begin{align*}
\rho_{k_{1}, k_{2}}^{*}&=\text{corr}\left(\frac{T^{+}_{k_{1}}-\mu^{+}_{k_{1}}}{\sigma^{+}_{k_{1}}},\ \frac{T^{+}_{k_{2}}-\mu^{+}_{k_{2}}}{\sigma^{+}_{k_{2}}} \mid  H_{0}, \mathcal{F},\mathcal{Z} \right)\\
&=\text{corr}\left(T_{k_{1}}^{+},T_{k_{2}}^{+}\mid H_{0}, \mathcal{F},\mathcal{Z}\right)\\
&=\frac{\sum_{i=1}^{I}q_{k_{1}, i}q_{k_{2},i}p_{I,i}^{+}(1-p_{I,i}^{+})}
{\sqrt{\sum_{i=1}^{I}q_{k_{1}, i}^{2}p_{I,i}^{+}(1-p_{I,i}^{+})}
\sqrt{\sum_{i=1}^{I}q_{k_{2},i}^{2}p_{I,i}^{+}(1-p_{I,i}^{+})}}.
\end{align*}
Let $\Sigma^{*}=(\rho_{k_{1}, k_{2}}^{*})_{K\times K}$ denote the $K\times K$ matrix whose $(k_{1},k_{2})$-th entry is $\rho_{k_{1}, k_{2}}^{*}$, and let
\begin{align*}
(X_{1},\ldots,X_{K})
\sim \mathcal{N}(\mathbf{0}_{1\times K}, \Sigma^{*}),
\end{align*}
and define $Q_{\Sigma^{*},\alpha}$ as the quantile satisfying
\begin{align*}
P\left(
X_{1}\leq Q_{\Sigma^{*},\alpha},
\ldots,
X_{K}\leq Q_{\Sigma^{*},\alpha}
\right)=1-\alpha.
\end{align*}
The adaptive procedure combining the $K$ candidate test statistics is then defined as follows:
\begin{equation}
\label{eqn: adaptive K tests}
\text{we reject $H_{0}$ iff }
\max_{1\leq k\leq K}
\frac{T_{k}-\mu_{k}^{+}}{\sigma_{k}^{+}}
\geq Q_{\Sigma^{*},\alpha}.
\end{equation}
When $K=2$, procedure (\ref{eqn: adaptive K tests}) reduces to the adaptive procedure presented in the main text. The validity argument used in the proof of Theorem~4.1 extends directly to this more general setting. In particular, the coupling argument in Appendix~A.5 applies to any fixed $K$ because all $K$ test statistics share the common term $\mathbbm{1}\{D_{i}>0\}$ and differ only in their pair-specific test scores $q_{k, i}$. It follows that procedure (\ref{eqn: adaptive K tests}) asymptotically controls the Type I error rate in sensitivity analysis. Likewise, the asymptotic argument used in the proof of Theorem~4.2 extends to any fixed $K$, showing that the resulting adaptive procedure attains the largest generalized design sensitivity and the largest Bahadur-Rosenbaum efficiency among the $K$ candidate test statistics.

\subsection*{D.6: Remark on the Trade-off Induced by Increasing Within-Pair Encouragement Dose Differences}

Previous studies have argued that increasing the within-pair difference in encouragement doses can improve robustness by increasing the within-pair compliance rate, defined as the probability that the individual receiving the higher encouragement dose takes the treatment while the individual receiving the lower encouragement dose does not. If the magnitude of within-pair unmeasured confounding remains fixed, a higher compliance rate can indeed improve robustness to unmeasured confounding. However, as pointed out by \citet{heng2023instrumental}, increasing the within-pair encouragement dose difference may also increase the within-pair difference in unmeasured confounders. In particular, unmeasured confounders associated with the encouragement dose may become increasingly imbalanced as the dose difference grows. A larger encouragement dose difference may therefore amplify the hypothetical unmeasured confounding bias considered in the sensitivity analysis. Consequently, increasing the within-pair difference in encouragement doses improves robustness only when the gain in compliance rate outweighs the accompanying increase in unmeasured confounding bias.

\subsection*{D.7: Relation to the Binary-Treatment Adaptive Approach in \citet{rosenbaum2012testing}}

The adaptive principle underlying the proposed procedure in Section~4 of the main text is inherited from \citet{rosenbaum2012testing}: multiple correlated sensitivity analyses are jointly calibrated, and the null hypothesis is rejected when any standardized component statistic exceeds a common critical value. Our contribution does not simply consist of allowing the test scores to depend on observed doses, as \citet{rosenbaum2012testing} also considered dose-weighted statistics. Rather, our contributions to adaptive testing are threefold. First, we extend the underlying sensitivity analysis construction from binary treatments to general treatment types, under which the observed treatment dose contrasts induce heterogeneous, pair-specific sensitivity bounds. Second, using the newly derived formulas for generalized design sensitivity and generalized Bahadur-Rosenbaum efficiency, we characterize the asymptotic robustness of the resulting adaptive procedure. Third, our theoretical and numerical investigations provide new insights into how the dose-response relationship affects a test's robustness to unmeasured confounding and how an adaptive procedure can reduce the risk of selecting a poorly performing test statistic.

In addition, we develop a coupling argument that provides a unified and rigorous justification for the asymptotic validity of the proposed adaptive procedure under general treatment types. For binary treatments, this argument also fills a technical gap in the original validity argument for the adaptive procedure of \citet{rosenbaum2012testing}. Specifically, the coupling places the component test statistics $T_{1}$ and $T_{2}$ and their corresponding upper-bounding test statistics $T^{+}_{1}$ and $T^{+}_{2}$ on a common probability space and establishes that $(T^{+}_{1},T^{+}_{2})$ jointly stochastically dominates $(T_{1},T_{2})$. Because the adaptive rejection region is coordinatewise non-decreasing, this joint stochastic domination is the key step in establishing asymptotic Type~I error rate  control. Appendix~A.5 provides the detailed argument.

\subsection*{D.8: Additional Remarks on the Sensitivity Parameters $\gamma$ and $\overline{\Gamma}$}

We first clarify the interpretations and complementary roles of $\gamma$ and $\overline{\Gamma}$. For matched pair $i$, let $\pi_{i1}=P(Z_{i 1}=z_{i}^{* *}, Z_{i 2}=z_{i}^{*} \mid \mathcal{F}, \mathcal{Z})$ denote the conditional probability that individual $i1$ receives the larger within-pair dose $z_i^{**}$ and individual $i2$ receives the smaller within-pair dose $z_i^*$, and let $\pi_{i2}=1-\pi_{i1}=P(Z_{i 1}=z_{i}^{*}, Z_{i 2}=z_{i}^{**} \mid \mathcal{F}, \mathcal{Z})$. Under the Rosenbaum sensitivity model, we have
$$
\log\left(\frac{\pi_{i1}}{\pi_{i2}}\right)
=\log\left(\frac{\pi_{i1}}{1-\pi_{i1}}\right)
=
\gamma\{\phi(z_i^{**})-\phi(z_i^*)\}(u_{i1}-u_{i2}),
$$
which implies that
$$
\frac{1}{\Gamma_i}
\leq
\frac{\pi_{i1}}{\pi_{i2}}
\leq
\Gamma_i,
\quad
\text{where $\Gamma_i=\exp\{\gamma|\phi(z_i^{**})-\phi(z_i^*)|\}\geq 1$}.
$$
Thus, $\Gamma_i$ retains a direct pair-specific sensitivity bound interpretation: it is the largest factor by which unmeasured confounding is allowed to tilt the conditional odds of assigning the two observed within-pair doses one way rather than the reverse within matched pair $i$. The proposed post-matching sensitivity parameter
$$
\overline{\Gamma}
=
\frac{1}{I}\sum_{i=1}^{I}\Gamma_i
=
\frac{1}{I}\sum_{i=1}^{I}
\exp\{\gamma|\phi(z_i^{**})-\phi(z_i^*)|\}
$$
is the mean value of these pair-specific sensitivity bounds in the realized matched dataset. Unlike the classical binary-treatment sensitivity parameter, $\overline{\Gamma}$ does not assume that the conditional assignment odds in every matched pair are bounded by the same factor. Instead, it provides a scalar summary of the heterogeneous pair-specific sensitivity bounds induced by a common value of $\gamma$ and the realized matched-pair dose differences. When the treatment is binary and the dose link function $\phi$ is the identity function, $|\phi(z_i^{**})-\phi(z_i^*)|=1$ for every pair, so that $\Gamma_i=\exp(\gamma)$ for all $i$ and $\overline{\Gamma}=\exp(\gamma)=\Gamma$;  therefore, the usual interpretation of the classical Rosenbaum sensitivity parameter is recovered exactly.

The population-level (pre-matching) sensitivity parameter $\gamma$ instead measures the maximum absolute tilt in the log odds of dose assignment per unit of the dose link contrast $|\phi(z_i^{**})-\phi(z_i^*)|$, given the normalization $u\in[0,1]$. Unlike $\overline{\Gamma}$, $\gamma$ is specified without reference to the realized matched-pair dose differences.

The two sensitivity parameters $\gamma$ and $\overline{\Gamma}$ therefore provide complementary information and interpretations: $\gamma$ characterizes the strength of unmeasured confounding in the population-level dose assignment model, while $\overline{\Gamma}$ summarizes how that level of confounding translates into pair-specific sensitivity bounds for the realized matched-pair treatment dose differences.

The distinction between $\gamma$ as a population-level sensitivity parameter and $\overline{\Gamma}$ as a post-matching, design-dependent sensitivity parameter also determines which parameter is most appropriate for different types of comparisons. When comparing different test statistics applied to the same matched dataset or under the same data-generating process, the bijective mapping (i.e., one-to-one correspondence) between $\gamma$ and $\overline{\Gamma}$ is common to all test statistics and is strictly increasing. Consequently, comparisons of sensitivity values, generalized design sensitivities, or generalized Bahadur-Rosenbaum efficiencies may be made on either the $\gamma$-scale or the $\overline{\Gamma}$-scale and yield the same ordering. In this setting, $\overline{\Gamma}$ is particularly useful for describing sensitivity to unmeasured confounding within a study and for comparing the robustness of competing test statistics applied to \textit{the same data}. Nevertheless, we also recommend reporting the corresponding value of $\gamma$ to provide a population-level, data-invariant calibration of the post-matching, data-dependent value of $\overline{\Gamma}$.

By contrast, when comparisons involve different studies or datasets, the mapping between $\gamma$ and $\overline{\Gamma}$ generally differs across analyses because it depends on the realized treatment dose differences. Consequently, the same value of $\overline{\Gamma}$ in two studies may correspond to different values of $\gamma$ and different vectors of pair-specific sensitivity parameters $(\Gamma_1,\ldots,\Gamma_I)$. Therefore, two studies with different dose differences but the same $\overline{\Gamma}$ should not be regarded as allowing the same strength of population-level unmeasured confounding. Instead, the value of $\overline{\Gamma}$ in each study should be mapped back to its corresponding value of $\gamma$ using that study's realized dose differences. For cross-study or cross-dataset comparisons, the resulting values of $\gamma$ are more appropriate, provided that the studies use the same dose link $\phi$ and the same treatment dose scale. We therefore recommend reporting both sensitivity parameters for each study: $\gamma$ facilitates comparisons on the population-level scale, while $\overline{\Gamma}$ describes how the population-level sensitivity parameter $\gamma$ translates into pair-specific sensitivity bounds in the realized matched sample.

Finally, because $\Gamma_i$ increases with $|\phi(z_i^{**})-\phi(z_i^*)|$, matched pairs with exceptionally large dose link differences may have correspondingly large $\Gamma_i$ values. For such pairs, the allowable within-pair dose assignment probabilities can approach the upper and lower bounds $\Gamma_i/(1+\Gamma_i)\approx 1$ and $1/(1+\Gamma_i)\approx 0$, respectively. These large pair-specific sensitivity bounds may also substantially influence $\overline{\Gamma}$. This behavior reflects the underlying mechanism of the Rosenbaum sensitivity model: for a fixed $\gamma$, matched pairs with larger dose link contrasts are allowed to exhibit greater bias in their within-pair dose assignment probabilities. 

Importantly, the generalized Rosenbaum-type sensitivity analysis does not replace the pair-specific sensitivity bounds $\Gamma_{i}$ with their average $\overline{\Gamma}$. Given $\overline{\Gamma}$ and the observed dataset, researchers can recover the corresponding $\gamma$ and the full vector of post-matching sensitivity parameters $(\Gamma_1,\ldots,\Gamma_I)$, as shown in Proposition~\ref{prop:sufficiency}. The worst-case $p$-value is then calculated using the pair-specific sensitivity bounds indexed by $(\Gamma_1,\ldots,\Gamma_I)$.

\subsection*{D.9: Rationale for Using the Mean Value of Pair-Specific Post-Matching Sensitivity Bounds $\overline{\Gamma}$ and Some Alternatives}

The post-matching sensitivity parameter $\overline{\Gamma}=I^{-1}\sum_{i=1}^{I}\Gamma_i$ is the arithmetic mean of the pair-specific worst-case odds-ratio bounds (i.e., pair-specific sensitivity bounds) $\Gamma_i$. It neither replaces these pair-specific bounds with a common average bound nor assumes that only the average magnitude of hidden bias is bounded. Given the matched design and dataset, $\overline{\Gamma}$ serves as a scalar index for the population-level sensitivity parameter $\gamma$ and, consequently, for the full vector of pair-specific bounds $(\Gamma_1,\dots,\Gamma_I)$. Thus, the worst-case nature of Rosenbaum-type sensitivity analysis lies in reporting the worst-case (i.e., largest) $p$-value over the post-matching dose assignment probabilities allowed by the vector of pair-specific sensitivity bounds $(\Gamma_1,\ldots,\Gamma_I)$, rather than in whether the vector $(\Gamma_1,\dots,\Gamma_I)$ is indexed by its mean or maximum.

We use the arithmetic mean $\overline{\Gamma}$ as a specific implementation of our framework because it incorporates information from all matched pairs, reduces exactly to the classical sensitivity parameter $\Gamma$ under binary treatment, and is not determined exclusively by a single pair with an extreme realized dose difference. Moreover, under our regularity conditions, it has the natural population-level limit $E[\exp\{\gamma|\phi(Z_{i1})-\phi(Z_{i2})|\}]$, facilitating the development and interpretation of our asymptotic theory.

In addition to $\overline{\Gamma}$, our framework allows researchers to use other summaries of $\Gamma_1,\dots,\Gamma_I$, such as their maximum or a fixed quantile, because these summaries are in one-to-one correspondence with $\overline{\Gamma}$ for a given matched design and dataset. Specifically, given the observed dataset and the population-level (pre-matching) sensitivity parameter $\gamma\geq 0$, let $\Gamma_i(\gamma)=\exp\{\gamma|\phi(z_i^{**})-\phi(z_i^*)|\}$ and, correspondingly, $\Gamma_{\text{max}}(\gamma)=\max_{i=1,\dots,I}\Gamma_i(\gamma)$. More generally, let $\Gamma_{\alpha}(\gamma)$ denote the $\alpha$-quantile of $\Gamma_1(\gamma),\dots,\Gamma_I(\gamma)$. When $\phi$ is nondegenerate after matching, in the sense that $\phi(z_i^{**})\neq\phi(z_i^*)$ for all matched pairs $i$, there is a pairwise one-to-one correspondence among $\gamma$, $\overline{\Gamma}(\gamma)$, $\Gamma_{\text{max}}(\gamma)$, and $\Gamma_{\alpha}(\gamma)$. Consequently, specifying any one of $\overline{\Gamma}(\gamma)$, $\Gamma_{\text{max}}(\gamma)$, or $\Gamma_{\alpha}(\gamma)$ uniquely determines the corresponding $\gamma$ and hence the full vector $(\Gamma_1,\ldots,\Gamma_I)$. The worst-case $p$-value is then calculated using all pair-specific sensitivity bounds indexed by $(\Gamma_1,\ldots,\Gamma_I)$. 

In addition, generalized design sensitivity and Bahadur-Rosenbaum efficiency on the $\gamma$-scale or the $\overline{\Gamma}$-scale can likewise be translated to the corresponding $\Gamma_{\text{max}}$-scale or $\Gamma_{\alpha}$-scale. For example, suppose that the continuous treatment dose $Z$ takes values in $[0,1]$, possibly after normalization, and that $\Gamma_{\max}(\gamma)\xrightarrow{a.s.}\Gamma^{\diamond}_{\max}(\gamma)$ and $\Gamma_{\alpha}(\gamma)\xrightarrow{a.s.}\Gamma^{\diamond}_{\alpha}(\gamma)$ as $I\rightarrow\infty$. If $\gamma_*$ is the generalized design sensitivity on the $\gamma$-scale derived in Theorem 3.4 in the main text, then the corresponding generalized design sensitivities on the $\Gamma_{\text{max}}$-scale and the $\Gamma_{\alpha}$-scale are $\Gamma^{\diamond}_{\max}(\gamma_*)$ and $\Gamma^{\diamond}_{\alpha}(\gamma_*)$, respectively.

\subsection*{D.10: Summary}

This paper develops a unified framework for conducting sensitivity analyses in matched observational studies with arbitrary treatment types. The framework accommodates two complementary sensitivity parameterizations: the population-level parameter $\gamma$ and the post-matching (data-specific) parameter $\overline{\Gamma}$. Their one-to-one correspondence allows sensitivity analyses to be conducted equivalently on either scale. Specifically, we show that dichotomizing a non-binary treatment can yield misleading sensitivity analyses, develop a principled approach that operates directly on the original treatment scale, and generalize design sensitivity and Bahadur-Rosenbaum efficiency to compare test statistics under both binary and non-binary treatments. We also propose an adaptive procedure that combines candidate tests and asymptotically attains the greatest robustness among them without requiring knowledge of the underlying data-generating process. Overall, our work fills a critical gap in the literature and extends the theoretical foundations of sensitivity analysis beyond the binary-treatment paradigm.

As an illustrative application of our framework, we address a longstanding question in matched observational studies with continuous treatments: Does incorporating treatment dose information into test statistics necessarily enhance robustness to unmeasured confounding? Contrary to conventional wisdom, our theoretical and numerical results reveal that incorporating dose information can either increase or decrease robustness, depending on the underlying dose-response relationship. Because this relationship is typically unknown in practice, selecting a suboptimal test statistic can substantially reduce robustness. To mitigate this vulnerability, our adaptive testing procedure combines competing test statistics and asymptotically retains the robustness of the best-performing candidate test statistic, thereby improving protection against unmeasured confounding in general non-binary treatment settings.

\newpage
\bibliographystyle{apalike}
\bibliography{reference}

\end{document}